\documentclass[preprint,12pt]{elsarticle}

\usepackage{amssymb}
\usepackage{amsmath}
\usepackage{graphicx}
\usepackage{subcaption}
\usepackage{float}
\usepackage{hyperref} 
\usepackage{color} 
\usepackage{scalerel,xparse}
\usepackage[normalem]{ulem}
\usepackage{amsthm}
\usepackage{ragged2e}
\newtheorem{Remark}{Remark}[section] 
\newcommand{\fr}{\displaystyle \frac} 

\journal{}

\begin{document}

\begin{frontmatter}


\title{Accurate contact discontinuity resolving Boltzmann schemes based on peculiar velocity}
\tnotetext[label1]{corresponding author}
\author[label2]{Balwinder Singh\corref{label1}}
\ead{balwinders@iisc.ac.in}
\author[label2]{S. V. Raghurama Rao}
\affiliation[label2]{organization={Department of Aerospace Engineering},
            addressline={Indian Institute of Science},
            city={Bengaluru},
            postcode={560012},
            country={India}}






\begin{abstract}
Starting from the robust foundation of the peculiar velocity based upwind (PVU) Boltzmann scheme, which fails to recognize contact discontinuities, this paper introduces new variants for the numerical solutions of equations of gas dynamics. A modified partial differential equation (MPDE) analysis of the original scheme reveals that the numerical diffusion coefficient corresponding to the peculiar velocity part does not vanish as the Mach number goes to zero. The numerical discretization of the peculiar velocity term is redesigned so that the numerical diffusion depends only on the bulk fluid velocity across the stationary contact discontinuities, enabling its exact capture in 1D and exact capture of slip surfaces in 2D. The resulting variants achieve low numerical diffusion while preserving the robustness of the original scheme, as demonstrated through several 1D and 2D benchmark compressible flow test cases.
\end{abstract}




\end{frontmatter}



\section{Introduction}
\label{sec1}

In the quest to develop robust and accurate numerical methods for inviscid gas dynamics, kinetic or Boltzmann schemes represent an elegant alternative to macroscopic schemes such as Riemann solvers. Boltzmann schemes are based on the fact that Euler equations can be obtained as suitable moments of the Boltzmann equation whenever the velocity distribution function is a Maxwellian. Because the advective terms in the molecular velocity-based Boltzmann equation are linear, upwinding is simpler to apply, and the resulting schemes do not rely on the eigen-structure of the macroscopic system of equations.  Classical kinetic schemes for the Euler equations, developed from the 1970s onward, include those described in \cite{Sanders_Prendergast}, \cite{Pullin}, \cite{Reitz}, \cite{SMD_NASA_TP_2613}, \cite{SMD_NASA_TP_2583}, \cite{Kaniel}, \cite{Perthame}, \cite{Prendergast_Xu}, \cite{JCMandal_SMD}, and references therein. Among these, the scheme with a compactly supported Maxwellian proved particularly effective, satisfying several important mathematical requirements \cite{Perthame_Book}.  However, a major drawback of these traditional kinetic schemes is their inability to capture steady contact discontinuities and slip surfaces accurately. The present paper focuses on developing new approaches to overcome this limitation.
\par
Most traditional kinetic schemes based on the continuous velocity Boltzmann equation are dynamically consistent, positive ($ L^1 $ stable), entropy stable, and robust in capturing nonlinear discontinuities at high Mach numbers. However, these schemes produce overly diffusive contact discontinuities in 1D and slip surfaces in 2D, and consequently they artificially broaden the laminar boundary layer \cite{vanLeer_1991}. Moreover, kinetic schemes suffer from spurious pressure and velocity oscillations near isolated contact discontinuities \cite{Chen_Jiang}. Thus, there is a clear need to develop a kinetic scheme that simultaneously offers sharp resolution of linear discontinuities, suppresses spurious oscillations near isolated contact discontinuities, and is completely free from the problem of shock instability and carbuncle phenomenon.
\par
For this purpose, we find it convenient to choose the peculiar velocity based upwind (PVU) Boltzmann scheme \cite{SVRRao_SMD} as the foundation. In the original work, the authors separate the fluid and peculiar velocities (and introduce upwinding based on each), thereby leading to a (Zha-Bilgen-type) convection-pressure splitting scheme at the macroscopic level. The original scheme is overly diffusive and cannot recognize any stationary discontinuities. However, the Boltzmann equation rewritten in terms of peculiar velocity provides an elegant framework to treat fluid velocity based flux independently of peculiar velocity based flux. A combination of upwind treatment for the convective flux (associated with fluid velocity) and central treatment for the pressure flux (associated with peculiar velocity) at the Boltzmann level allows an isolated stationary contact discontinuity to be resolved exactly. However, such a discretization is not stable across strong nonlinear discontinuities. In the original PVU method, the peculiar velocity part is also discretized in a fully upwind manner, leading to a stable and robust scheme but lacking the ability to recognize any discontinuities. Thus, the motivation behind this work is to re-design the peculiar velocity based upwind numerical fluxes such that the numerical diffusion corresponding to them switches off across stationary contact discontinuities and remains low but non-zero otherwise. Such schemes are expected to perform better in simulating boundary layers in viscous flows, thereby making traditional kinetic schemes competitive with macroscopic Riemann solvers (such as \cite{Toro_HLLC}) in accuracy.
\par
The remainder of this paper is structured as follows. Section 2 contains a detailed overview of the kinetic model for polyatomic gases. In Section 3, we describe in detail the numerical formulation of the one-dimensional PVU scheme and perform the MPDE analysis to understand the cause of its overly diffusive prediction of stationary contact discontinuities. In Section 4, we select general forms of numerical fluxes for fluid and peculiar velocity terms while keeping a newly introduced diffusion control parameters unspecified. We then propose two variants of the modified PVU scheme by specifying the diffusion control parameters. The second variant captures stationary contacts and slip surfaces without any numerical diffusion. We further enhance the performance of the first variant across stationary contacts by enforcing the isentropy conditions in numerical flux based on the peculiar velocity. In Section 5, we extend the modified PVU scheme to two dimensions. In Section 6 we present a way to extend the modified PVU scheme to second-order accuracy in space and time. Section 7 contains the extension of the proposed scheme for numerical solutions of the system of Navier-Stokes-Fourier equations. In Section 8, the scheme is tested on benchmark one- and two-dimensional gas dynamics test cases. Finally, we arrive at the conclusion that variants 2 and 3 of the modified PVU scheme preserve the robustness of the original scheme while improving accuracy, apart from exactly capturing the grid-aligned steady contact discontinuities.


\section{From kinetic theory to macroscopic fluid dynamics}
\label{sec2}

The kinetic theory of gases describes a flow using the molecular velocity distribution function $ f(\textbf{x}, t, \textbf{v}, I) $ of particles, where $ \textbf{x} $ is the position, $ t $ is time, $ \textbf{v} $ is the molecular velocity, and $ I $ accounts for internal energy corresponding to non-translational degrees of freedom. The additional variable $ I $ in the phase space of the velocity distribution function is necessary to enforce the correct value of $ \gamma $ (ratio of specific heats) for polyatomic gases \cite{Morse}. The Boltzmann equation then governs the evolution of the distribution function, which in $k$ spatial dimensions reads

\begin{equation} \label{Boltzmann_equation}
    \frac{\partial f}{\partial t} + \sum_{i = 1}^{k} v_i \frac{\partial f}{\partial x_i} = Q(f,f) \quad t, I \in \Re^+, \textbf{x}, \textbf{v} \in \Re^k.
\end{equation}
$ Q(f,f) $ represents a nonlinear collision operator whose properties are

\begin{equation} \label{moment_collision_operator}
    \int_{I} \, dI \int_{\textbf{v}} \, d\textbf{v} \, \boldsymbol{\psi} \, Q(f,f)  = 0 \quad \forall f,
\end{equation}

\begin{equation} \label{collision_term_when_f_tends_to_feq}
    Q(f,f) = 0 \iff f = f^{eq},
\end{equation}

\begin{equation} \label{collision_term_multiplied_entropy_variable}
    \int_{I} \, dI \int_{\textbf{v}} \, d\textbf{v} \,\ln f \,Q(f,f) \leq 0 \quad \forall f,
\end{equation}
where $ \boldsymbol{\psi} $ denotes a moment vector (representing the conservation of mass, momentum, and energy during molecular collisions) defined as
\begin{equation} \label{moment_vector}
    \boldsymbol{\psi} (\boldsymbol{v}, I) \equiv \left[1, v_j, I + \sum_{i = 1}^k \frac{v_i^2}{2} \right]^{T}, \; j \in \{ 1, 2, ..., k \}.
\end{equation}
Here, $ f^{eq} $ denotes the equilibrium distribution function, namely the Maxwellian associated with local thermodynamic equilibrium at the macroscopic level. Equation (\ref{moment_collision_operator}) is a direct consequence of the conservation of mass, momentum, and energy during molecular collisions. Substituting equation (\ref{Boltzmann_equation}) into (\ref{moment_collision_operator}) yields
\begin{equation} \label{kinetic_equation_for_conservation_laws}
    \int_{I} \, dI \int_{\textbf{v}} \, d\textbf{v} \, \boldsymbol{\psi} \left( \frac{\partial f}{\partial t} + \sum_{i = 1}^{k} v_i \frac{\partial f}{\partial x_i} \right) = 0 \quad \forall f.
\end{equation}
The macroscopic conserved variables density $ \rho $, momentum $ \rho \boldsymbol{u} $, and total energy $ \rho E $ are related to the velocity distribution function as
\begin{equation} \label{definition_of_conserved_variable_vector}
    \textbf{U} = \int_{I} \, dI \int_{\textbf{v}} \, d\textbf{v} \, \, \boldsymbol{\psi} \, f,
\end{equation}
where $ \boldsymbol{U} $ denotes the conserved variable vector, defined by 
\begin{equation} \label{macroscopic_conserved_variable_vector}
        \textbf{U} = \begin{bmatrix}
        \rho & \rho u_j & \rho E 
        \end{bmatrix}^T. \\
\end{equation}
Here, 
\begin{equation} \label{definition_of_E_e}
    E = e + \sum_{i = 1}^{k} \frac{u_i^2}{2}, \; e = \frac{p}{\rho (\gamma - 1)}  
\end{equation} 
and $ p $ denotes the thermodynamic pressure. The shear stress tensor $ \boldsymbol{\tau} $ and the heat flux vector $ \boldsymbol{q} $ are defined with 
\begin{subequations} \label{defintions_of_shear_stress_heat_flux}
    \begin{align}
        p \, \delta_{ij} - \tau_{ij} &\equiv \int_I \, dI \int_{\boldsymbol{v}} \, d\boldsymbol{v} \, c_i c_j f, \\
        q_{i} &\equiv - \int_I \, dI \int_{\boldsymbol{v}} \, d\boldsymbol{v} \, \left( I + \dfrac{|\boldsymbol{c}|^2}{2} \right) c_i f,
    \end{align}
\end{subequations}
where $ \delta_{ij} $ denotes the Kronecker delta and $ \boldsymbol{c} \equiv \boldsymbol{v} - \boldsymbol{u} $ represents the peculiar velocity (relative velocity of the molecules with respect to the macroscopic fluid velocity). Using equations (\ref{definition_of_conserved_variable_vector}) and (\ref{defintions_of_shear_stress_heat_flux}) in (\ref{kinetic_equation_for_conservation_laws}) gives
\begin{equation} \label{tensor_equation_for_macroscopic_conservation_laws}
    \frac{\partial \textbf{U}}{\partial t} + \sum_{i = 1}^{k} \frac{\partial \textbf{G}_i}{\partial x_i} = \sum_{i = 1}^{k} \frac{\partial \textbf{G}^{(v)}_i}{\partial x_i},
\end{equation}
where $ \textbf{G}_i $ and $ \textbf{G}^{(v)}_i $ represent the $ i^{th} $ component of the $k$ dimensional tensor of convective and viscous fluxes, respectively, yielding 
\begin{subequations} 
    \begin{equation} \label{definition_of_advective_flux}
        \textbf{G}_i = \begin{bmatrix}
        \rho u_i & \rho u_i u_j + p \delta_{i j} & \rho u_i H
        \end{bmatrix}^T 
    \end{equation} 
    and 
    \begin{equation} \label{definition_of_viscous_flux}
        \textbf{G}^{(v)}_i = \begin{bmatrix}
        0 & \tau_{ij} & \tau_{ij} \, u_j - q_i
        \end{bmatrix}^T,
    \end{equation}
\end{subequations}
with
\begin{equation} \label{defintion_H}
    H = E + \frac{p}{\rho}.
\end{equation}
The system of partial differential equations for the inviscid fluid flow, which are the Euler equations, can be obtained as the suitable moments ($ \boldsymbol{\psi} \text{-moments} $) of the Boltzmann equation of the kinetic theory of gases in the limit of the appropriate local thermodynamic equilibrium.  
\begin{equation} \label{Boltzmann_to_Euler}
    \int_{I} \, dI \int_{\textbf{v}} \, d\textbf{v} \, \boldsymbol{\psi} \left( \frac{\partial f}{\partial t} + \sum_{i = 1}^{k} v_i \frac{\partial f}{\partial x_i} = 0, f = f^{eq} \right) \Leftrightarrow \frac{\partial \textbf{U}}{\partial t} + \sum_{i = 1}^{k} \frac{\partial \textbf{G}_i}{\partial x_i} = 0,
\end{equation}
where $ f^{eq}(\textbf{x}, t, \textbf{v}, I) $ represents a Maxwellian distribution function defined as
\begin{equation} \label{Maxwellian}
    f^{eq} = \frac{\rho}{I_0} \left(\frac{\beta}{\pi} \right)^{D_{f}/2} \exp \left( -\beta (\textbf{v} - \textbf{u})^2 - \frac{I}{I_0} \right),
\end{equation}
\begin{equation} \label{I_0_and_beta}
    I_0 = \frac{(2 + D_f) - \gamma D_f}{4 (\gamma - 1) \beta}, \, \beta = \frac{1}{2 R T}.
\end{equation}
Here, $ D_f $ denotes the number of translational degrees of freedom, $ T $ denotes the temperature, and $ R $ is the gas constant. The shear stress tensor and the heat flux vector vanish in the limit $ f \rightarrow f^{eq} $. If one chooses the Chapman-Enskog distribution function $ f^{CE} $ in equation (\ref{kinetic_equation_for_conservation_laws}), then the $ \boldsymbol{\psi} \text{-moments} $ of the Boltzmann equation yield the system of Navier-Stokes-Fourier equations, \textit{i.e.,} 
\begin{equation} \label{Boltzmann_to_NSF}
    \int_{I} \, dI \int_{\textbf{v}} \, d\textbf{v} \, \boldsymbol{\psi} \left( \frac{\partial f}{\partial t} + \sum_{i = 1}^{k} v_i \frac{\partial f}{\partial x_i} = 0, f = f^{CE} \right) \Leftrightarrow \frac{\partial \textbf{U}}{\partial t} + \sum_{i = 1}^{k} \frac{\partial G^{T}_{i}}{\partial x_i} = 0, 
\end{equation} 
with 
\begin{equation} 
G^{T}_{i} = \left( \textbf{G}_i - \textbf{G}^{(v)}_i \right)
\end{equation} 
being the total (inviscid + viscous) fluxes.
Here,
\begin{equation}
    f^{CE} = f^{eq} (1 + P_{CE}) = f^{eq} + f^{neq},
\end{equation}
and $ f^{neq} = f^{eq} P_{CE} $ denotes the first correction to the leading order term in the asymptotic expansion of the distribution function. The shear stress tensor and the heat flux vector assume the following form in equation (\ref{definition_of_viscous_flux}):
\begin{equation}
    \tau_{ij} = \mu \left( \dfrac{\partial u_j}{\partial x_i} + \dfrac{\partial u_i}{\partial x_j} \right) - \dfrac{2}{3} \mu \dfrac{\partial u_m}{\partial x_m}, \, q_i = -k \dfrac{\partial T}{\partial x_i},
\end{equation}
where $ \mu, k $ denote the coefficients of dynamic viscosity and thermal conductivity, respectively. We refer to \cite{SMD_NASA_TP_2613} for more details on the Chapman-Enskog distribution function. In addition, the $ H $-function inequality augments the Boltzmann equation to ensure the uniqueness of the solution. For a monatomic gas, the $ H $-function inequality can be derived by multiplying the Boltzmann equation (\ref{Boltzmann_equation}) by $ \ln f $, \textit{i.e.}, 
\begin{equation} \label{entropy_inequality}
    \int_{I} \, dI \int_{\textbf{v}} \, d\textbf{v} \, \ln f \left( \frac{\partial f}{\partial t} + \sum_{i = 1}^{k} v_i \frac{\partial f}{\partial x_i} \right) = \int_{I} \,dI \int_{\textbf{v}} \,d\textbf{v} \ln f \, Q(f,f) \leq 0.
\end{equation}
The inequality above assumes equality in the limit $ f \to f^{eq} $, and its moment gives the macroscopic conservation law for the physical entropy $ s $, with the following relation between the entropy \& entropy flux pair:
\begin{align}
    H &= \int_{I} \, dI \int_{\textbf{v}} \, d\textbf{v} \, f^{eq} \, \ln f^{eq}, = \rho s\\
    H_{v_{i}} &= \int_{I} \, dI \int_{\textbf{v}} \, d\textbf{v} \, v_i \, f^{eq} \, \ln f^{eq} = \rho u_i s \quad \forall i.
\end{align}
Thus, (\ref{entropy_inequality}) leads to the $ H $-inequality 
\begin{equation} 
\frac{\partial H}{\partial t} + \sum_{i} \frac{\partial H_{v_{i}}}{\partial x_i} \leq 0
\end{equation}
which corresponds to the macroscopic entropy conservation equation in smooth regions and the macroscopic entropy inequality when discontinuous regions are included. We refer to \cite{SMD_NASA_TP_2583} for a detailed derivation of the $ H $-function inequality for polyatomic gases.


\section{A peculiar velocity based upwind Boltzmann scheme}
\label{sec3}

Kinetic schemes for the numerical solution of the Euler equations of gas dynamics are constructed by taking $ \boldsymbol{\psi}- $moments of an appropriately discretized Boltzmann equation in the local thermodynamic equilibrium limit, referred to as the {\em moment method strategy} \cite{SMD_NASA_TP_2613}. The peculiar velocity based upwind Boltzmann scheme belongs to the class of kinetic numerical methods. It is derived by taking $ \boldsymbol{\psi} $-moments of the upwind-discretized Boltzmann equation written in terms of peculiar velocity. At the macroscopic level, this yields an upwind scheme based on {\em convection-pressure splitting}.

\subsection{A peculiar velocity based Boltzmann equation}
\label{subsec3.1}

The peculiar velocity $ \boldsymbol{c} $ (also called as {\em thermal velocity} or {\em random velocity}) is defined as the relative velocity of the molecules with respect to the bulk fluid velocity $ \textbf{u} $, \textit{i.e.},
\begin{equation} \label{peculiar_velocity_definition}
    \boldsymbol{c} \equiv \boldsymbol{v} - \boldsymbol{u}.  
\end{equation}
A peculiar velocity based Boltzmann equation is obtained by replacing the molecular velocity with a sum of peculiar and bulk fluid velocities in Equation (\ref{Boltzmann_equation}), which in $ k $ spatial dimensions reads
\begin{equation} \label{peculiar_velocity_based_Boltzmann_equation}
    \frac{\partial f}{\partial t} + \sum_{i = 1}^{k} \frac{\partial}{\partial x_i} (u_i f) + \sum_{i = 1}^{k} \frac{\partial}{\partial x_i} (c_i f) = Q(f,f) \quad t, I \in \Re^+, \boldsymbol{x}, \boldsymbol{c}, \boldsymbol{u} \in \Re^k.
\end{equation}
We take note of some interesting properties of the peculiar velocity based Boltzmann equation. We redefine the moment vector in terms of the peculiar and bulk fluid velocity as
\begin{equation} \label{moment_vector_in_terms_of_peculiar_velocity}
    \boldsymbol{\psi} \left(\boldsymbol{u} (\boldsymbol{x}, t), \boldsymbol{c} (\boldsymbol{x}, t), I\right) = \left[ 1, u_j + c_j, I + \dfrac{1}{2} \sum_{i = 1}^{k} (u_i + c_i)^2 \right]^{T},
\end{equation}
and take moments of (\ref{peculiar_velocity_based_Boltzmann_equation}) in the limit $ f \to f^{eq} $ to obtain
\begin{equation}
    \int_{I} \, dI \int_{\textbf{c}} \, d\textbf{c} \, \left( \frac{\partial}{\partial t} \left( \boldsymbol{\psi} f \right) + \sum_{i = 1}^{k} \frac{\partial}{\partial x_i} (\boldsymbol{\psi} \, u_i f) + \sum_{i = 1}^{k} \frac{\partial}{\partial x_i} (\boldsymbol{\psi} \, c_i f), \, f = f^{eq} \right) = 0.
\end{equation}
Note that $ d\textbf{v} = d\textbf{c} $. The evaluation of integrals in the above equation leads to a system of split Euler equations as
\begin{equation} \label{Zha-Bilgen_type_split_Euler_equations}
    \frac{\partial \textbf{U}}{\partial t} + \sum_{i = 1}^{k} \frac{\partial}{\partial x_i} \textbf{G}_i^{(a)} + \sum_{i = 1}^{k} \frac{\partial}{\partial x_i} \textbf{G}_i^{(p)} = 0,
\end{equation}
where $ \textbf{G}_i^{(a)} $ and $ \textbf{G}_i^{(p)} $ represent the $ i^{th} $ component of the $ k $ dimensional tensor of the advective (or convective) and pressure fluxes, respectively. These are defined by the following moment relations:
\begin{subequations} \label{definitions_of_convective_pressure_fluxes}
    \begin{align}
        \textbf{G}_i^{(a)} &= \int_{I} \, dI \int_{\textbf{c}} \, d\textbf{c} \, \boldsymbol{\psi} \, u_i \, f^{eq} = \begin{bmatrix}
        \rho u_i & \rho u_i u_j & \rho u_i E 
        \end{bmatrix}^T \quad \forall i, \\
        \textbf{G}_i^{(p)} &= \int_{I} \, dI \int_{\textbf{c}} \, d\textbf{c} \, \boldsymbol{\psi} \, c_i \, f^{eq} = \begin{bmatrix}
        0 & p \delta_{i j} & p u_i
        \end{bmatrix}^T \quad \forall i.
    \end{align}
\end{subequations}
This is precisely the {\em convection-pressure splitting} of the macroscopic flux vector, utilized as the basis for numerical schemes by Deshpande {\em et al.} \cite{SMD_Balak_SVRRao} and, with some variations at the macroscopic level, also by Zha \& Bilgen \cite{Zha_Bilgen}, Meng-Sing Liou \cite{Meng-Sing_Liou_Review, Meng-Sing_Liou_Review_2}, Jameson \cite{Jameson_CUSP}, and Toro \& Vazquez-Cendón \cite{Toro_Vazquez}. Thus, the peculiar velocity based Boltzmann equation provides the physical motivation for such flux vector splittings. The peculiar velocity based Boltzmann equation separates the fluid velocity based uni-directional (along the tangent to a streamline) information propagation from the random velocity based multi-directional information propagation (corresponding to the jiggling of the molecules in random directions). It is worth noting here that the entropy flux pertaining to the peculiar velocity is zero, \textit{i.e.}, 
\begin{equation}
    \int_{I} \, dI \int_{\textbf{c}} \, d\textbf{c} \, c_i \, f^{eq} \, \ln f^{eq} = 0 \quad \forall i.
\end{equation}
The peculiar velocity based Boltzmann equation, its associated PVU scheme, and the physical basis of convection-pressure splitting are discussed in detail in \cite{SVRRao_SMD}.

\subsection{Operator splitting strategy}
\label{subsec3.2}
For simplicity, let us consider the 1-D case.  The 1-D Euler equations, in moment form and rewritten with peculiar velocity, are given by  
\begin{equation} \label{1D_PVBE}
    \int_{0}^{\infty} \int_{-\infty}^{\infty} \left( \fr{\partial f}{\partial t} 
    + \fr{\partial \left(u f\right)}{\partial x} 
    + \fr{\partial \left(c f\right)}{\partial x} = Q(f,f)  \right) \ dc \ dv
\end{equation} 
Using the popular Bhatnagar-Gross-Krook (BGK) model \cite{BGK}, the collision term can be simplified as 
\begin{equation} \label{BGK}
    Q(f,f) = - \fr{1}{\epsilon} \left[ f - f^{eq} \right]
\end{equation} 
with the distribution function approaching its equilibrium state over a characteristic relaxation time, $\epsilon$ ($\epsilon \rightarrow 0$). The moment relations (\ref{1D_PVBE}) then take the form 
\begin{equation} \label{PVBE_BGK}
    \int_{0}^{\infty} \int_{-\infty}^{\infty} \left( \fr{\partial f}{\partial t} 
    + \fr{\partial \left(u f\right)}{\partial x} 
    + \fr{\partial \left(c f\right)}{\partial x} = - \fr{1}{\epsilon} \left[ f - f^{eq} \right] 
    \right) \ dc \ dv,
\end{equation} 
which may be expressed as
\begin{equation} \label{Unsplit_PVBE}
    \int_{0}^{\infty} \int_{-\infty}^{\infty} \left( \fr{\partial f}{\partial t} 
    + \mathcal{O}_{S}\left(f\right) = \mathcal{O}_{C} \left(f\right)   
    \right),
\end{equation}
where the {\em convection operator} and the {\em collision operator} are given by 
\begin{equation} 
{\rm \bf Convection \ Operator:} \ \  \mathcal{O}_{S}\left(f\right) = \fr{\partial}{\partial x} \left(u f\right)
    + \fr{\partial}{\partial x} \left(c f\right),
\end{equation} 
and 
\begin{equation}
{\rm \bf Collision \ Operator:} \ \  \mathcal{O}_{C}\left(f\right) = - \fr{1}{\epsilon} \left[ f - f^{eq} \right].
\end{equation} 
We now introduce operator splitting in (\ref{Unsplit_PVBE}) as 
\begin{equation} \label{OS_PVBE}
    \int_{0}^{\infty} \int_{-\infty}^{\infty} \left( 
    \underbrace{\fr{\partial f}{\partial t} 
    + \mathcal{O}_{S}\left(f\right) = 0}_{\rm Convection \ step}, \ 
    \underbrace{\fr{d f}{d t} = \mathcal{O}_{C} \left(f\right)}_{\rm Collision \ step}   
    \right) \ d c \ d I 
\end{equation} 
We can introduce a further simplification in the collision step, by introducing the approximation of  {\em instantaneous relaxation to equilibrium}, {\em i.e.}, taking $\epsilon = 0$. Therefore, $\mathcal{O}_{C}\left(f\right)$ yields $f = f^{eq}$.  Thus, the moment equation (\ref{OS_PVBE})  becomes 
\begin{equation} 
    \int_{0}^{\infty} \int_{-\infty}^{\infty} \left( 
    \underbrace{\fr{\partial f}{\partial t} 
    + \mathcal{O}_{S}\left(f\right) = 0}_{\rm Convection \ step}, \ 
    \underbrace{f = f^{eq}}_{\rm Instantaneous \ relaxation \ step}   
    \right) \ d c \ d I 
\end{equation} 
or 
\begin{equation} \label{OS_IR}
    \int_{0}^{\infty} \int_{-\infty}^{\infty} \left( 
    \underbrace{\fr{\partial f}{\partial t} 
    + \fr{\partial \left(u f\right)}{\partial x} 
    + \fr{\partial \left(c f\right)}{\partial x} = 0}_{\rm Convection \ step}, \ 
    \underbrace{f = f^{eq}}_{\rm Instantaneous \ relaxation \ step}   
    \right) \ d c \ d I 
\end{equation}
Evaluating the moments of Equation (\ref{OS_IR}) yields the one‑dimensional Euler equations in vector form, exhibiting the natural convection–pressure splitting (also referred to as advective–pressure splitting) that arises from the reformulation based on peculiar velocity. These equations, with the convective and pressure fluxes defined in (\ref{Zha-Bilgen_type_split_Euler_equations}), (\ref{definitions_of_convective_pressure_fluxes}), are not repeated here for brevity.

\subsection{Numerical details of the 1D PVU Boltzmann scheme}
\label{subsec3.3}
In this subsection, we outline the numerical details of the one-dimensional, first order peculiar velocity based upwind Boltzmann scheme for the numerical solution of the Euler equations of gas dynamics. A one-dimensional domain is discretized into a set of $ N $ finite volumes (also called cells), with cell centers located at $ x_{j} $, $ j=1, \cdots, N $. Cell boundaries are labeled as $ x_{j \pm \frac{1}{2}} $. Initially, the bulk fluid velocity and the thermodynamic state variables are assumed to be piecewise constant in each finite volume cell, \textit{i.e.},
\begin{subequations} \label{constant_state_distribution}
    \begin{align}
        u(x,0) &= u_j \quad \forall x \in  (x_{j-\frac{1}{2}}, x_{j+\frac{1}{2}}), \\
        \rho(x,0) &= \rho_j \quad \forall x \in  (x_{j-\frac{1}{2}}, x_{j+\frac{1}{2}}), \\
        T(x,0) &= T_j \quad \forall x \in (x_{j-\frac{1}{2}}, x_{j+\frac{1}{2}}).
    \end{align}
\end{subequations}

Consequently, a Maxwell distribution function is constructed within each finite volume cell at time $ t = 0 $, \textit{i.e.},

\begin{subequations} \label{Maxwellian_defined_in_each_cell}
    \begin{align}
        f^{eq}_j(x, c, I) = \frac{\rho_j}{I_{0,j}} \left( \frac{\beta_j}{\pi} \right)^{1/2} &\exp \left( -\beta_j c_j^2 - \frac{I}{I_{0,j}} \right) \quad \forall x \in (x_{j-\frac{1}{2}}, x_{j+\frac{1}{2}}), \\
        I_{0,j} = &\frac{(3 - \gamma)}{4 (\gamma - 1) \beta_j}, \, \beta_j = \frac{1}{2 R T_j}.
    \end{align}
\end{subequations}
The molecular velocity distribution function within each cell is advanced in time using an operator splitting strategy, consisting of two sequential stages: a convection step and a collision step.  
\par
\textbf{Convection step:} In this step, the underlying model represents the movement of particles from one cell to their neighbor cells, through the mathematical representation of the convection terms. We start with the one-dimensional peculiar velocity based Boltzmann equation (without the collision term), which reads 
\begin{equation} \label{1D_collisionless_peculiar_velocity_based_Boltzmann_equation}
    \frac{\partial f}{\partial t} + \frac{\partial}{\partial x} (u f) + \frac{\partial}{\partial x} (c f) = 0 
\end{equation} 
or 
\begin{equation} \label{1D_collisionless_PVBE}
    \frac{\partial f}{\partial t} + \frac{\partial h^{(u)}}{\partial x} + \frac{\partial h^{(c)}}{\partial x} = 0, \ \textrm{where} \ h^{(u)} = u f \ \textrm{and} \ h^{(c)} = c f. 
\end{equation} 
Thus, we obtain 
\begin{equation}
    \frac{\partial f}{\partial t} + \frac{\partial h}{\partial x} = 0, \ \textrm{where} \ h = h^{u} + h^{c}.   
\end{equation} 
We then use a classical upwind scheme for the numerical discretization of convection terms to obtain
\begin{equation} \label{discretized_PV_Boltzmann_equation}
    f_j^{(n + 1)-} = f_j^{(n)} + \sigma \left[ \hat{h}_{j + \frac{1}{2}}^{(n)} - \hat{h}_{j - \frac{1}{2}}^{(n)} \right],
\end{equation} 
\noindent where $ \sigma = \Delta t/\Delta x $ and $ \hat{h}_{j \pm \frac{1}{2}}^{(n)} $ are the numerical fluxes (defined at the kinetic level) at the cell interfaces, which read
\begin{subequations}
    \begin{align}
        \hat{h}_{j - \frac{1}{2}} &= (u_{j - 1}^+ + c_{j - 1}^+) f^{eq}_{j - 1} + (u_{j}^- + c_{j}^-) f^{eq}_{j}, \\
        \hat{h}_{j + \frac{1}{2}} &= (u_j^+ + c_j^+) f^{eq}_j + (u_{j + 1}^- + c_{j + 1}^-) f^{eq}_{j + 1},
    \end{align}
\end{subequations}
where
\begin{equation} 
    \ u^{\pm} = \fr{1}{2} (u\pm |u|) \ \textrm{and} \ c^{\pm} = \fr{1}{2} (c\pm |c|).
\end{equation}
\par
\textbf{Collision step:} After the convection step, the molecular velocity distribution function becomes a union of all the Maxwellians in the considered stencil of points (in a first order scheme, it becomes a union of three Maxwellians on a 3-point stencil [$x_{j-1}, x_{j}, x_{j+1}$]). Further, as it changes in time and space satisfying the convection terms, the distribution function is no longer an equilibrium one. In the collision stage, the effect of particle collisions is taken into account by instantaneously relaxing this distribution function to the new equilibrium distribution function without changing the mass, momentum, and total energy in the cell. It is equivalent to solving the following ordinary differential equation, with the BGK model:
\begin{equation} \label{equation_equivalent_collision_step}
    \frac{d f}{d t} = Q(f,f) = - \fr{1}{\epsilon} \left[ f  - f^{eq} \right] 
\end{equation}
We thus obtain the following.
\begin{equation} \label{essence_of_collision_step}
    \int_{c} \, dc \, \boldsymbol{\psi}_j \, f_j^{eq, (n+1)} = \int_{c} \, dc \, \boldsymbol{\psi}_j \, f_j^{(n + 1)-}.
\end{equation}
Substituting Equation (\ref{discretized_PV_Boltzmann_equation}) into Equation (\ref{essence_of_collision_step}) yields a first-order finite volume discretization of the Euler equations, given by
\begin{equation} \label{1D_finite_volume_update_formula}
    \bar{\boldsymbol{U}}_j^{n + 1} = \bar{\boldsymbol{U}}_j^{n} - \sigma \left( \hat{\textbf{G}}_{j + \frac{1}{2}} - \hat{\textbf{G}}_{j - \frac{1}{2}} \right),
\end{equation}
where $ \bar{\boldsymbol{U}} $ denotes the cell-averaged value of the conserved variable vector in each cell and $ \hat{\textbf{G}}_{j \pm \frac{1}{2}} $ denote the cell-interface numerical fluxes, which are defined as the sum of advective ($ G^{(a)} $) and pressure ($ G^{(p)} $) numerical fluxes. The numerical advective and pressure fluxes are defined, based on upwinding, as follows:
\begin{subequations} \label{definition_advective_pressure_numerical_flux}
    \begin{align}
        \hat{\textbf{G}}_{j + \frac{1}{2}}^{(a)} &= \int_{0}^{\infty} dI \, \int_{-\infty}^{\infty} \, dc \, \left( \boldsymbol{\psi} \, u^+ \, f^{eq} \right)_j + \int_{0}^{\infty} dI \, \int_{-\infty}^{\infty} \, dc \, \left( \boldsymbol{\psi} \, u^- \, f^{eq} \right)_{j + 1}, \\
        \hat{\textbf{G}}_{j + \frac{1}{2}}^{(p)} &= \int_{0}^{\infty} dI \, \int_{-\infty}^{\infty} \, dc \, \left( \boldsymbol{\psi} \, c^+ \, f^{eq} \right)_j + \int_{0}^{\infty} dI \, \int_{-\infty}^{\infty} \, dc \, \left( \boldsymbol{\psi} \, c^- \, f^{eq} \right)_{j + 1},
    \end{align}
\end{subequations}
where 
\begin{equation} \label{Definition_of_PV_psi}
    \boldsymbol{\psi} = \begin{bmatrix}
        1 & \left( u + c \right) & I + \left( \dfrac{u^2}{2} + \dfrac{c^2}{2} + u c \right)
    \end{bmatrix}^T.
\end{equation}
We refer to \cite{SVRRao_SMD} for more details.  The final expressions for the macroscopic split fluxes are given by 
\begin{eqnarray} 
G^{(a)\pm} = u^{\pm} U \ \textrm{and} \ 
G^{(p)\pm} = \left[ \begin{array}{c} 
\pm \fr{\rho}{2 \sqrt{\pi \beta}} \\ 
\fr{p}{2} \pm \fr{\rho u}{2 \sqrt{\pi \beta}} \\ 
\fr{p u}{2} \pm \fr{\rho}{2 \sqrt{\pi \beta}} \left( \fr{p}{2 \rho} + E\right) 
\end{array} \right] 
\end{eqnarray}
\par
\noindent
Note that 
\begin{equation}
    \sqrt{\beta} = \fr{1}{\sqrt{2 R T}} = \fr{\sqrt{\gamma}}{\sqrt{2}} \fr{1}{\sqrt{\gamma R T}} 
    = \fr{\sqrt{\gamma}}{\sqrt{2}} \fr{1}{a} \ \textrm{where} \ a \ \textrm{is the sound speed.}  
\end{equation} 
Therefore, 
\begin{equation}
    \fr{1}{\sqrt{\beta}} = a \fr{\sqrt{2}}{\sqrt{\gamma}}.  
\end{equation}
Thus, the split fluxes can be written as 
\begin{eqnarray} 
G^{(a)\pm} = u^{\pm} U \ \textrm{and} \ 
G^{(p)\pm} = \left[ \begin{array}{c} 
\pm \fr{\rho a}{\sqrt{2 \gamma \pi}} \\[5mm]  
\fr{p}{2} \pm \fr{\rho a u}{\sqrt{2 \gamma \pi}} \\ 
\fr{p u}{2} \pm \fr{\rho a}{\sqrt{2 \gamma \pi}} \left( \fr{p}{2 \rho} + E\right) 
\end{array} \right].   
\end{eqnarray} 
Noting further that 
\begin{equation}
    a^{2} = \fr{\gamma p}{\rho},  
\end{equation} 
we can write 
\begin{equation}
    \fr{\rho a}{\sqrt{2 \gamma \pi}} = \fr{\gamma p}{a^{2}}  \fr{a}{\sqrt{2 \gamma \pi}} 
    = \fr{p}{a} \fr{\sqrt{\gamma}}{\sqrt{2 \pi}}.  
\end{equation}
Thus, the split fluxes become 
\begin{eqnarray} 
G^{(a)\pm} = u^{\pm} U \ \textrm{and} \ 
G^{(p)\pm} = \left[ \begin{array}{c} 
\pm \fr{p}{a} \fr{\sqrt{\gamma}}{\sqrt{2 \pi}} \\[5mm]  
\fr{p}{2} \pm \fr{p u}{a} \fr{\sqrt{\gamma}}{\sqrt{2 \pi}} \\ 
\fr{p u}{2} \pm \fr{p}{a} \fr{\sqrt{\gamma}}{\sqrt{2 \pi}} \left( \fr{p}{2 \rho} + E\right) 
\end{array} \right].   
\end{eqnarray} 
\begin{Remark}
Clearly, while $ \Delta G^{(a \pm)} $ vanishes as $ u\rightarrow 0 $ and $ \Delta p \rightarrow 0 $, $\Delta G^{(p \pm)}$ does not (where $\Delta (\cdot) = (\cdot)_{j} - (\cdot)_{j-1}$).  Thus, the original PVU scheme cannot recognize a steady contact discontinuity. 
\end{Remark} 
\begin{Remark} 
Notice that the collision step introduces a temporal change in the velocity distribution function but the conserved variable vector (moment) remains the same, \textit{i.e.}, 
\begin{equation}
    f^{(n+1)-} \neq f^{(n+1)}, \, \Bar{\textbf{U}}^{(n+1)-} = \Bar{\textbf{U}}^{(n+1)}.
\end{equation} 
\end{Remark} 
\par 
\noindent 
\begin{Remark} 
Appendix C establishes that, when based on a compactly supported distribution function, the PVU Boltzmann scheme maintains non-negative density and pressure provided a CFL-type condition is satisfied. This result does not extend to the Maxwellian-based PVU scheme, as the Maxwellian has infinite support in velocity space. Although in the moment method strategy, the velocity integrals formally extend from $ -\infty $ to $ +\infty $, but contributions from velocities outside a finite interval become negligibly small due to rapid exponential decay of the Maxwellian. Specifically, integrating the Maxwellian (first moment) over a $ 3 \sqrt{R T} $ interval in velocity space, where $ \sqrt{R T} $ denotes the standard deviation of the Maxwellian distribution function, captures approximately $ 99.73 \% $ of its total mass, thus providing an accurate approximation of the full-range moments. Mandal and Deshpande found through numerical experiments that the second, third, fourth and fifth moments of the $ 3 \sqrt{R T} $ limit cover approximately $ 99.73 $, $ 97.92 $, $ 97.43 $ and $ 93.2 \% $ of the total area, respectively \cite{JCMandal_SMD}. Accordingly, we adopt a $ 3 \sqrt{R T} $ velocity cut-off (for obtaining the stability condition), beyond which the Maxwellian is assumed to be zero. Thus, the stability condition for the peculiar velocity based upwind Boltzmann scheme (defined using the Maxwellian) is given by
\begin{equation} \label{stability_condition}
    (\max_{j} \{ |u_j| + 3 \sqrt{R T_j} \}) \Delta t \leq \Delta x.
\end{equation}
\end{Remark}

\subsection{Modified Partial Differential Equation analysis}
\label{subsubsec 3.3}

In this subsection, we attempt to understand the reason behind the overly diffusive performance of the 1D PVU scheme across the stationary contact discontinuities. We perform the modified partial differential equation (MPDE) analysis on the semi-discrete form of the convection step of the PVU scheme, which reads
\begin{equation} \label{semi-discrete_form_of_the_PVU_scheme}
    \frac{\partial f}{\partial t} + \frac{1}{\Delta x} \left( |u_j| f^{eq}_j + |c_j| f^{eq}_j + h_{j-1}^+ + h_{j+1}^- \right) = 0,
\end{equation}
where
\begin{subequations} \label{definition_h}
    \begin{align}
        h_{j-1}^+ &= u_{j-1}^+ f^{eq}_{j-1} + c_{j-1}^+ f^{eq}_{j-1}, \\
        h_{j+1}^- &= -u_{j+1}^- f^{eq}_{j+1} - c_{j+1}^- f^{eq}_{j+1}.
    \end{align}
\end{subequations}
We expand the terms in $ h_{j \pm 1}^\pm $ about $ x_j $ using the Taylor series expansion as
\begin{subequations} \label{Taylor_series_expansion_of_h}
    \begin{align}
        (u^\pm f^{eq})_{j \pm 1} &= (u^\pm f^{eq})_j \pm \left( \frac{\partial}{\partial x} (u^\pm f) \right)_j \Delta x + \left( \frac{\partial^2}{\partial x^2} (u^\pm f) \right)_j \frac{(\Delta x)^2}{2}, \\
        (c^\pm f^{eq})_{j \pm 1} &= (c^\pm f^{eq})_j \pm \left( \frac{\partial}{\partial x} (c^\pm f) \right)_j \Delta x + \left( \frac{\partial^2}{\partial x^2} (c^\pm f) \right)_j \frac{(\Delta x)^2}{2}. 
    \end{align}
\end{subequations}
We then substitute (\ref{Taylor_series_expansion_of_h}) into (\ref{semi-discrete_form_of_the_PVU_scheme}) to obtain
\begin{equation}
    \frac{\partial f}{\partial t} + \frac{\partial}{\partial x} (u f) + \frac{\partial}{\partial x} (c f) = \frac{\partial^2}{\partial x^2} \left( (|u| + |c|) \frac{\Delta x}{2} f \right).
\end{equation}
\begin{Remark} 
We note that the peculiar velocity based coefficient of numerical diffusion does not vanish as the Mach number tends to zero.   This is the primary reason behind the diffusive prediction of the stationary contact discontinuities by the PVU Boltzmann scheme.
\end{Remark} 


\section{A contact discontinuity capturing Boltzmann scheme in 1D}
\label{sec4}

The peculiar velocity based upwind Boltzmann scheme is positive, entropy stable and robust, but is unable to recognize stationary contact discontinuities. In addition to the diffusive prediction of stationary contacts, the PVU scheme also suffers from spurious pressure and velocity oscillations near isolated contact discontinuities. However, the Boltzmann equation based on peculiar velocity (\ref{peculiar_velocity_based_Boltzmann_equation}) provides an elegant framework, with the flexibility to treat the peculiar velocity based flux independently of the bulk fluid velocity based flux. The goal of this section is to design a peculiar velocity based numerical diffusion that vanishes as the Mach number goes to zero and remains low but nonzero otherwise. This improves scheme's performance across linear discontinuities without compromising the robustness of the original scheme.

\subsection{Numerical formulation}
\label{subsec 4.1}

We choose a general flux difference splitting form of the numerical flux for each of the advective and pressure parts. In addition, we introduce diffusion control parameters in the diffusion terms. Thus, we obtain     
\begin{subequations} \label{defintion_numerical_flux_pvu_plus}
    \begin{align}
        \hat{\textbf{G}}_{j + \frac{1}{2}}^{(a)} &= \frac{1}{2} \left\langle (u f^{eq})_j + (u f^{eq})_{j+1} - \Delta_{j + \frac{1}{2}} \left( \phi^{(u)}_{j + \frac{1}{2}} |u| f^{eq} \right) \right\rangle, \\
        \hat{\textbf{G}}_{j + \frac{1}{2}}^{(p)} &= \frac{1}{2} \left\langle (c f^{eq})_j + (c f^{eq})_{j+1} - \Delta_{j + \frac{1}{2}} \left( \phi^{(c)}_{j + \frac{1}{2}} |c| f^{eq} \right) \right\rangle,
    \end{align}
\end{subequations}
with
\begin{equation}
    \Delta_{j + \frac{1}{2}} (\cdot) = (\cdot)_{j + 1} - (\cdot)_{j},
\end{equation}
and
\begin{equation}
    \left\langle \cdot \right\rangle = \int_{I} \, dI \int_{c} \, dc \,\boldsymbol{\psi} \, (\cdot).
\end{equation}
The diffusion control parameters, $ \phi^{(u)} $ and $ \phi^{(c)} $, obey the dynamical consistency condition, \textit{i.e.},
\begin{equation} \label{dynamical_consistency_of_phi's}
    \textbf{U}_j = \textbf{U}_{j \pm 1} \implies \left( \phi^{(u), (c)}_{j \pm \frac{1}{2}} \right)_j = \left( \phi^{(u), (c)}_{j \pm \frac{1}{2}} \right)_{j \pm 1}.
\end{equation}

\subsection{A few choices for the diffusion control parameters}
\label{subsubsec 4.2}

\subsubsection{Variant 1}
\label{subsubsec 4.2.1}

We choose
\begin{subequations} \label{phi's_for_variant_1}
    \begin{align}
        \left( \phi^{(u)}_{j + \frac{1}{2}} \right)_j &= 1 \quad \forall j, \\
        \left( \phi^{(c)}_{j + \frac{1}{2}} \right)_j &= \frac{a_j}{2 \bar{a}},
    \end{align}
\end{subequations}
where
\begin{equation}
    \bar{a} = \frac{a_j + a_{j + 1}}{2}.
\end{equation}
$ a $ denotes the sound speed. The details of the 1D numerical flux corresponding to the chosen $ \phi $'s are given in Appendix A.

\subsubsection{Variant 2}
\label{subsubsec 4.2.2}

We choose
\begin{subequations} \label{phi's_for_variant_2}
    \begin{align}
        \left( \phi^{(u)}_{j + \frac{1}{2}} \right)_j &= \frac{1}{| u_j |} \left( \max \{ |u_j|, |u_{j + 1}| \} + \left | \frac{\sqrt{\rho_j} u_j + \sqrt{\rho_{j + 1}} u_{j + 1}}{\sqrt{\rho_j} + \sqrt{\rho_{j + 1}}} \right | \right) \quad \forall j, \label{phi_1} \\
        \left( \phi^{(c)}_{j + \frac{1}{2}} \right)_j &= \exp \left( -\frac{c_j^2}{u_j^2} \right) \quad \forall j.
    \end{align}
\end{subequations}
The details of the 1D numerical flux corresponding to the chosen $ \phi $'s are given in Appendix A.

\begin{Remark} 
    We note that, in variant 2, the coefficient of numerical diffusion pertaining to peculiar velocity based numerical flux, \textit{i.e.}, $ \left( \phi^{(c)}_{j + \frac{1}{2}} \right)_j = 0 $ for $ |u_j| = 0 $. Hence, the numerical flux corresponding to these $\phi$'s reproduces the exact solution of the Riemann problem that connects the left and right states through an isolated stationary contact discontinuity.  
\end{Remark}

\subsubsection{Variant 3}
\label{subsec 4.3}
In this variant, we attempt to further improve the accuracy of variant 1 at the stationary contact discontinuity. We note that across the stationary contact discontinuity, the numerical flux obtained in variant 1 reads
\begin{equation}
    \hat{\textbf{G}}_{j + \frac{1}{2}} = \begin{bmatrix}
        0 & p & \dfrac{(\gamma + 1)}{4 (\gamma - 1)} \sqrt{\dfrac{2}{\gamma \pi}} \dfrac{\left( (p a^2)_j - (p a^2)_{j + 1} \right)}{2 \Bar{a}}
    \end{bmatrix}^{T},
\end{equation}
which clearly reveals the nonzero term in the numerical energy flux that is responsible for the diffusive prediction of the 1D stationary contact discontinuity. We take inspiration from \cite{Sun_Takayama, Mandal_Panwar} and numerically approximate the density in the non-zero term as
\begin{equation}
    \rho_j = \frac{p_j}{\Bar{a}^2} \quad \forall j.
\end{equation}
The nonzero term in the energy flux then becomes the following:
\begin{equation}
    \dfrac{(\gamma + 1)}{4 (\gamma - 1)} \sqrt{\dfrac{\gamma}{2 \pi}} \left( p_j - p_{j + 1} \right) \Bar{a}, 
\end{equation}
which automatically shuts off whenever $ p_j = p_{j + 1} $, leading to a numerical scheme that recognizes the one-dimensional stationary contact discontinuity. We refer to Appendix A for more details on the 1D numerical flux obtained using variant 3.


\section{A 2D formulation of the modified PVU Boltzmann scheme}
\label{sec5}
This section presents the modified PVU Boltzmann scheme in two dimensions. 

\subsection{Numerical formulation}
\label{subsec 5.1}
Let $ T_h $ be a tessellation of the $ \Re^2 $ space. Let $ m(K) $ be a measure in $ \Re^2 $ of the cell $ K $, $ e $ be a side of $ K $, $ K_e $ be a neighboring cell of $ K $ along the edge $ e $, $ \hat{\textbf{n}}_e $ be an outward unit normal vector to the edge $ e $ making an angle $ \theta_e $ with the $ x- $axis, $ \hat{\textbf{t}}_e $ be a unit tangent vector to the edge, $ m(e) $ be a measure in $ \Re $ of the edge $ e $. Initially, the fluid velocity vector and the thermodynamic state variables are assumed to be piecewise constant in each finite volume cell. The Maxwellian distribution function is then constructed inside each cell, which reads
\begin{equation} \label{2D_Maxwellian}
    f^{eq}_K = \frac{\rho_K}{I_{0, K}} \frac{\beta_K}{\pi} \exp \left( -\beta_K \textbf{c}_K^2 - \frac{I}{I_{0, K}} \right),
\end{equation}
with
\begin{equation} \label{defintions_I_beta}
    I_{0, K} = \frac{(2 - \gamma)}{2 (\gamma - 1) \beta_K}, \, \beta_K = \frac{1}{2 R T_K}.
\end{equation}
The velocity distribution function in each finite volume is evolved in time by solving the Boltzmann equation in two stages:

\textbf{Convection step:} We start with a two-dimensional collision-less peculiar velocity-based Boltzmann equation:
\begin{equation} \label{2D_collisionless_PV_Boltzmann_equation}
    \frac{\partial f}{\partial t} + \sum_{i = 1}^{2} \frac{\partial}{\partial x_i} (u_i f) + \sum_{i = 1}^{2} \frac{\partial}{\partial x_i} (c_i f) = 0.
\end{equation}
Upon integrating Equation (\ref{2D_collisionless_PV_Boltzmann_equation}) over the finite volume and applying the 2-D equivalent of Gauss's divergence theorem, the following result follows.
\begin{equation}
    \int_{K} \, dA \, \dfrac{\partial f}{\partial t} = -\oint \, dl \, \hat{h},
\end{equation}
where
\begin{equation}
    \hat{h} = \left( \boldsymbol{u} \cdot \hat{\boldsymbol{n}} + \boldsymbol{c} \cdot \hat{\boldsymbol{n}} \right) f  
\end{equation}
with $ l $ representing the periphery of the finite volume. Summing the locally one-dimensional fluxes evaluated at the midpoints of all line segments gives the total boundary flux, \textit{i.e.},
\begin{equation}
    \int_{K} \, dA \, \dfrac{\partial f}{\partial t} = -\sum_e \hat{h}_e \, m(e),
\end{equation}
with
\begin{equation}
    \hat{h}_e = \left( \boldsymbol{u}_e \cdot \hat{\boldsymbol{n}}_e + \boldsymbol{c}_e \cdot \hat{\boldsymbol{n}}_e \right) f_e. 
\end{equation}
The forward-in-time discretization of the temporal derivative leads to
\begin{equation} \label{discretized_2D_PV_Boltzmann_equation}
    \int_{K} \, dA \, f_K^{(n + 1)-} = \int_{K} \, dA \, f_K^{(n)} - \Delta t \, \sum_e \hat{h}^{(n)}_e \, m(e).
\end{equation}

\textbf{Collision step:} The collision step restores equilibrium in each finite volume cell, \textit{i.e.},
\begin{equation} \label{essence_of_collision_step_2D}
    \int_{I} \, dI \int_{\textbf{c}} \, d\textbf{c} \, \boldsymbol{\psi}_K \, f^{eq, (n + 1)}_K = \int_{I} \, dI \int_\textbf{c} \, d\textbf{c} \, \boldsymbol{\psi}_K \, f^{(n + 1)-}_K.
\end{equation}
Using equation (\ref{discretized_2D_PV_Boltzmann_equation}) in equation (\ref{essence_of_collision_step_2D}) leads to a first-order finite volume approximation of the two-dimensional Euler equations, which reads
\begin{equation} \label{2D_finite_volume_update_formula}
    \bar{\boldsymbol{U}}_K^{n + 1} = \bar{\boldsymbol{U}}_K^{n} - \sigma \sum_e \hat{\textbf{G}}_e,
\end{equation}
where $ \bar{\boldsymbol{U}} $ denotes the cell-averaged value of the conserved variable vector in each cell, $ \hat{\textbf{G}}_e $ denotes the numerical flux in the direction normal to the edge $ e $, which is represented by the sum of the numerical advective and pressure fluxes, \textit{i.e.},
\begin{equation} \label{2D_definition_of_total_flux}
    \hat{\textbf{G}}_e = \hat{\textbf{G}}^{(a)}_e + \hat{\textbf{G}}^{(p)}_e.
\end{equation} 
We choose
\begin{subequations} \label{definitions_2D_advective_pressure_numerical_flux}
    \begin{align}
        \hat{\textbf{G}}^{(a)}_e &= \frac{1}{2} \left\langle (u_n f^{eq})_K + (u_n f^{eq})_{K_e} - \left( (\phi^{(u)}_e |u_n| f^{eq})_{K_e} - (\phi^{(u)}_e |u_n| f^{eq})_K \right) \right\rangle, \\
        \hat{\textbf{G}}^{(p)}_e &= \frac{1}{2} \left\langle (c_n f^{eq})_K + (c_n f^{eq})_{K_e} - \left( (\phi^{(c)}_e |c_n| f^{eq})_{K_e} - (\phi^{(c)}_e |c_n| f^{eq})_K \right) \right\rangle,
    \end{align}
\end{subequations}
where
\begin{equation}
    \left\langle \cdot \right\rangle = \int_{I} \, dI \int_{c_t} \, dc_t \int_{c_n} \, dc_n \,\boldsymbol{\psi}_e \,(\cdot),
\end{equation}
and
\begin{subequations} \label{normal_bulk_fluid_and_peculiar_velocity}
    \begin{align}
        u_n &= \textbf{u} \cdot \hat{\textbf{n}}_e, \, u_t = \textbf{u} \cdot \hat{\textbf{t}}_e, \\
        c_n &= \textbf{c} \cdot \hat{\textbf{n}}_e, \, c_t = \textbf{c} \cdot \hat{\textbf{t}}_e.
    \end{align}
\end{subequations}
$ \boldsymbol{\psi}_e $ denotes the moment vector in the transformed coordinate system, \textit{i.e.},
\begin{equation} \label{2D_defintion_of_moment_vector}
    \boldsymbol{\psi}_e = \left( 1, \, v_n \cos \theta_e - v_t \sin \theta_e, \, v_n \sin \theta_e + v_t \cos \theta_e, \, I + \frac{(v_n^2 + v_t^2)}{2} \right),
\end{equation}
where
\begin{subequations}
    \begin{align}
        v_n &= u_n + c_n, \\
        v_t &= u_t + c_t.
    \end{align}
\end{subequations}
The diffusion control parameters $ \phi^{(u)}, \phi^{(c)} $ obey the dynamical consistency condition, \textit{i.e.},
\begin{equation}
    \textbf{U}_K = \textbf{U}_{K_e} \implies \left( \phi^{(u), (c)}_e \right)_K = \left( \phi^{(u), (c)}_e \right)_{K_e}.
\end{equation}

\subsection{Choices for the diffusion control parameters}
\label{subsec 5.2}

\subsubsection{Variant 1}
\label{subsubsec 5.2.1}
We choose
\begin{subequations} \label{defintion_of_diffusion_control_parameters_for_variant_1}
    \begin{align}
        \left( \phi^{(u)}_e \right)_K &= \frac{\max \{ |(u_n)_K|, |(u_n)_{K_e}| \}}{|(u_n)_K|}, \\
        \left( \phi^{(c)}_e \right)_{K_e} &= \frac{a_K}{2 \Bar{a}},
    \end{align}
\end{subequations}
where
\begin{equation} \label{definition_a_bar}
    \Bar{a} = \frac{a_K + a_{K_e}}{2}.
\end{equation}
$ a $ denotes the sound speed. The details of the 2D numerical flux corresponding to the chosen $ \phi $'s are given in Appendix B.

\subsubsection{Variant 2}
\label{subsubsec 5.2.2}
We choose
\begin{subequations} \label{defintion_of_diffusion_control_parameters_for_variant_2}
    \begin{align}
        \left( \phi^{(u)}_e \right)_K &= \frac{1}{|(u_n)_K|} \left( \max \{ |u_n|_K, |u_n|_{K_e} \} + \frac{\sqrt{\rho_K} (u_n)_K + \sqrt{\rho_{K_e}} (u_n)_{K_e}}{\sqrt{\rho_K} + \sqrt{\rho_{K_e}}} \right), \label{2D_phi_1} \\
        \left( \phi^{(c)}_e \right)_{K_e} &= \exp \left( -\frac{(c_n)_K^2}{(u_n)_K^2} \right).
    \end{align}
\end{subequations}
The details of the 2D numerical flux corresponding to the chosen $ \phi $'s are given in Appendix B.

\subsection{A two-dimensional extension of variant 3}
\label{subsec 5.3}
We extend variant 3 of the modified PVU scheme to two dimensions by approximating the density in the numerical energy flux as
\begin{equation}
    \rho_K = \frac{p_K}{\Bar{a}^2} \quad \forall K,
\end{equation}
where $ \Bar{a} $ is defined in (\ref{definition_a_bar}). The modified numerical pressure flux reads
\begin{equation} \label{2D_pressure_numerical_flux_variant_3}
    \hat{\textbf{G}}_{e}^{(p)} = \frac{1}{2} \left( \textbf{G}_{n, K}^{(p)} + \textbf{G}_{n, K_e}^{(p)} \right) - \textbf{d}_{e}^{(p)},
\end{equation}
where
\renewcommand{\arraystretch}{2.0}
\begin{equation} \label{2D_Numerical_diffustion_pressure_flux_variant_3}
    \textbf{d}_{e}^{(p)} = \begin{bmatrix}
        \left( \dfrac{\gamma}{2 \pi} \right)^{\frac{1}{2}} \dfrac{1}{2 \Bar{a}} \Delta_{e} \left( p \right) \\
        \left( \dfrac{\gamma}{2 \pi} \right)^{\frac{1}{2}} \dfrac{1}{2 \Bar{a}} \Delta_{e} \left( p u_n \right) \\
        \left( \dfrac{\gamma}{2 \pi} \right)^{\frac{1}{2}} \dfrac{1}{2 \Bar{a}} \Delta_{e} \left( p u_t \right) \\
        \left( \dfrac{\gamma}{2 \pi} \right)^{\frac{1}{2}} \dfrac{1}{4 \Bar{a}} \Delta_{e} (p | \boldsymbol{u} |^2) + \dfrac{(\gamma + 1)}{4 (\gamma - 1)} \left( \dfrac{\gamma}{2 \pi} \right)^{\frac{1}{2}} \Delta_{e} (p \Bar{a})
    \end{bmatrix}.
\end{equation}

\vspace{1\baselineskip} 
\begin{Remark} 
We note that the numerical diffusion defined in (\ref{2D_Numerical_diffustion_pressure_flux_variant_3}) does not completely switch off across the stationary slip surfaces, leading to their poor resolution relative to variant 2.
\end{Remark} 

\vspace{1 em}
\textbf{Note:} Appendix D proves that the two-dimensional PVU Boltzmann scheme with compact support in the distribution function preserves density and pressure positivity under a CFL-type condition. Motivated by this result, the two-dimensional stability condition applicable to all variants of the modified PVU Boltzmann scheme reads
\begin{equation} \label{2D_stability_condition_modified_PVU}
    \frac{\Delta t}{2 m(K)} \max_{K} \left\{ \sum_{e} \left( |(u_n)_K| + 3 \sqrt{R T_K} \right) m(e) \right\} \leq 1.
\end{equation}


\section{Extension of the scheme to second order accuracy}
\label{sec6}

In this section, we present a way, inspired by \cite{Khobalatte_Perthame}, to extend the modified PVU Boltzmann scheme to second-order accuracy in space and time. We select the three-stage Strong Stability Preserving Runge-Kutta (SSPRK-3) method for time integration. In each stage of the R-K method, a piecewise linear approximation of the primitive variables is assumed in each finite volume cell, \textit{i.e.},
\begin{equation} \label{2D_2ndOrder_Reconstruction}
    \left.
    \begin{aligned}
        \rho(\boldsymbol{x}) &= \rho_K + \nabla \rho \Big|_K (\boldsymbol{x} - \boldsymbol{x}_K) \\
        \boldsymbol{u} (\boldsymbol{x}) &= \boldsymbol{u}_K + \nabla \boldsymbol{u} \Big|_K (\boldsymbol{x} - \boldsymbol{x}_K) \\
        p(\boldsymbol{x}) &= p_K + \nabla p \Big|_K (\boldsymbol{x} - \boldsymbol{x}_K)
    \end{aligned}
    \right\}
    \quad\quad \forall \boldsymbol{x} \in K,
\end{equation}
where $ \nabla ( \cdot ) $ denotes the gradient of the quantity in parentheses and $ \boldsymbol{x}_K $ denote the coordinates of the cell centroid. In each cell, the gradient is evaluated using the approximate form of the following exact relation:
\begin{equation}
    \int_{K} \, dA \, \nabla ( \cdot) = \oint_{\partial K} \, ds \, ( \cdot) \, \boldsymbol{n}.
\end{equation}
Thus, the primitive variables are reconstructed at each cell interface $ e $ as
\begin{equation} \label{2D_2ndOrder_Reconstruction_interface}
\begin{aligned}
    \rho^L_{e} &= \rho_K + \nabla \rho \big|_{K} \bigl( \boldsymbol{x}_e - \boldsymbol{x}_K \bigr), \quad  \quad \rho^R_{e} = \rho_{K_e} + \nabla \rho \big|_{K_e} \bigl( \boldsymbol{x}_e - \boldsymbol{x}_{K_e} \bigr), \\
    \boldsymbol{u}^L_{e} &= \boldsymbol{u}_K + \nabla \boldsymbol{u} \big|_{K} \bigl( \boldsymbol{x}_e - \boldsymbol{x}_K \bigr), \quad \quad \boldsymbol{u}^R_{e} = \boldsymbol{u}_{K_e} + \nabla \boldsymbol{u} \big|_{K_e} \bigl( \boldsymbol{x}_e - \boldsymbol{x}_{K_e} \bigr), \\
    p^L_{e} &= p_{K} + \nabla p \big|_{K} \bigl( \boldsymbol{x}_e - \boldsymbol{x}_K \bigr), \quad \quad p^R_{e} = p_{K_e} + \nabla p \big|_{K_e} \bigl( \boldsymbol{x}_e - \boldsymbol{x}_{K_e} \bigr).
\end{aligned}
\end{equation}
$ ( \cdot )^L_{e}, \, ( \cdot)^R_{e} $ denote the primitive variables reconstructed on the left and right of the interface, respectively. A semi-discrete finite volume update formula of the two-dimensional collisionless Boltzmann equation reads
\begin{equation} \label{eq:2D_semi_discrete_update}
    \int_{A} \, \frac{\partial f}{\partial t} \, dA = -\sum_{e} \left( \boldsymbol{u}_e \cdot \boldsymbol{n}_e  + \boldsymbol{c}_e \cdot \boldsymbol{n}_e \right) \, f_e \, m(e).
\end{equation}
The Maxwell distribution function is constructed at each interface using the reconstructed primitive variables, which reads
\begin{subequations}
    \begin{align}
        f^L_{e} &= f^{eq} (\rho^L_{e}, \, u^L_{e}, \, p^L_{e}), \\
        f^R_{e} &= f^{eq} (\rho^R_{e}, \, u^R_{e}, \, p^R_{e}).
    \end{align}
\end{subequations}
The kinetic numerical flux normal to each interface is defined on the basis of the reconstructed Maxwellian, \textit{i.e.},
\begin{multline} \label{definitions_of_2D_advective_flux}
    (\boldsymbol{u}_e \cdot \boldsymbol{n}_e) \, f_e = \left( \dfrac{(\boldsymbol{u}_e^{L} \cdot \boldsymbol{n}_e) + |\boldsymbol{u}_e^{L} \cdot \boldsymbol{n}_e| \, \phi_e^{(u)}}{2} \right) f^L_{e} \\
    +  \left( \dfrac{(\boldsymbol{u}_e^{R} \cdot \boldsymbol{n}_e) - |\boldsymbol{u}_e^{R} \cdot \boldsymbol{n}_e| \, \phi_e^{(u)}}{2} \right) f^R_{e},
\end{multline} 
\begin{multline} \label{definitions_of_2D_pressure_flux}
     (\boldsymbol{c}_e \cdot \boldsymbol{n}_e) \, f_e = \left( \dfrac{(\boldsymbol{c}_e^{L} \cdot \boldsymbol{n}_e) + |\boldsymbol{c}_e^{L} \cdot \boldsymbol{n}_e| \, \phi_e^{(c)}}{2} \right) f^L_{e} \\
    +  \left( \dfrac{(\boldsymbol{c}_e^{R} \cdot \boldsymbol{n}_e) - |\boldsymbol{c}_e^{R} \cdot \boldsymbol{n}_e| \, \phi_e^{(c)}}{2} \right) f^R_{e}.
\end{multline}
With second-order accuracy in the spatial dimension, we choose the three-stage Strong Stability Preserving Runge-Kutta (SSPRK-3) method for explicit time discretization of the temporal term in Equation (\ref{eq:2D_semi_discrete_update}) which reads
\begin{subequations}
    \begin{equation}
        f^{(1)}_K = f^{eq}_K - \frac{\Delta t}{m(K)} \, R_K (f^{eq})
    \end{equation}
    \begin{equation}
        f^{(2)}_K = \frac{3}{4} f^{eq}_K + \frac{1}{4} f^{(1)}_K - \frac{1}{4} \frac{\Delta t}{m(K)} \, R_K (f^{(1)})
    \end{equation}
    \begin{equation}
        f^{(n + 1)}_K = \frac{1}{3} f^{eq}_K + \frac{2}{3} f^{(2)}_K -\frac{2}{3} \frac{\Delta t}{m(K)} \, R_K (f^{(2)}),
    \end{equation}
\end{subequations}
where $ R $ denotes the summation of the numerical fluxes at all the edges $ e $ of the cell $ K $, termed residual. Taking moments of the above equations yields a macroscopic evolution for the conserved variable vector, namely,
\begin{subequations}
    \begin{equation}
        \bar{\boldsymbol{U}}^{(1)}_K = \bar{\boldsymbol{U}}^{n}_K - \frac{\Delta t}{m(K)} R_K (\bar{\boldsymbol{U}}^n)
    \end{equation}
    \begin{equation}
        \bar{\boldsymbol{U}}^{(2)}_K = \frac{3}{4} \bar{\boldsymbol{U}}^n_K + \frac{1}{4} \bar{\boldsymbol{U}}^{(1)}_K - \frac{1}{4} \frac{\Delta t}{m(K)} R_K (\bar{\boldsymbol{U}}^{(1)})
    \end{equation}
    \begin{equation}
        \bar{\boldsymbol{U}}^{(n+1)}_K = \frac{1}{3} \bar{\boldsymbol{U}}^n_K + \frac{2}{3} \bar{\boldsymbol{U}}^{(2)}_K - \frac{2}{3} \frac{\Delta t}{m(K)} R_K (\bar{\boldsymbol{U}}^{(2)}).
    \end{equation}
\end{subequations}
where the residual now represents the summation of macroscopic flux vectors, obtained after taking moments.

\vspace{1 em}
\textbf{Note 1:} It is to be noted that in each cell, the collision step instantaneously re-establishes the equilibrium without changing the mass, momentum, and energy densities after each Runge-Kutta stage for the convection step.
\par
\vspace{1 em}
\textbf{Note 2:} The CFL-like condition in (\ref{2D_stability_condition_modified_PVU}) is used to determine the time step $ \Delta t $, which is taken with a lower CFL number than that used for a first-order scheme.
\par
\vspace{1 em}
\textbf{Note 3:} We consider several slope limiters for the reconstruction, including the Venkatakrishnan limiter, which is known for its good steady-state convergence characteristics, as well as the Barth-Jespersen and R5 limiters. While the Venkatakrishnan limiter drives the slope to zero near large gradients, the latter two are less diffusive and are therefore employed to capture subtle flow features. Details on the numerical implementation of these limiters can be found in \cite{Venkatakrishnan, Barth_Jespersen, Nishikawa}.


\section{A modified PVU Boltzmann scheme for viscous flows}
\label{sec7}

It is a well-established fact that the $ \boldsymbol{\psi} \text{-moments} $ of the Boltzmann equation recover the Navier-Stokes-Fourier equations for viscous flows in the limit $ f \rightarrow f^{CE} $, \textit{i.e.}, with the Chapman-Enskog distribution function. Consequently, the moments of an appropriately discretized Boltzmann equation yield a numerical scheme for the Navier-Stokes-Fourier system when the Chapman-Enskog distribution function is utilized. In each stage of the R-K method, the convective flux is defined in equations (\ref{definitions_of_2D_advective_flux})-(\ref{definitions_of_2D_pressure_flux}) and the viscous flux pertaining to the bulk fluid and peculiar velocities is defined at each edge as
\begin{subequations} \label{2D_definition_viscous_flux}
    \begin{align}
        \hat{\textbf{G}}^{(v), (a)}_e &= \frac{1}{2} \left\langle (u_n P_{CE} f^{eq})_K + (u_n P_{CE} f^{eq})_{K_e} - \Delta_e \left( \varphi^{(u)} |u_n| P_{CE} f^{eq} \right) \right\rangle, \\
        \hat{\textbf{G}}^{(v), (p)}_e &= \frac{1}{2} \left\langle (c_n P_{CE} f^{eq})_K + (c_n P_{CE} f^{eq})_{K_e} - \Delta_e \left( \varphi^{(c)} |c_n| P_{CE} f^{eq} \right) \right\rangle,
    \end{align}
\end{subequations}
where $ \varphi^{(u), (c)} $ denote the coefficients of numerical diffusion for the advective and pressure components of the viscous flux. We choose here
\begin{equation}
    \varphi^{(u), (c)} = 0,
\end{equation}
so that the viscous fluxes reduce to
\begin{subequations}
    \begin{align}
        \hat{\textbf{G}}^{(v), (a)}_e &= \frac{1}{2} \left\langle (u_n P_{CE} f^{eq})_K + (u_n P_{CE} f^{eq})_{K_e} \right\rangle, \\
        \hat{\textbf{G}}^{(v), (p)}_e &= \frac{1}{2} \left\langle (c_n P_{CE} f^{eq})_K + (c_n P_{CE} f^{eq})_{K_e} \right\rangle.
    \end{align}
\end{subequations} 
The total flux is thus defined as
\begin{equation}
    \hat{\textbf{G}}_e = \hat{\textbf{G}}^{(a)}_e - \hat{\textbf{G}}^{(v), (a)}_e + \hat{\textbf{G}}^{(p)}_e - \hat{\textbf{G}}^{(v), (p)}_e,
\end{equation}
which is then used in the convection steps of the three-stage Strong Stability Preserving Runge-Kutta 3 method.

\vspace{1 em}
\textbf{Note:} The stability condition for the numerical scheme corresponding to viscous flows reads
\begin{equation} \label{2D_stability_condition_for_NS_equations}
    \frac{\Delta t}{2 m(K)} \max_K \left\{ \sum_{e} \left( \lambda_K + \lambda_{visc, K} \right) m(e) \right\} \leq 1,
\end{equation}
where
\begin{subequations} \label{definitions_of_lambdas}
    \begin{align}
        \lambda_K \, m(e) &= \left( |(u_n)_K| + 3 \sqrt{R T_K} \right) m(e), \\
        \lambda_{visc, K} \, m(e) &= 4 \max \left\{ \dfrac{4}{3 \rho_K}, \dfrac{\gamma}{\rho_K} \right\} \dfrac{\mu_K}{Pr} \dfrac{\left( m(e_1)^2 + m(e_2)^2 \right)}{m(K)}.
    \end{align}
\end{subequations}


\section{Numerical results}
\label{sec8}

\subsection{Empirical Order of Convergence}
\label{subsec 8.1}
The experimental order of convergence, or EOC, is a useful way to assess the convergence order of any scheme by testing it on a problem for which the initial conditions consist of a sinusoidal variation of the density, while the velocity and pressure remain constant in the domain $ x \in [0, 1] $. The true solution for the problem considered here takes the form
\begin{equation} \label{exact_solution_of_density_for_EOC_test}
    \rho(x,t) = \rho_0 (x - u t) = 1.0 + 0.2 \sin{\{ 2 \pi (x - 0.1 t) \}},
\end{equation}
\begin{equation} \label{exact_solution_of_velocity_pressure_for_EOC_test}
    p(x, t) = 0.5, \, u(x, t) = 0.1.
\end{equation}
The numerical solution is allowed to evolve until t = 0.5 units for the grid of sizes $ N_x $ = 10, 20, 40, 80, 160, 320, 640, 1280 with periodic boundary conditions at the left and right ends of the domain. The $ L_1 $ and $ L_2 $ norms of the error are evaluated as
\begin{equation}
    \bigl|\bigl| \varepsilon \bigl|\bigl|_{L_1} = {\dfrac{1}{N_x}} \sum^{N}_{j=1} \bigl| \rho_j - \rho^{exact}_j \bigl|,
\end{equation}
\begin{equation}
    \bigl|\bigl| \varepsilon\bigl |\bigl|_{L_2} = \sqrt{{\dfrac{1}{N_x}} \sum^{N}_{j=1} \bigl( \rho_j - \rho^{exact}_j \bigl)^2},
\end{equation}
where $ \rho_j $ and $ \rho^{exact}_j $ are the numerical and exact densities at the cell center $ j $. Pressure and velocity remain constant, with no contribution to the error.
The experimental order of convergence is then evaluated as
\begin{equation}
    EOC = log_2 \Biggl( \dfrac{\bigl|\bigl| {\varepsilon}_{{N_x}/2} \bigl|\bigl|}{\bigl|\bigl| {\varepsilon}_{N_x} \bigl|\bigl|} \Biggl).
\end{equation}
The $ L_1 $ and $ L_2 $ error norms and the EOC evaluated for the respective grid sizes are listed in tables \ref{tab:1D_L1_L2_variant2} and \ref{tab:1D_L1_L2_variant3} for variants 2 and 3, respectively.

\vspace{-0.2 cm}
\begin{table}[h]
    \centering
    \setlength{\tabcolsep}{10pt}
    \renewcommand{\arraystretch}{1.25}
    \begin{tabular}{ l l c c c c }
    \hline
        $ N_x $ & $ \Delta x $ & $ L_1 $ error & EOC & $ L_2 $ error & EOC \\
    \hline
        10 & 0.1  & 0.0015 & & 0.0018 & \\
        20 & 0.05 & 3.46E-04 & 2.11 & 3.87E-04 & 2.22 \\
        40 & 0.025 & 8.34E-05 & 2.05 & 9.27E-05 & 2.06 \\
        80 & 0.0125 & 2.06E-05 & 2.01 & 2.29E-05 & 2.02 \\
        160 & 0.00625 & 5.14E-06 & 2.00 & 5.72E-06 & 2.00 \\
        320 & 0.003125 & 1.29E-06 & 2.00 & 1.43E-06 & 2.00 \\
        640 & 0.0015625 & 3.21E-07 & 2.00 & 3.57E-07 & 2.00 \\
        1280 & 0.00078125 & 8.03E-08 & 2.00 & 8.92E-08 & 2.00 \\
    \hline
    \end{tabular}
    \caption{EOC study for the second-order modified PVU scheme: variant 2}
    \label{tab:1D_L1_L2_variant2}
\end{table}

\begin{table}[H]
    \centering
    \setlength{\tabcolsep}{10pt}
    \renewcommand{\arraystretch}{1.25}
    \begin{tabular}{ l l c c c c }
    \hline
        $ N_x $ & $ \Delta x $ & $ L_1 $ error & EOC & $ L_2 $ error & EOC \\
    \hline
        10 & 0.1  & 0.0015 & & 0.0017 & \\
        20 & 0.05 & 3.41E-04 & 2.14 & 3.81E-04 & 2.16 \\
        40 & 0.025 & 8.30E-05 & 2.04 & 9.24E-05 & 2.05 \\
        80 & 0.0125 & 2.06E-05 & 2.01 & 2.29E-05 & 2.01 \\
        160 & 0.00625 & 5.14E-06 & 2.00 & 5.71E-06 & 2.00 \\
        320 & 0.003125 & 1.29E-06 & 2.00 & 1.43E-06 & 2.00 \\
        640 & 0.0015625 & 3.21E-07 & 2.00 & 3.57E-07 & 2.00 \\
        1280 & 0.00078125 & 8.03E-08 & 2.00 & 8.92E-08 & 2.00 \\
    \hline
    \end{tabular}
    \caption{EOC study for the second-order modified PVU scheme: variant 3}
    \label{tab:1D_L1_L2_variant3}
\end{table}

\subsection{1D results and discussions}
\label{subsec 8.2}

\begin{figure}[H]
    \centering
    \includegraphics[width=1\linewidth]{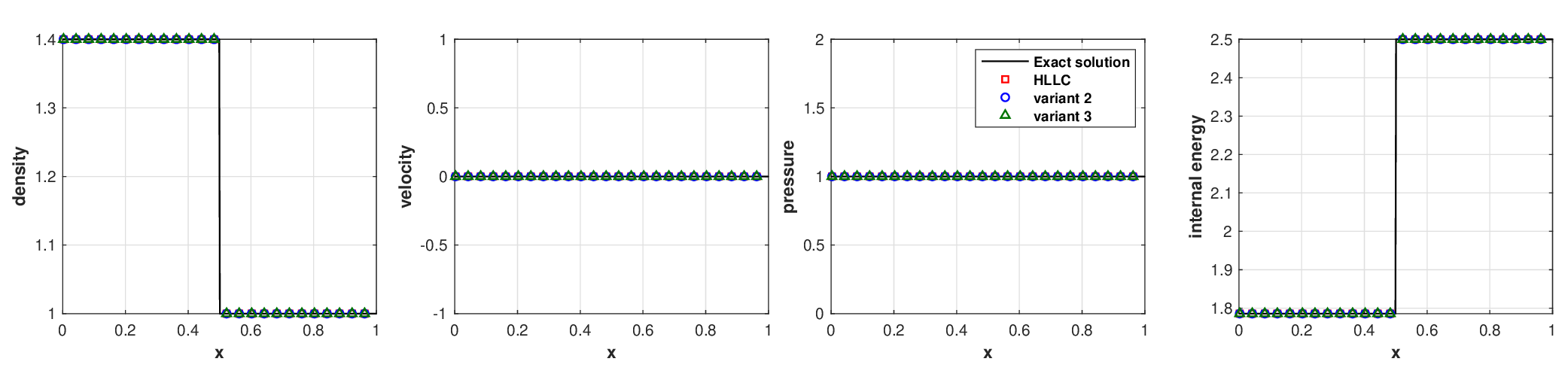}
    \vspace{-0.8 cm}
    \caption{Stationary contact discontinuity}
    \label{1D_figure1}
\end{figure}

In this subsection, we evaluate the performance of modified PVU Boltzmann schemes on several one-dimensional Riemann problems for which analytical solutions are available. Each test consists of constant left and right states ($ \textbf{U}_L, \textbf{U}_R $) separated at the point $ x_0 $. Computations are performed on the domain $ x \in [0, 1] $ discretized with $ 200 $ cells. CFL numbers of 0.4 and 0.8 are used for variant 2 and variant 3, respectively, to obtain first-order results. For second-order results, both variants are run with a CFL number of 0.2. The initial conditions for all numerical tests are cataloged in the table \ref{tab:1D_Riemann_problems}.
\begin{table}[h]
    \centering
    \begin{tabular}{|c|c|c|c|c|c|c|c|c|}
        \hline
        Test & $ \rho_L $ & $ u_L $ & $ p_L $ & $ \rho_R $ & $ u_R $ & $ p_R $ & runtime & $ x_0 $ \\
        \hline
        1 & 1.4 & 0.0 & 1.0 & 1.0 & 0.0 & 1.0 & 2.0 & 0.5 \\
        \hline
        2 & 1.4 & 0.1 & 1.0 & 1.0 & 0.1 & 1.0 & 1.0 & 0.5 \\
        \hline
        3 & 3.86 & -0.81 & 10.33 & 1.0 & -3.44 & 1.0 & 4.0 & 0.1 \\
        \hline
        4 & 1.0 & 0.75 & 1.0 & 0.125 & 0.0 & 0.1 & 0.2 & 0.3 \\
        \hline
        5 & 1.0 & -2.0 & 0.4 & 1.0 & 2.0 & 0.4 & 0.15 & 0.5 \\
        \hline
        6 & 5.999 & 19.5975 & 460.894 & 5.992 & -6.1963 & 46.095 & 0.035 & 0.4 \\
        \hline
    \end{tabular}
    \caption{1D Riemann problems}
    \label{tab:1D_Riemann_problems}
\end{table}
Numerical test 1 is designed to assess the performance of the scheme across an isolated stationary contact discontinuity. The results in Figure \ref{1D_figure1} show that variants 2 and 3 of the modified PVU scheme reduce to exact solution in case of the stationary isolated contact discontinuity. Also, both variants of the modified PVU Boltzmann scheme do not suffer from spurious pressure and velocity oscillations near the contact discontinuity.

\begin{figure}[H]
    \centering
    \includegraphics[width=1\linewidth]{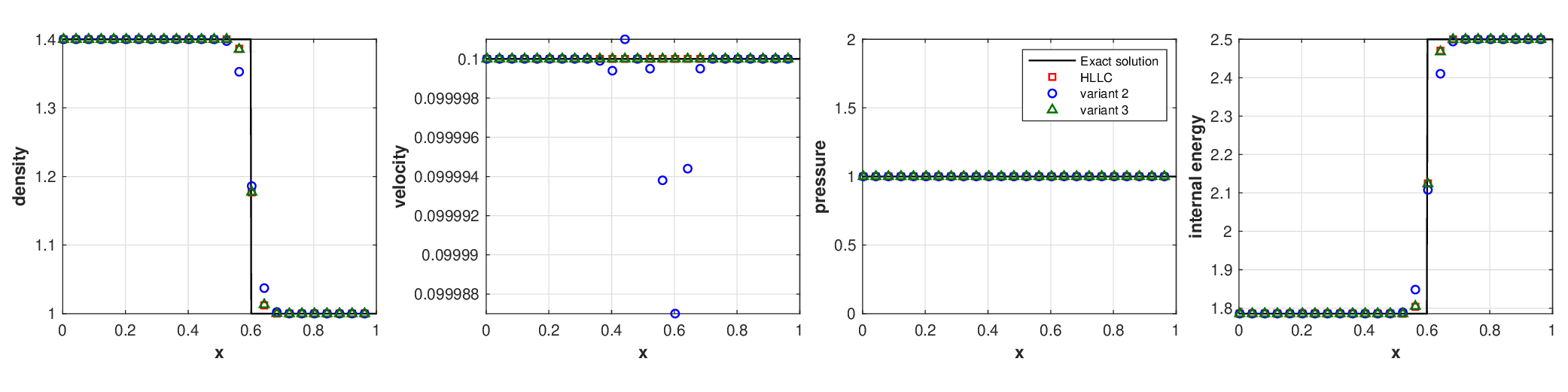}
    \vspace{-0.8 cm}
    \caption{Slowly moving contact discontinuity}
    \label{1D_figure2}
\end{figure}

Numerical test 2 assesses the performance of the scheme across an isolated, slowly moving contact discontinuity. The solution consists of a right-moving contact discontinuity propagating at a constant speed. As shown in Figure \ref{1D_figure2}, the results obtained with variant 3 and the HLLC scheme exhibit a similar level of diffusion, with neither scheme producing pressure or velocity oscillations. However, variant 2 of the modified PVU scheme exhibits slightly greater diffusion and suffers from velocity oscillations of $ \mathcal{O} \left( 10^{-5} \right) $.

\begin{figure}[H]
    \centering
    \includegraphics[width=1\linewidth]{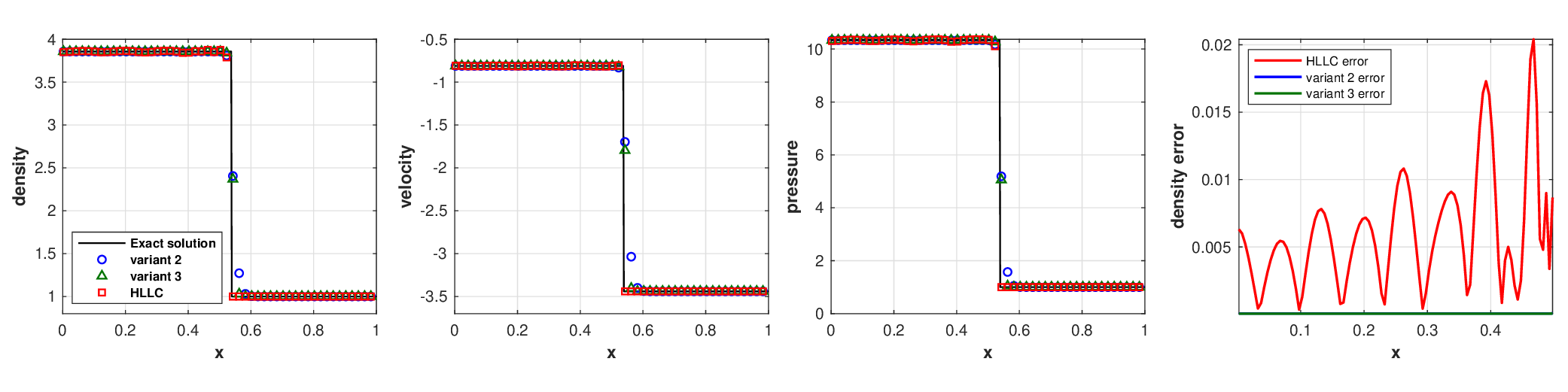}
    \vspace{-0.8 cm}
    \caption{Slowly moving shock}
    \label{1D_figure3}
\end{figure}

\begin{figure}[H]
    \centering
    \includegraphics[width=1\linewidth]{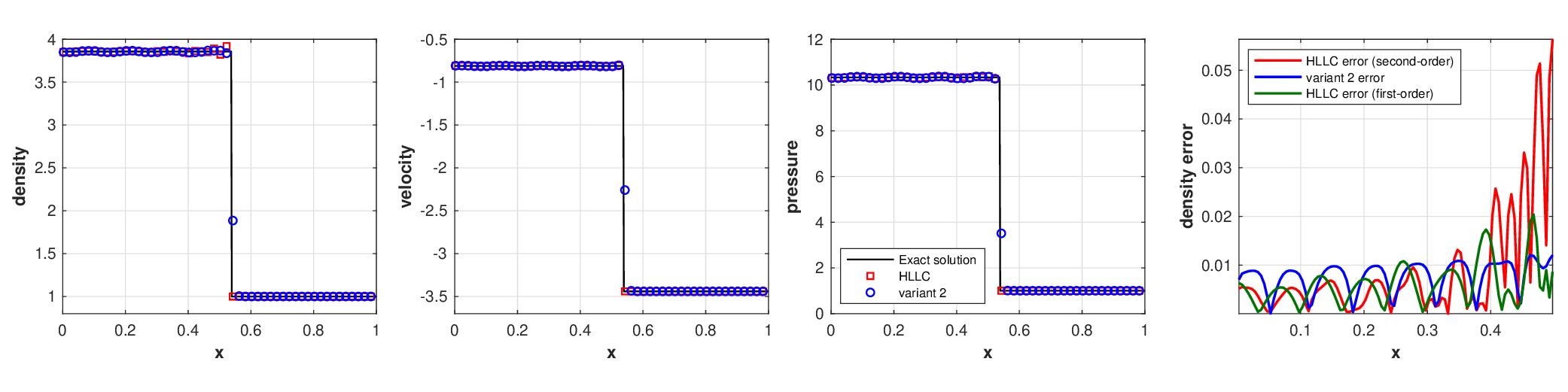}
    \vspace{-0.8 cm}
    \caption{Slowly moving shock computed using second-order variant 2}
    \label{1D_figure4}
\end{figure}

\begin{figure}[H]
    \centering
    \includegraphics[width=1\linewidth]{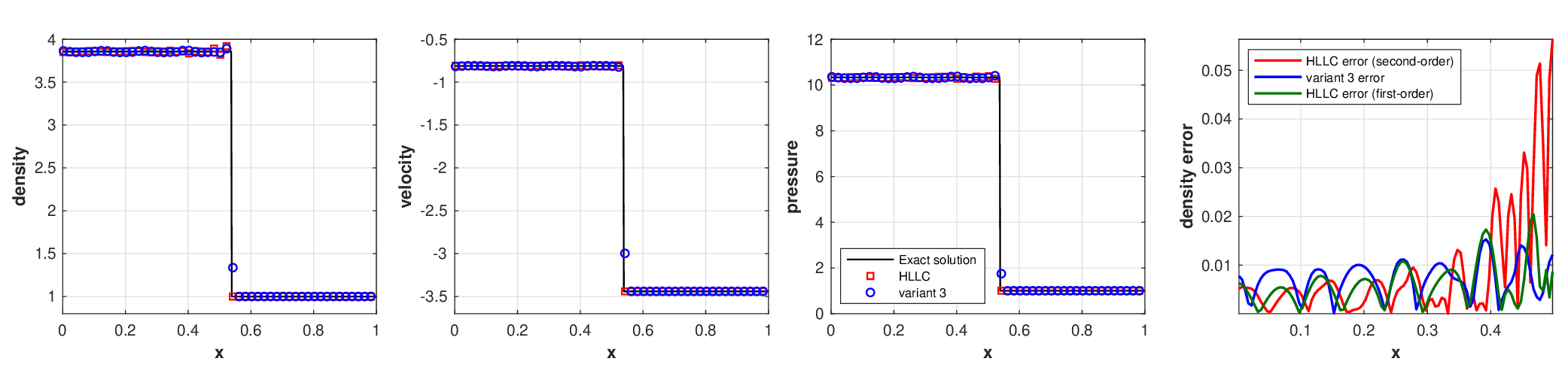}
    \vspace{-0.8 cm}
    \caption{Slowly moving shock computed using second-order variant 3}
    \label{1D_figure5}
\end{figure}

The solution of test 3 consists of a slowly moving shock propagating to the right. This test is designed to check the scheme's ability to avoid low-frequency post-shock oscillations. As seen in Figure \ref{1D_figure3}, neither variant of the modified PVU scheme exhibits post-shock oscillations, in contrast to the HLLC scheme, which develops noticeable oscillations behind the shock. Figures \ref{1D_figure4} and \ref{1D_figure5} compare the first- and second-order results obtained with the HLLC scheme, and variants 2 \& 3 of the modified PVU scheme, respectively. The second-order results for both variants exhibit subtle post-shock oscillations; however, their amplitudes are smaller than those observed with the first- and second-order HLLC scheme.

\vspace{-0.6 cm}
\vspace{-0.5 cm}

\begin{figure}[H]
    \centering
    \includegraphics[width=1\linewidth]{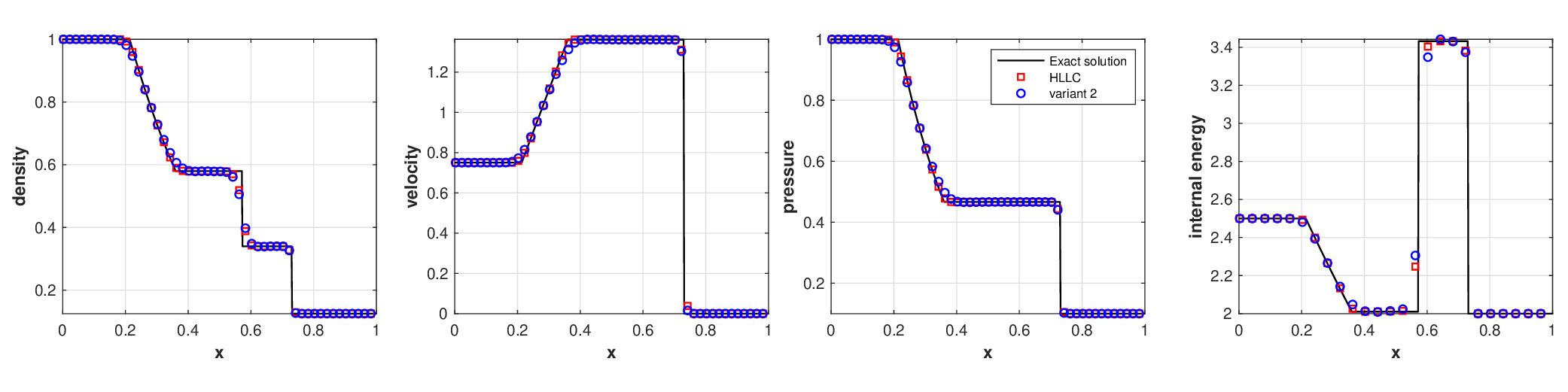}
    \vspace{-1 cm}
    \caption{Sod's shock tube problem with variant 2}
    \label{1D_figure6}
\end{figure}

\begin{figure}[H]
    \centering
    \includegraphics[width=1\linewidth]{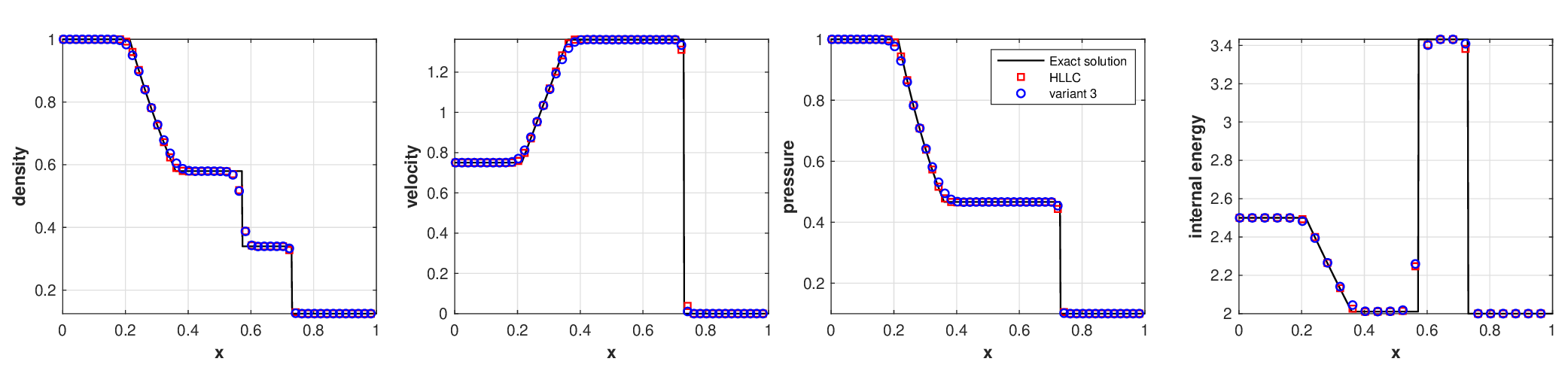}
    \vspace{-0.8 cm}
    \caption{Sod's shock tube problem with variant 3}
    \label{1D_figure7}
\end{figure}

\begin{figure}[H]
    \centering
    \includegraphics[width=1\linewidth]{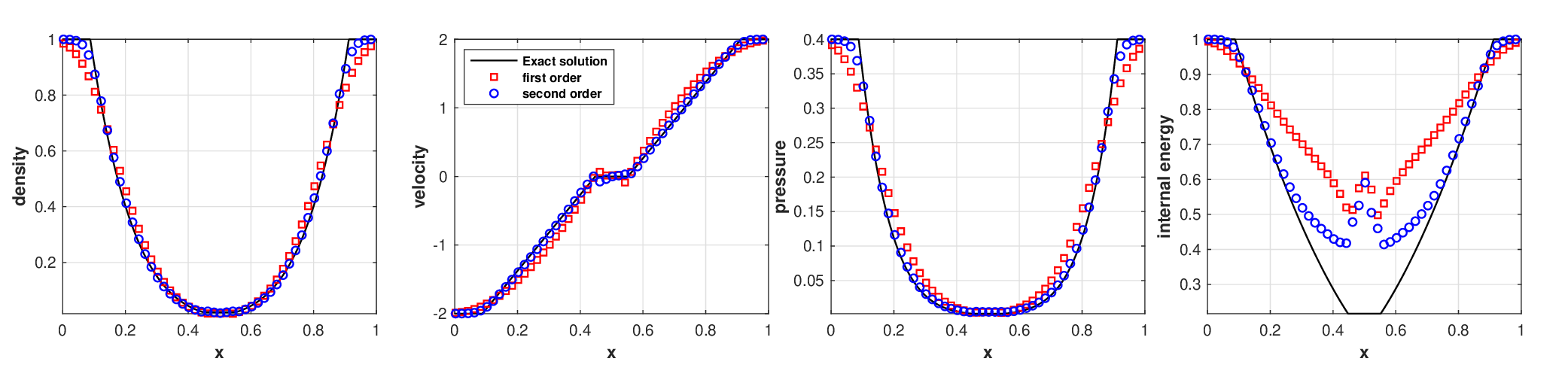}
    \vspace{-0.8 cm}
    \caption{Overheating problem with variant 2}
    \label{1D_figure8}
\end{figure}

\begin{figure}[H]
    \centering
    \includegraphics[width=1\linewidth]{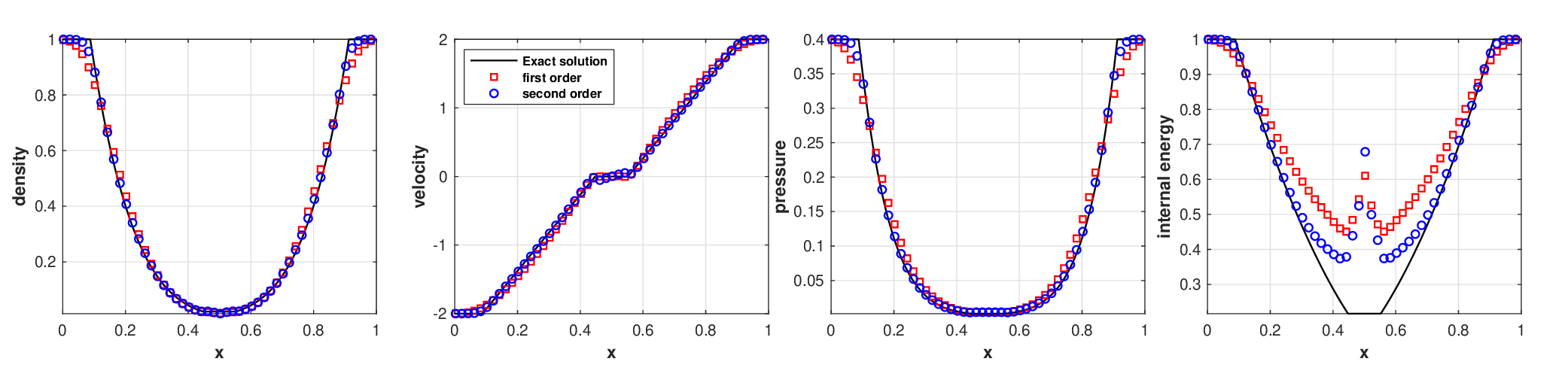}
    \vspace{-0.8 cm}
    \caption{Overheating problem with variant 3}
    \label{1D_figure9}
\end{figure}

Numerical test 4 is the modified version of the Sod's shock tube problem; its analytical solution consists of a left-sonic expansion, a right-moving contact discontinuity, and a right-moving shock. This test assesses the entropy satisfaction property of a numerical method. As seen in Figures \ref{1D_figure6} and \ref{1D_figure7}, the results of HLLC and modified PVU schemes are coinciding, and no entropy violating solutions are obtained.
\vspace{-0.5 cm}
The exact solution of test 5 consists of left and right expansion fans connected by a stationary contact discontinuity. This test helps assess the performance of a numerical scheme near low-density regions. As seen in Figures \ref{1D_figure8} and \ref{1D_figure9}, both variants of the modified PVU Boltzmann scheme do not fail near low densities. Nevertheless, both variants exhibit an overshoot in the internal energy near the center, a feature typical of most upwind schemes.
\vspace{-0.5 cm}
Numerical test 6 is also designed to test the robustness of a scheme; the solution consists of three strong discontinuities moving to the right. Both variants of the modified PVU scheme capture the flow features reasonably accurately.
\vspace{-0.4 cm}
\begin{figure}[H]
    \centering
    \includegraphics[width=1\linewidth]{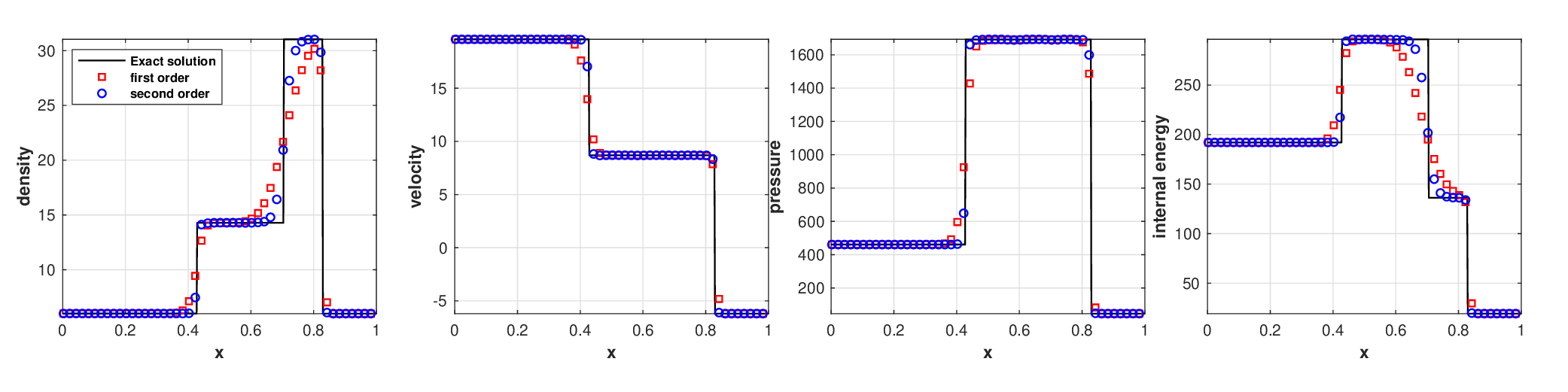}
    \vspace{-1 cm}
    \caption{Colliding strong shocks with variant 2}
    \label{1D_figure10}
\end{figure}
\vspace{-0.8 cm}
\begin{figure}[H]
    \centering
    \includegraphics[width=1\linewidth]{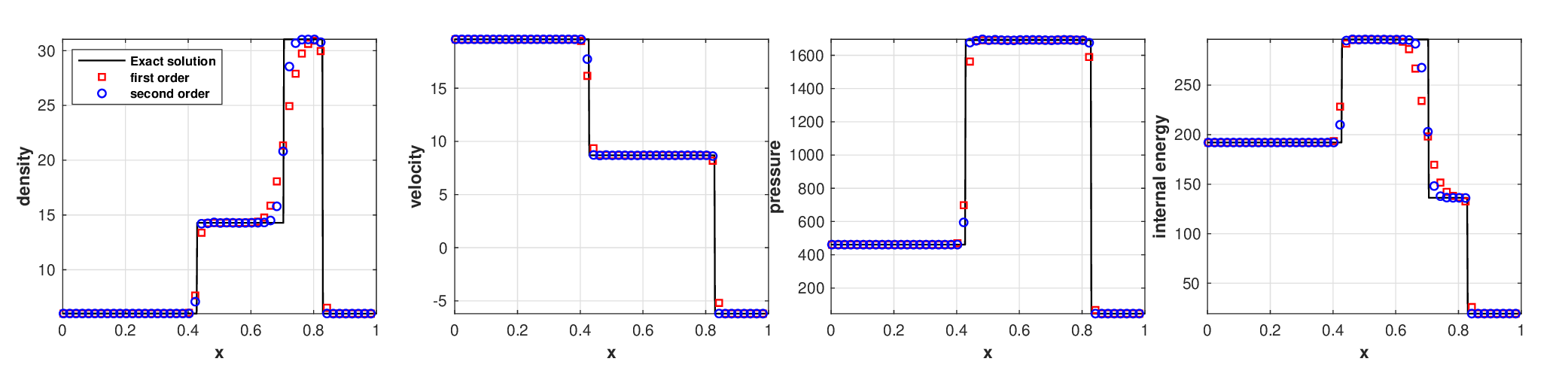}
    \vspace{-1 cm}
    \caption{Colliding strong shocks with variant 3}
    \label{1D_figure11}
\end{figure}

\subsection{2D results and discussions}
\label{subsec 8.3}
In this section, we test both variants of the modified PVU scheme on standard 2D inviscid gas dynamics problems to assess their robustness, accuracy, and positivity-preserving properties. We use the kinetic characteristic boundary condition (KCBC) at the walls to ensure flow tangency. Unless otherwise mentioned, the CFL numbers of 0.4 and 0.2 are used to generate first-order and second-order results, respectively. For steady-state convergence, the first-order solution is advanced until the density residual reaches at least $ 10^{-8} $ is achieved, while the second-order solution is run until the residual reaches at least $ 10^{-6} $ is achieved.  We set K (a free parameter in the Venkatakrishnan limiter) to $ 0.3 $ for unsteady problems and $ 5 $ for steady problems, unless otherwise mentioned.

\subsubsection{Oblique shock reflection}
\label{subsubsec 8.3.1}
\begin{figure}[!h]
    \centering
    \begin{subfigure}{0.45\textwidth}
        \centering
        \includegraphics[width=\textwidth]{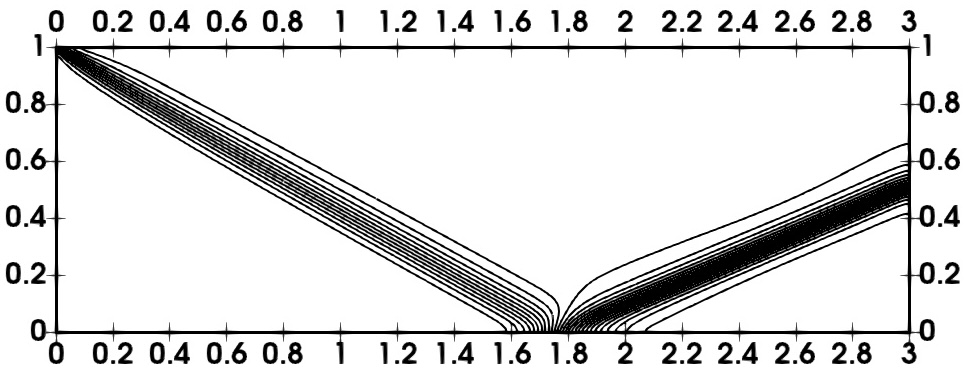}
    \end{subfigure}
    \begin{subfigure}{0.45\textwidth}
        \centering
        \includegraphics[width=\textwidth]{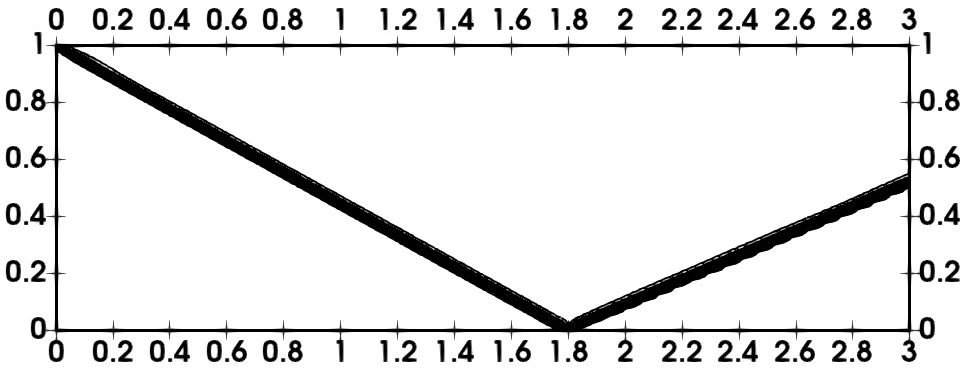}
    \end{subfigure}
    
    \begin{subfigure}{0.45\textwidth}
        \centering
        \includegraphics[width=\textwidth]{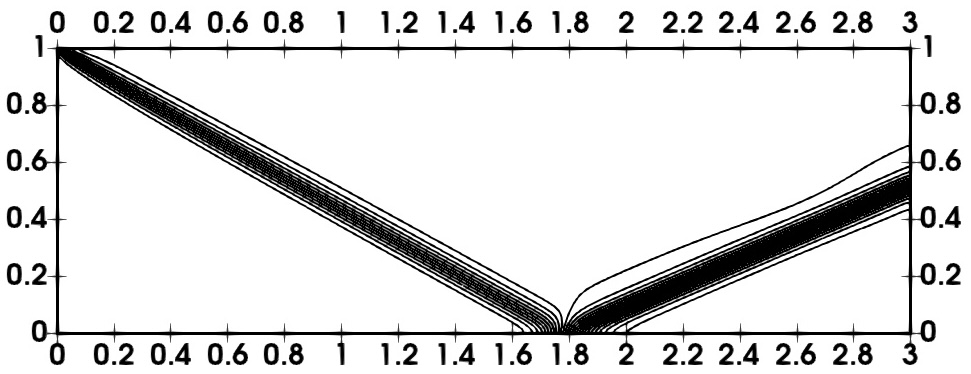}
    \end{subfigure}
    \begin{subfigure}{0.45\textwidth}
        \centering
        \includegraphics[width=\textwidth]{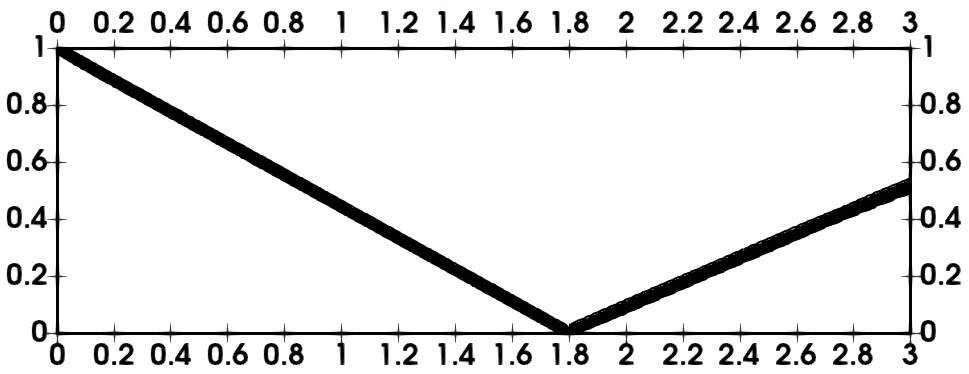}
    \end{subfigure}
    \caption{Oblique shock reflection; density contours (1:0.053:2.6); (Top) variant 2, (Bottom) variant 3, (Left) First-order, (Right) Second-order}
    \label{2D_figure1}
\end{figure}

In this test, an oblique shock impinges on a solid wall and reflects from it. The undisturbed flow is at Mach 2.9. The computational domain is a rectangle of dimensions $ [3, 0] \times [0, 1] $. A Cartesian mesh comprising $ 240 \times 80 $ cells is employed over the computational domain. The left boundary is prescribed with a supersonic inlet condition, while the kinetic characteristic boundary condition (KCBC) is imposed at the bottom wall, post-shock states, obtained via the inviscid relations for gas dynamics, are applied at the top wall, and the zero gradient boundary condition is used at the right boundary. Free-stream conditions are used to initialize the domain. Figure \ref{2D_figure1} presents the density contours obtained using the first- and second-order variants of the modified PVU scheme. As expected, the shock profile is sharper in second-order accurate results.
\begin{figure}[!h]
    \centering
    \begin{subfigure}{0.49\textwidth}
        \centering
        \includegraphics[width=\textwidth]{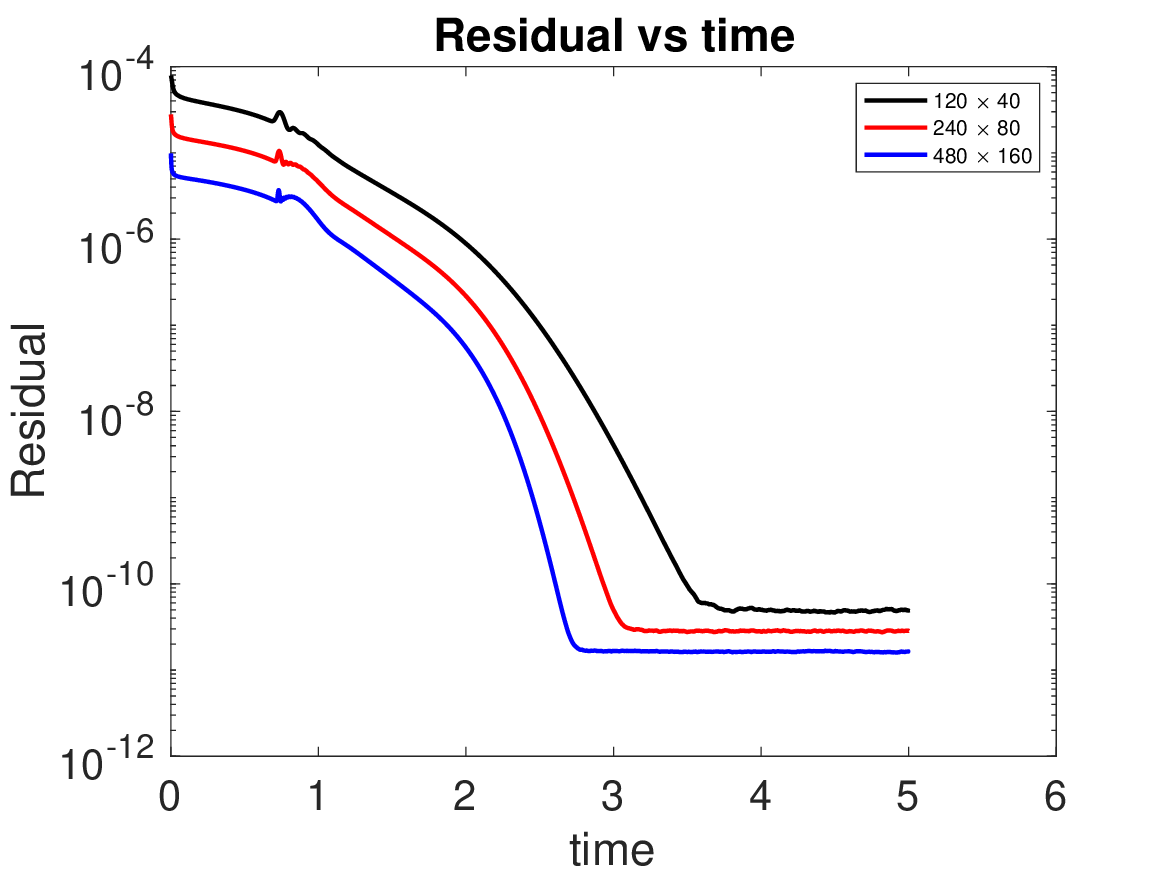}
    \end{subfigure}
    \begin{subfigure}{0.49\textwidth}
        \centering
        \includegraphics[width=\textwidth]{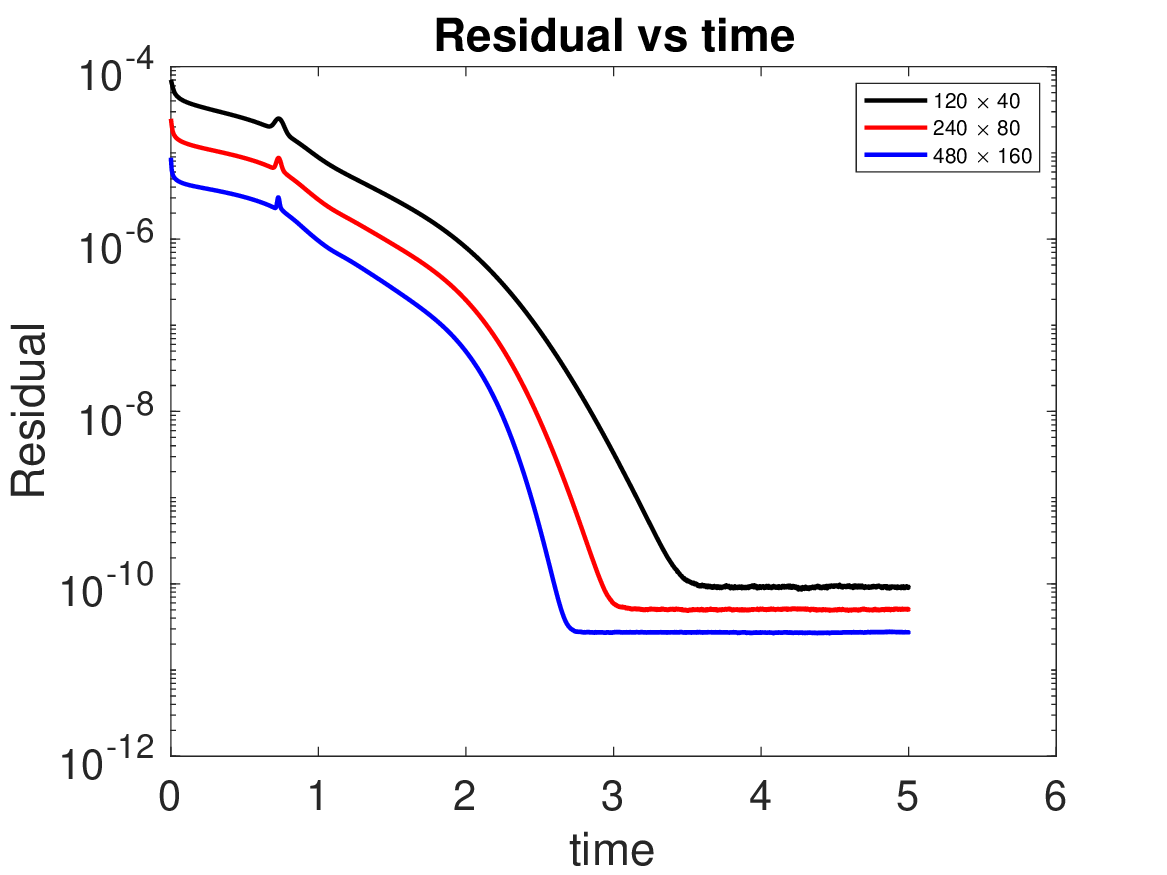}
    \end{subfigure}
    \begin{subfigure}{0.49\textwidth}
        \centering
        \includegraphics[width=\textwidth]{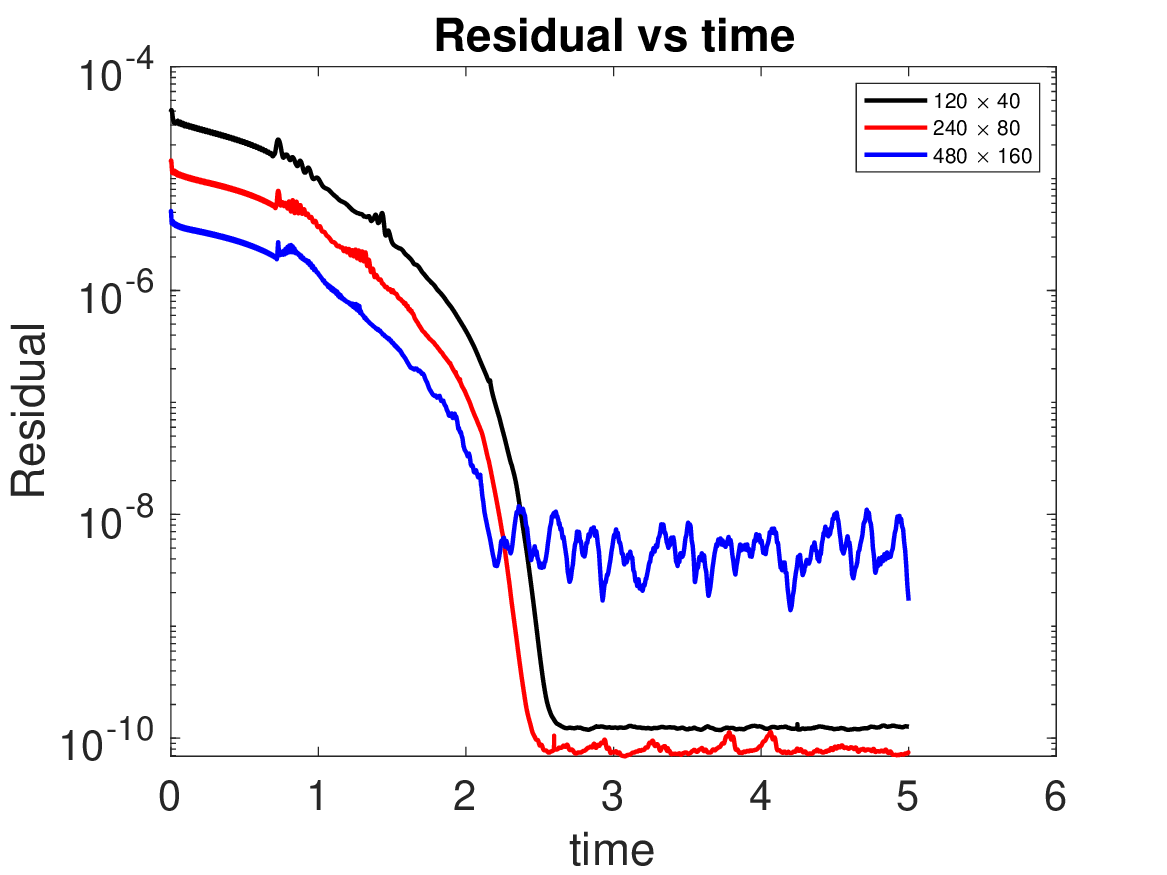}
    \end{subfigure}
    \begin{subfigure}{0.49\textwidth}
        \centering
        \includegraphics[width=\textwidth]{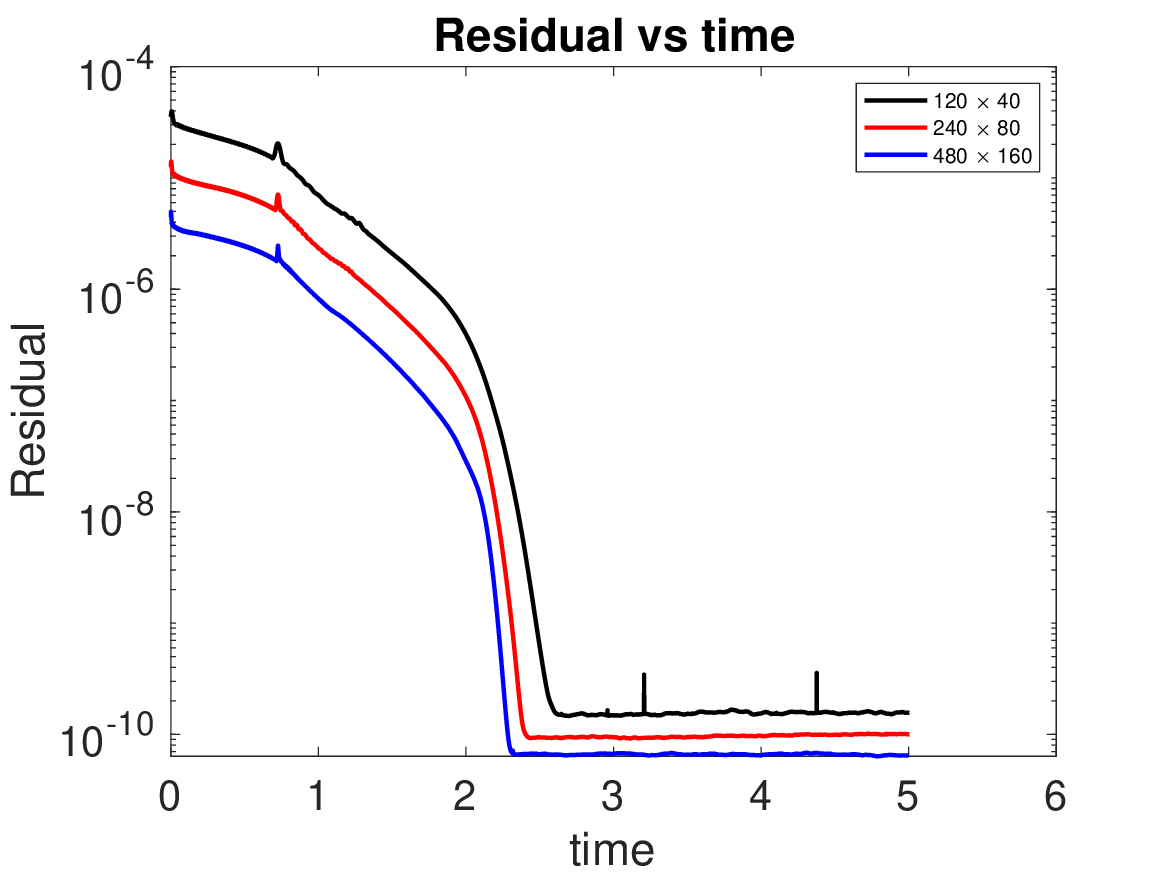}
    \end{subfigure}
    \caption{RMS error in density over time for the oblique shock reflection test case; (Top) first-order result, (Bottom) second-order result, (Left) variant 2, (Right) variant 3}
    \label{2D_figure2}
\end{figure}
The convergence behavior of the modified PVU Boltzmann scheme (variants 2 and 3) is assessed for the oblique shock reflection test case on three successively refined meshes: $ 120 \times 40 $, $ 240 \times 80 $ and $ 480 \times 160 $. Figure \ref{2D_figure2} presents the residuals as functions of time (representing time-marching iterations) for first-order and second-order accurate results. For first-order accurate results, the residuals decrease with mesh refinement, although not strictly monotonically; a characteristic non-monotonic bump appears around time $ 0.8 $, which is typical for many schemes. The second-order results exhibit faster residual reduction and achieve similarly low values, confirming the expected accuracy and robustness of both schemes. Notably, second-order accurate solutions corresponding to variant 2 and variant 3 achieve density convergences of at least $ 10^{-8} $ and $ 10^{-10} $, respectively, on all meshes.

\subsubsection{Supersonic flow over a compression ramp}
\label{subsubsec 8.3.2}

\begin{figure}[h]
    \centering
    \begin{subfigure}{0.45\textwidth}
        \centering
        \includegraphics[width=\textwidth]{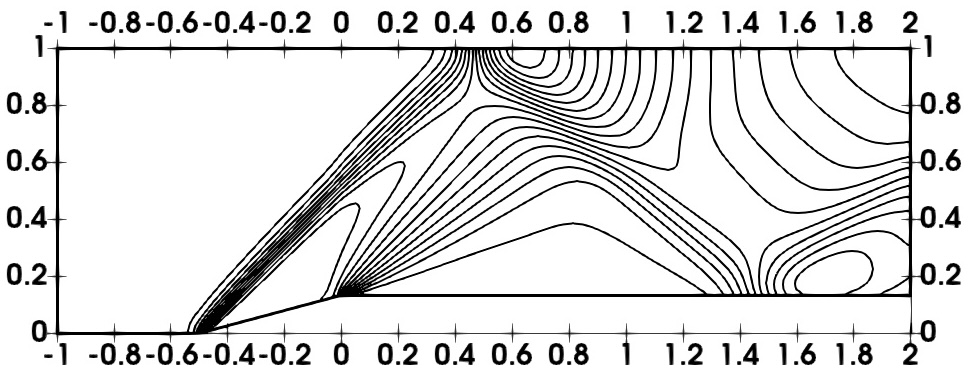}
    \end{subfigure}
    \begin{subfigure}{0.45\textwidth}
        \centering
        \includegraphics[width=\textwidth]{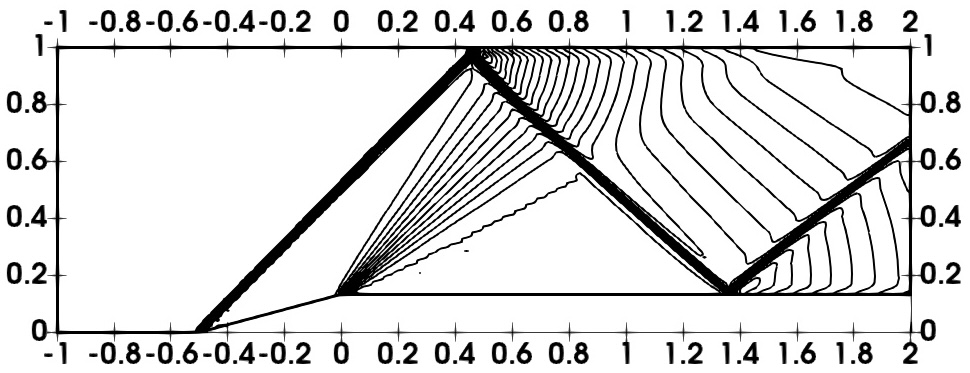}
    \end{subfigure}

    \begin{subfigure}{0.45\textwidth}
        \centering
        \includegraphics[width=\textwidth]{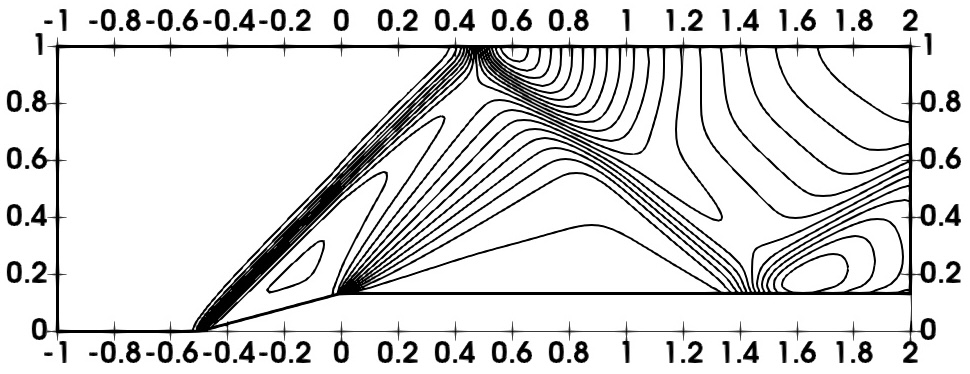}
    \end{subfigure}
    \begin{subfigure}{0.45\textwidth}
        \centering
        \includegraphics[width=\textwidth]{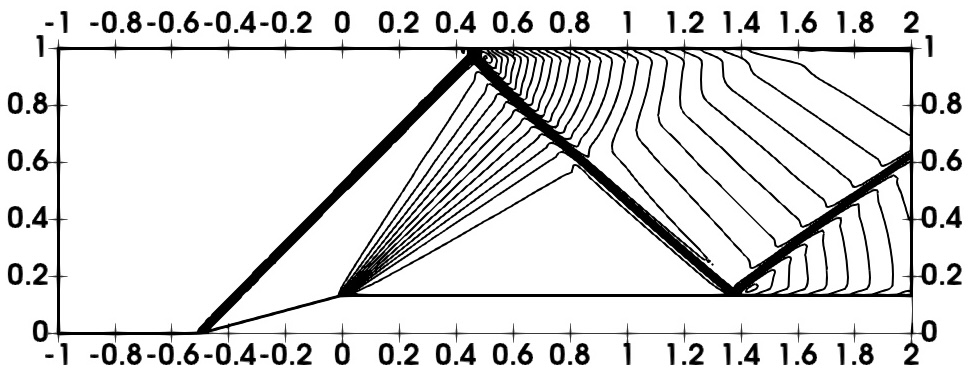}
    \end{subfigure}
    \caption{Steady supersonic flow over a compression wedge; density contours (1.2:0.096:3.8); (Top) variant 2, (Bottom) variant 3, (Left) First-order, (Right) Second-order}
    \label{2D_figure3}
\end{figure}

This test consists of a Mach 2 flow over a $ 15^{\circ} $ compression ramp placed inside a wind tunnel. The dimensions of the computational domain are $ [-1, 2] \times [0, 1] $, with a $ 15^{\circ} $ ramp at the bottom that starts from $ x = -0.5 $ to $ x = 0 $. The computational domain comprises $ 240 \times 80 $ cells. The supersonic inlet boundary condition is used at the left boundary of the domain. The kinetic characteristic boundary condition is used on the top and bottom walls. The zero-gradient boundary condition is used at the right boundary of the domain. Free-flow conditions are assumed everywhere within the computational domain at time $ t = 0 $. Figure \ref{2D_figure3} shows the steady-state results obtained by both variants of the modified PVU scheme. The steady-state solution consists of an oblique shock that emanates from the concave corner of the ramp and an expansion fan originating from the convex corner of the ramp. The oblique shock is reflected from the top wall and interacts with an expansion fan. The shock subsequently hits the bottom wall, reflects, interacts further with the reflected expansion, and finally exits the domain through the right boundary.

\subsubsection{Supersonic flow over a forward-facing step}
\label{subsubsec 8.3.3}
\begin{figure}[!h]
    \centering
    \begin{subfigure}{0.45\textwidth}
        \centering
        \includegraphics[width=\textwidth]{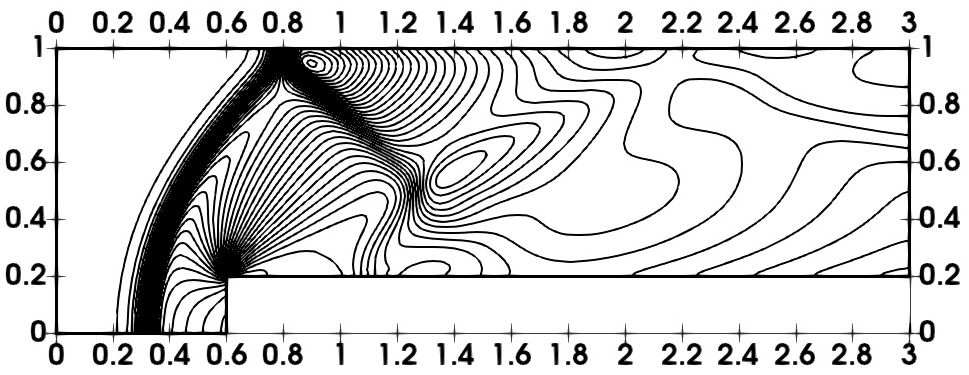}
    \end{subfigure}
    \begin{subfigure}{0.45\textwidth}
        \centering
        \includegraphics[width=\textwidth]{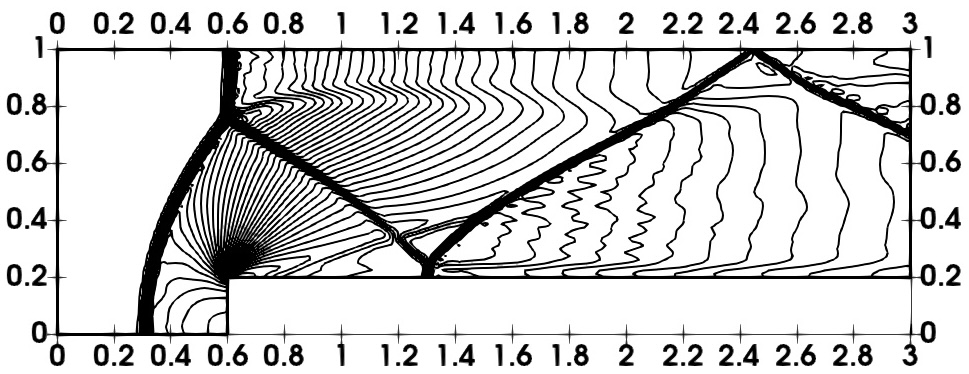}
    \end{subfigure}
    
    \begin{subfigure}{0.45\textwidth}
        \centering
        \includegraphics[width=\textwidth]{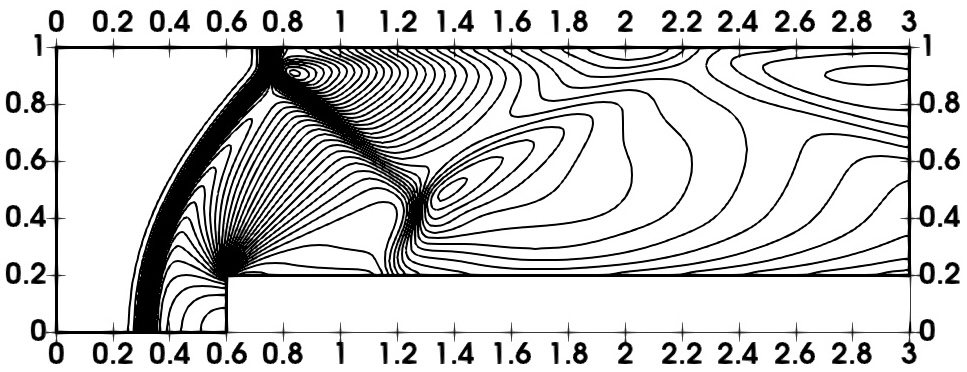}
    \end{subfigure}
    \begin{subfigure}{0.45\textwidth}
        \centering
        \includegraphics[width=\textwidth]{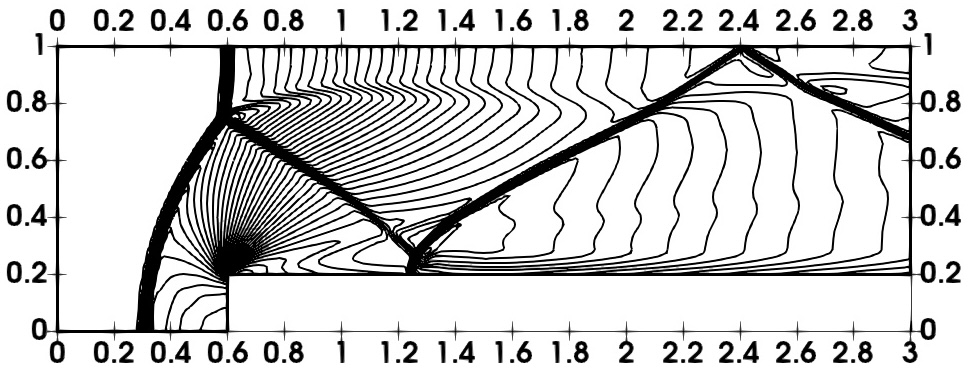}
    \end{subfigure}
    \caption{Supersonic flow over a forward-facing step; density contours (0.2:0.15:7.0); (Top) variant 2, (Bottom) variant 3, (Left) First-order, (Right) Second-order}
    \label{2D_figure4}
\end{figure}

This problem consists of a supersonic flow of Mach 3 in a wind tunnel with a step. The wind tunnel is $ 1 $ length units wide and $ 3 $ length units long. The step is $ 0.2 $ length units high and is placed $ 0.6 $ length units from the left end of the tunnel. The computational domain is divided into $ 240 \times 80 $ Cartesian cells. The left boundary is prescribed with a supersonic inlet condition, a kinetic characteristic boundary condition is applied along the upper and lower walls, and a zero gradient boundary condition is prescribed at the tunnel's right boundary. The initial conditions corresponding to the free stream are assumed everywhere. The flow evolves until $ t = 4 $ time units. Second-order results are obtained using the R5 limiter with $ K = 0.3 $. Figure \ref{2D_figure4} compares the first- and second-order results obtained for this test case. In the first-order results, the second reflected shock obtained using variant 2 is observed to be relatively more kinked compared to that obtained using variant 3. This behavior, however, is not observed in the corresponding second-order results, where both variants yield comparably well-resolved shock structures. Notably, both variants yield sharp resolution of the primary bow shock, Mach stem, and reflected shocks at second order.

\subsubsection{Horizontal slip flow}
\label{subsubsec 8.3.4}

\begin{figure}[!h]
    \centering
    \begin{subfigure}{0.45\textwidth}
        \centering
        \includegraphics[width=\textwidth]{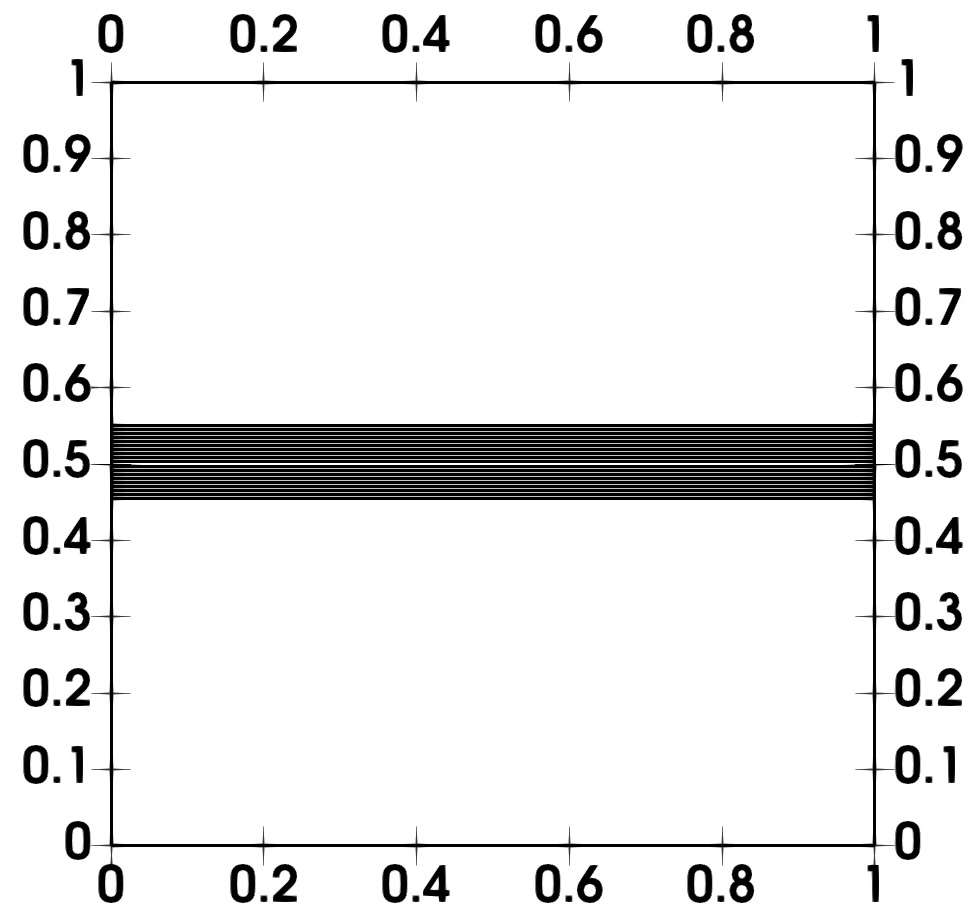}
    \end{subfigure}
    \begin{subfigure}{0.45\textwidth}
        \centering
        \includegraphics[width=\textwidth]{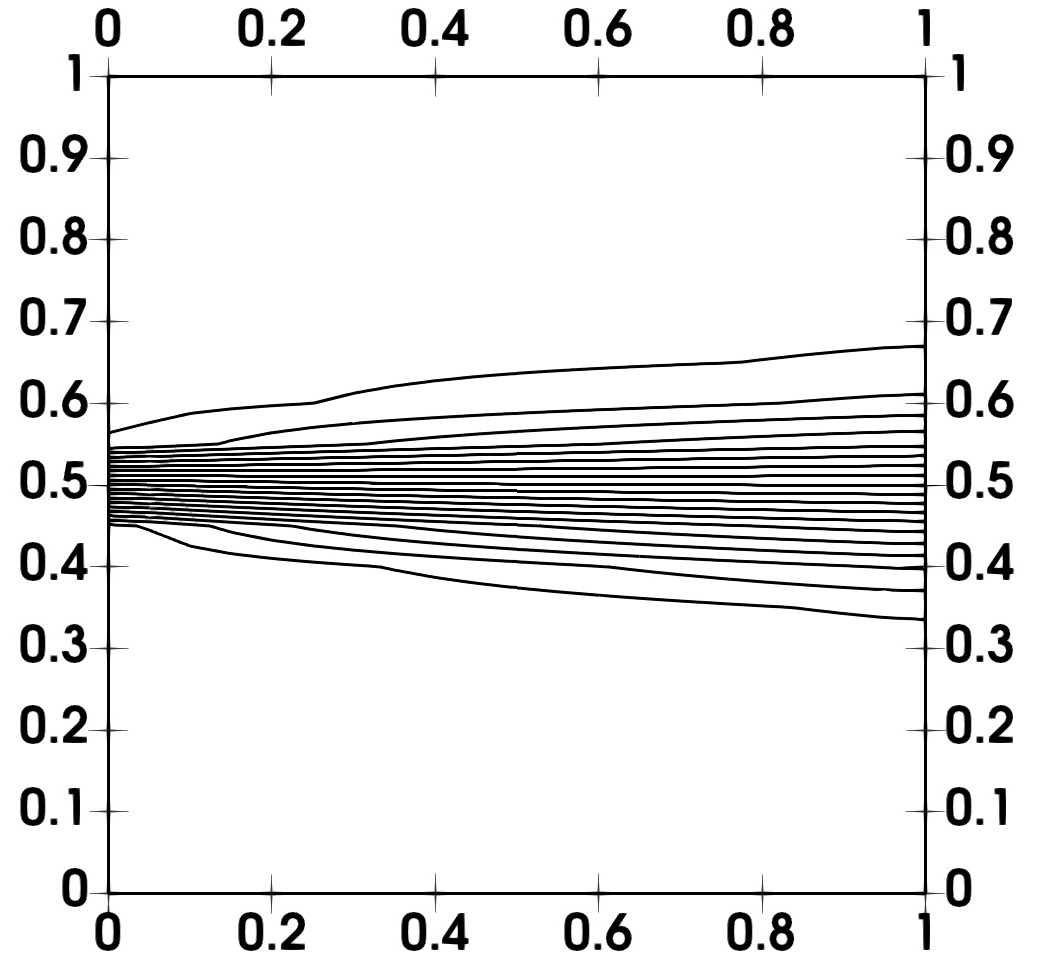}
    \end{subfigure}
    \caption{Horizontal slip flow; Horizontal velocity contours (2.0:0.05:3.0); variant 2 (Left), variant 3 (Right)}
    \label{2D_figure5}
\end{figure}
A Mach $ 3.0 $ stream slides over a Mach $ 2.0 $ stream in this test. The computational domain is a unit square spanning $ [1, 0] \times [0, 1] $. The domain is discretized using $ 20 \times 20 $ Cartesian cells. Initially, a horizontal Mach $ 3.0 $ flow occupies the upper half ($ y > 0.5 $), while the lower half ($ y < 0.5 $) contains a Mach $ 2.0 $ flow, with the same density and pressure throughout. The supersonic inflow and outflow boundary conditions are used at the left and right boundaries, respectively. The zero-gradient boundary condition is employed at the top and bottom boundaries. Figure \ref{2D_figure5} displays the horizontal velocity contours obtained using the first-order variant 2 and variant 3 of the modified PVU scheme. As the results indicate, variant 2 reproduces exactly a grid-aligned, stationary, isolated slip surface, free from any numerical diffusion. This test case also demonstrates that a scheme that accurately captures the one-dimensional stationary contact discontinuity need not necessarily capture the two-dimensional slip surface with the same precision.

\subsubsection{Odd-even decoupling}
\label{subsubsec 8.3.5}

\begin{figure}[h!]
    \centering
    \begin{subfigure}{0.45\textwidth}
        \centering
        \includegraphics[width=\textwidth]{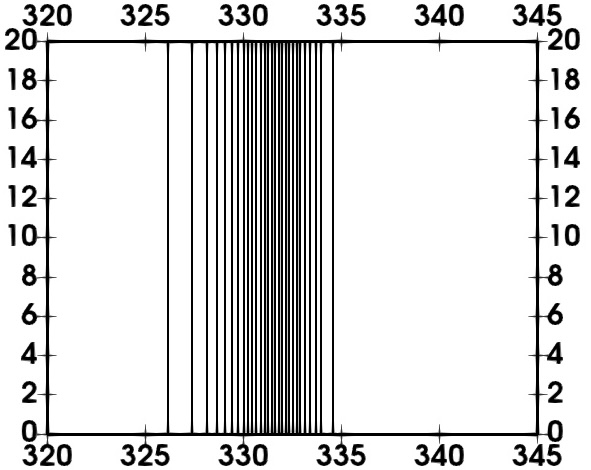}
    \end{subfigure}
    \begin{subfigure}{0.45\textwidth}
        \centering
        \includegraphics[width=\textwidth]{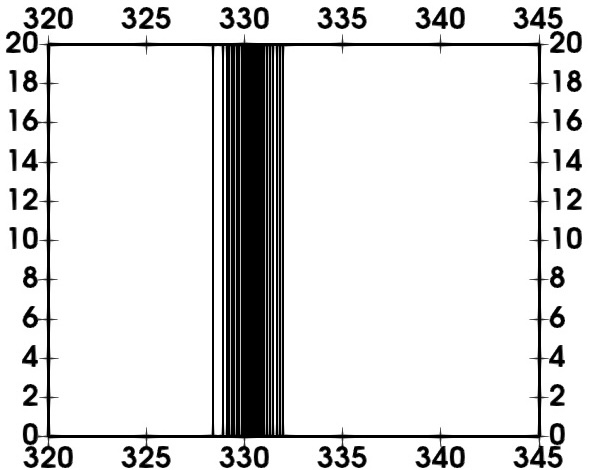}
    \end{subfigure}
    \caption{Odd-even decoupling; density contours (1.4:0.2:7.4); variant 2, (Left) First-order, (Right) Second-order}
    \label{2D_figure6}
\end{figure}
\begin{figure}
    \begin{subfigure}{0.45\textwidth}
        \centering
        \includegraphics[width=\textwidth]{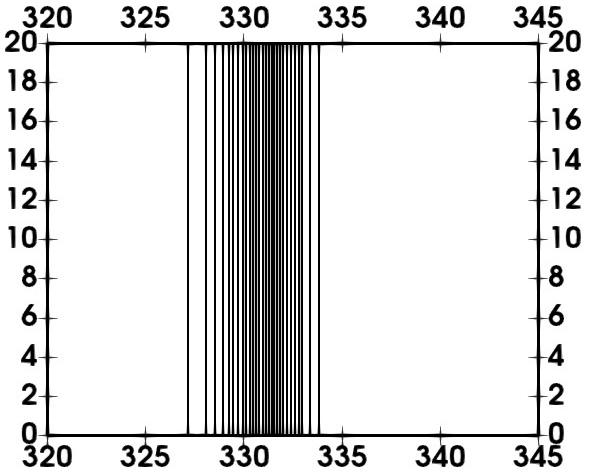}
    \end{subfigure}
    \begin{subfigure}{0.45\textwidth}
        \centering
        \includegraphics[width=\textwidth]{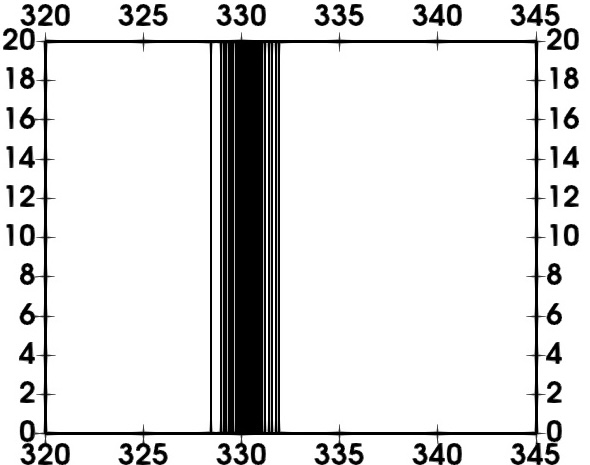}
    \end{subfigure}
    \caption{Odd-even decoupling; density contours (1.4:0.2:7.4); variant 3, (Left) First-order, (Right) Second-order}
    \label{2D_figure7}
\end{figure}

This test is designed to assess the performance of a numerical scheme concerning a shock instability popularly known as odd-even decoupling in the literature \cite{Quirk}. In this test, a normal shock with a Mach number of $ 6 $ moves through the rectangular channel. The computational domain is divided into $ 800 \times 20 $ square cells of unit length, the center-line of which is perturbed as 
\begin{subequations}
    \begin{align}
        y_{i, jmid} &= y_{jmid} + 10^{-3} \quad \text{for i even}, \\
        y_{i, jmid} &= y_{jmid} - 10^{-3} \quad \text{for i odd}.
    \end{align}
\end{subequations}
The left boundary admits the supersonic inflow, while the top and bottom boundaries are treated using the kinetic characteristic condition. A zero-gradient condition is prescribed at the downstream outlet. The planar shock evolves until $ 55 $ time units. The first- and second-order solutions corresponding to both variants of the modified PVU scheme are shown in Figures \ref{2D_figure6} and \ref{2D_figure7}. As seen in the figures, both variants are free from this form of instability (which results in the complete breakdown of the solution in some schemes like HLLC \cite{Mandal_Panwar}).

\subsubsection{Positivity test case}
\label{subsubsec 8.3.6}

\begin{figure}[h!]
    \centering
    \begin{subfigure}{0.45\textwidth}
        \centering
        \includegraphics[width=\textwidth]{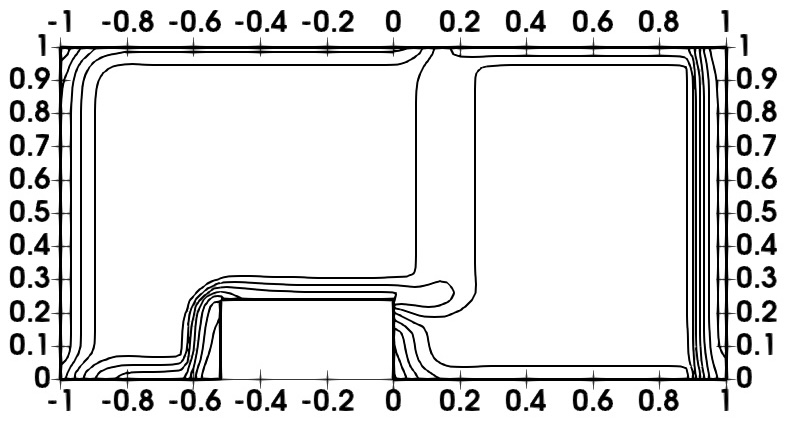}
    \end{subfigure}
    \begin{subfigure}{0.45\textwidth}
        \centering
        \includegraphics[width=\textwidth]{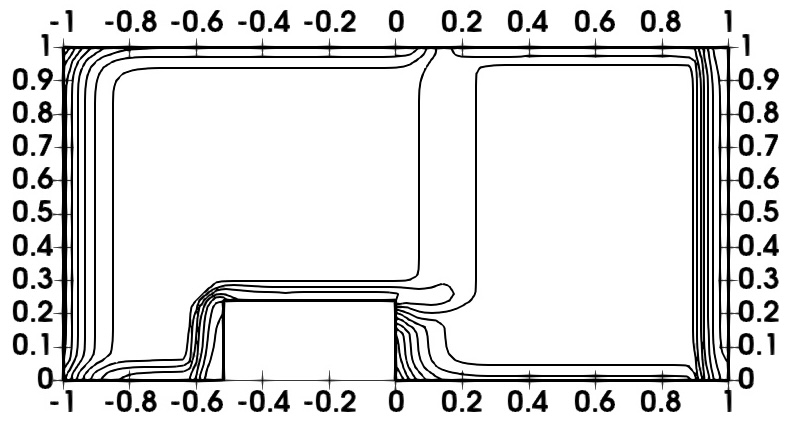}
    \end{subfigure}
    \caption{2D Enclosure test; 40 logarithmic pressure contours ($ 10^{-6} $ - $ 10^6 $); (Left) variant 2, (Right) variant 3}
    \label{2D_figure8}
\end{figure}

This test case is taken from Parent \cite{Parent} and is designed to assess the positivity-preserving property of a numerical scheme. This test consists of large initial Mach numbers, which leads to strong expansions in the first few iterations. It is an excellent test to empirically determine the positivity-preserving property of a scheme (as non-positive schemes fail) and both variants of the modified PVU Boltzmann scheme successfully simulate the test case without breaking down. The computational region spans $ x \in [-1, 1] $ and $ y \in [0, 1] $, with a cutout of height $ 0.24 $ located at $ x = -0.52 $. We partition the computational domain into square Cartesian cells of dimensions $ 0.02 \times 0.02 $ units. The kinetic characteristic boundary condition is employed on all walls to ensure the flow tangency. The initial conditions are as follows:
\begin{subequations}
    \begin{align}
        M_x &= 10, \; M_y = -3, \; p = 0.1 \; \text{bar}, \; T = 35 \; \text{K} \quad &\forall x \leq 0 \\
        M_x &= 10, \; M_y = 2, \; p = 0.1 \; \text{bar}, \; T = 35 \; \text{K} \quad &\forall x \geq 0
    \end{align}
\end{subequations}
The CFL number is fixed at 0.25, and the flow is allowed to evolve for 100 numerical iterations. Figure \ref{2D_figure8} shows the pressure contours corresponding to the first-order accuracy in both variants of the modified PVU scheme. As expected, both variants of the modified PVU scheme preserve the positivity of pressure and density.

\subsubsection{Shock diffraction}
\label{subsubsec 8.3.7}
\begin{figure}[H]
    \centering
    \begin{subfigure}{0.45\textwidth}
        \centering
        \includegraphics[width=\textwidth]{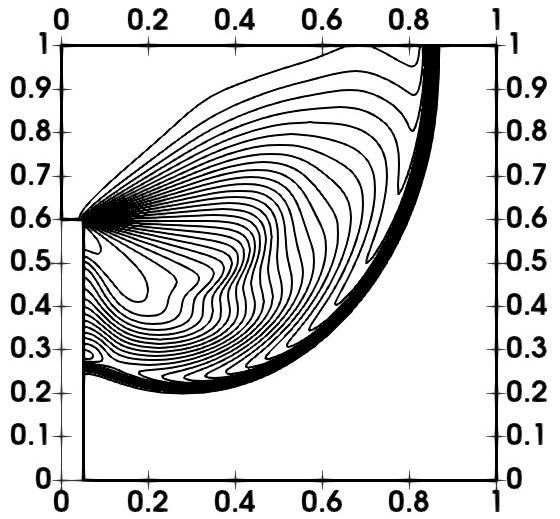}
    \end{subfigure}
    \begin{subfigure}{0.45\textwidth}
        \centering
        \includegraphics[width=\textwidth]{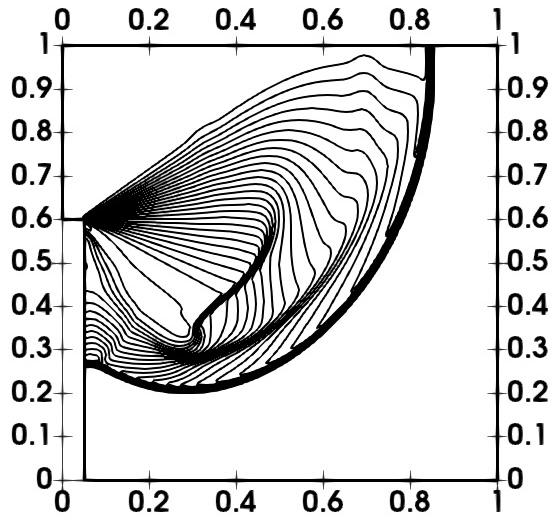}
    \end{subfigure}
    \caption{Shock diffraction; density contours (0.095:0.233:7.1); variant 2, (Left) First-order, (Right) Second-order}
    \label{2D_figure9}
\end{figure}

\begin{figure}[ht]
    \centering
    \begin{subfigure}{0.45\textwidth}
        \centering
        \includegraphics[width=\textwidth]{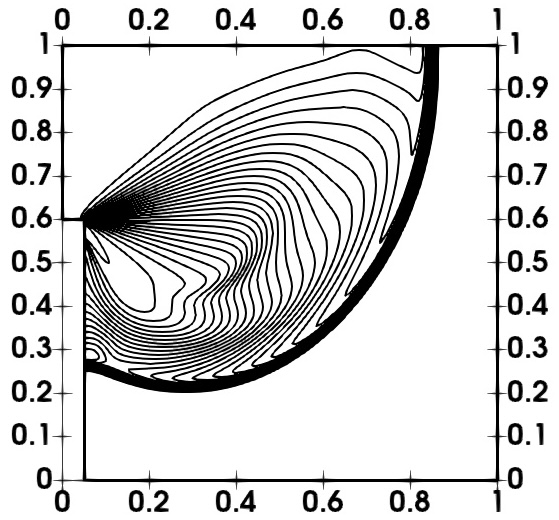}
    \end{subfigure}
    \begin{subfigure}{0.45\textwidth}
        \centering
        \includegraphics[width=\textwidth]{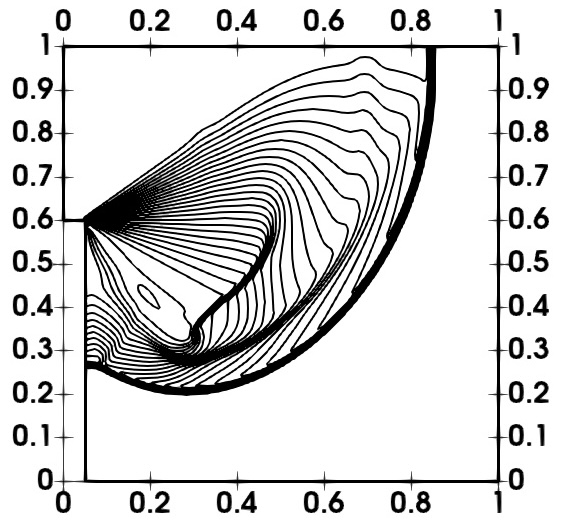}
    \end{subfigure}
    \caption{Shock diffraction; density contours (0.095:0.233:7.1); variant 3, (Left) First-order, (Right) Second-order}
    \label{2D_figure10}
\end{figure}

In this test case, a normal shock of Mach 5.09 diffracts around a $ 90^{\circ} $ corner. The computational domain is a square of $ 1 $ length units, which is divided into $ 200 \times 200 $ Cartesian cells. The corner is located at $ x = 0.05 $ and $ y = 0.45 $ length units. The initial location of the shock is at the corner. The domain to the right of the shock is initialized with the pre-shock conditions of $ \rho = 1.4, \, u_1 = 0.0, \, u_2 = 0.0, \, p = 1.0 $. The domain to the left of the shock is initialized using the post-shock conditions. The boundary conditions used are the supersonic inlet boundary condition at the left boundary, the kinetic characteristic boundary condition at the walls, and the zero gradient boundary condition at the right boundary of the domain. The solution evolves until $ 0.1561 $ time units. Results obtained with first-order and second-order accuracy for both variants of the modified PVU Boltzmann scheme are shown in Figures \ref{2D_figure9} and \ref{2D_figure10}. 

\vspace{-0.4 cm}

As seen in the figures, the numerical results are free from any entropy violating solutions and oscillations. The first-order solution obtained using the HLLC scheme exhibits an instability ahead of the shock near the upper boundary \cite{Mandal_Panwar}. This oscillatory behavior becomes more pronounced in the higher-order results obtained using the HLLC scheme.

\subsubsection{Double Mach reflection}
\label{subsubsec 8.3.8}

\begin{figure}[h]
    \centering
    \begin{subfigure}{0.45\textwidth}
        \centering
        \includegraphics[width=\textwidth]{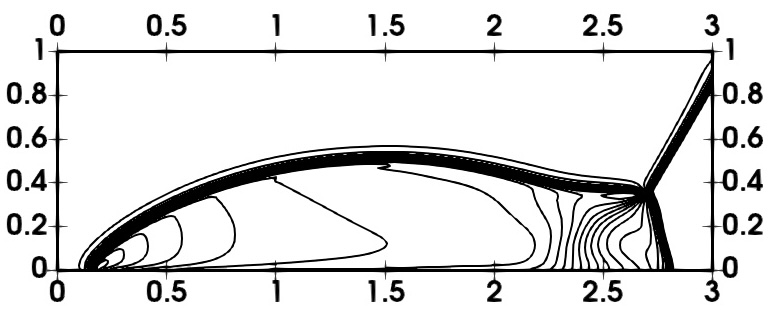}
    \end{subfigure}
    \begin{subfigure}{0.45\textwidth}
        \centering
        \includegraphics[width=\textwidth]{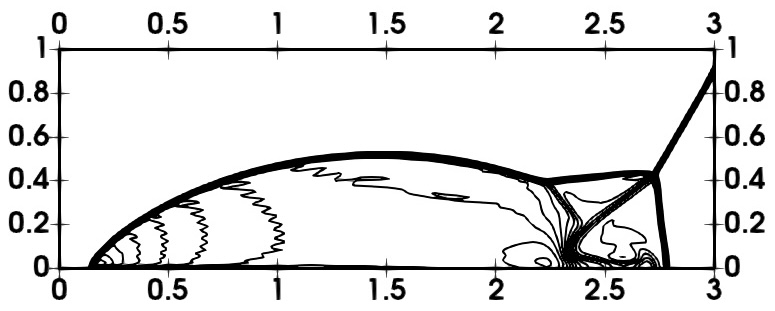}
    \end{subfigure}

    \begin{subfigure}{0.45\textwidth}
        \centering
        \includegraphics[width=\textwidth]{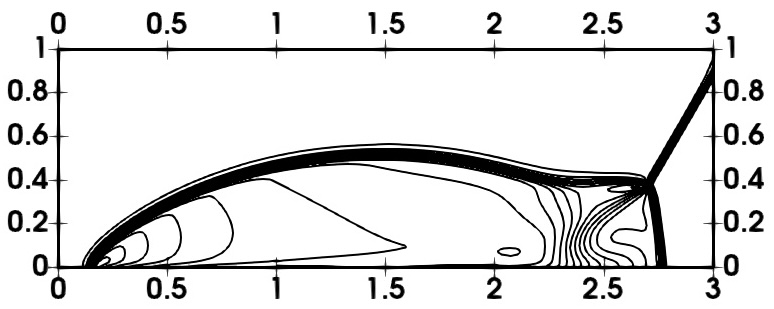}
    \end{subfigure}
    \begin{subfigure}{0.45\textwidth}
        \centering
        \includegraphics[width=\textwidth]{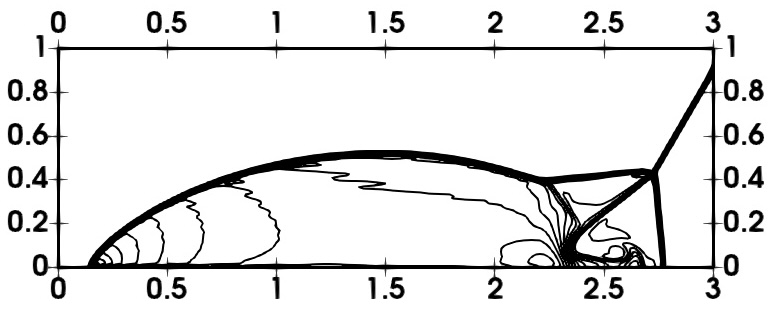}
    \end{subfigure}
    \caption{Double Mach reflection; density contours (1.4:0.67:21.5); (Top) variant 2, (Bottom) variant 3, (Left) First-order, (Right) Second-order}
    \label{2D_figure11}
\end{figure}
This test case is designed to assess the performance of a numerical scheme with respect to another shock instability known as a kinked Mach stem. The problem is solved by keeping the domain rectangular and forcing an oblique shock through it \cite{Woodward_Colella}. The computational domain is four units long and one unit wide. It is divided into $ 480 \times 160 $ Cartesian cells. An oblique Mach 10 shock that initially has an angle of $ 60^{\circ} $ with the bottom wall at $ x = 1/6 $ is propagated in a channel. The domain to the right of the shock is initialized with pre-shock values of $ \rho = 1.4, u_1 = 0.0, u_2 = 0.0, p = 1.0 $, and to the left is initialized using post-shock values. The inlet boundary condition is set to post-shock values, and the outlet boundary is set to zero gradient. The top boundary condition is set to simulate the actual shock movement. The bottom wall up to $ x = 1/6 $ is set to post-shock values and the kinetic characteristic boundary condition is used for the remaining wall. The flow evolves until $ 0.2 $ time units. Some low diffusive numerical schemes are plagued by the unphysical kinked Mach stems. The HLLC scheme is known to exhibit such an instability \cite{Mandal_Panwar}, whereas neither variant of the modified PVU scheme produces an unphysical kink in the Mach stem, as shown in Figure \ref{2D_figure11}.
\par
\begin{figure}[h]
    \centering
    \begin{subfigure}{0.58\textwidth}
        \centering
        \includegraphics[width=\textwidth]{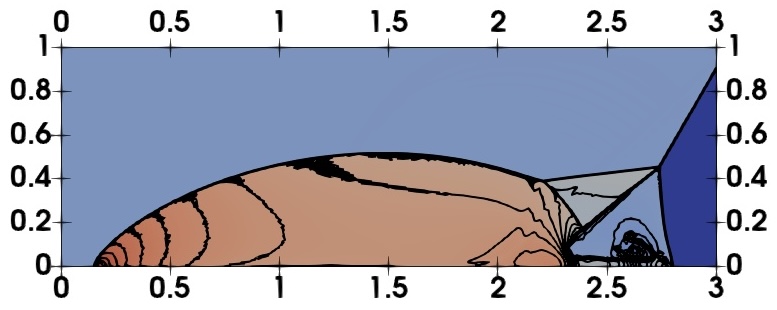}
    \end{subfigure}
    \hspace{0.1 em}
    \begin{subfigure}{0.38\textwidth}
        \centering
        \includegraphics[width=\textwidth]{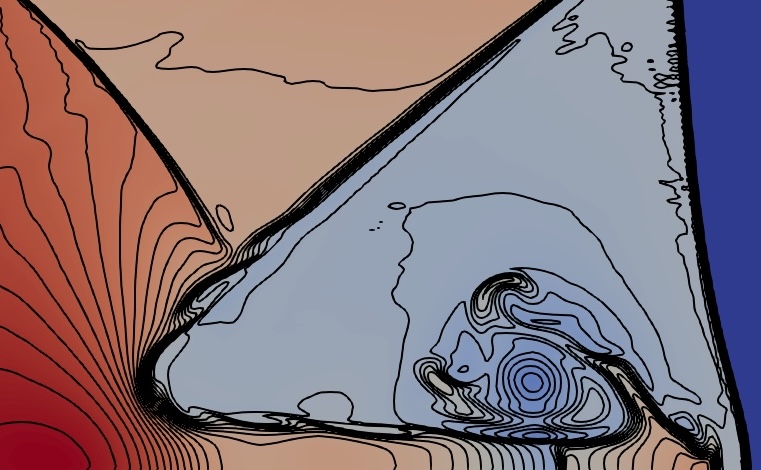}
    \end{subfigure}
    \vspace{-0.2 cm}
    \caption{Double Mach reflection test case using variant 2 of the second-order modified PVU Boltzmann scheme; density contours (1.77:0.41:22.44)}
    \label{2D_figure12}
\end{figure}
To further demonstrate the capability of the scheme to resolve fine-scale features, the test is repeated on a fine mesh of $ 1920 \times 640 $. Note that the Barth-Jespersen limiter has been used in conjunction with primitive variable-based reconstruction to generate this result on the fine mesh. The results for variant 2, presented in Figure \ref{2D_figure12}, show that the scheme cleanly captures the complex vortical structure along the primary slip line, underscoring its suitability for high-resolution simulations of demanding flow problems.

\subsubsection{Hypersonic flow over a half cylinder}
\label{subsubsec 8.3.9}

\begin{figure}[H]
    \centering
    \begin{subfigure}{0.45\textwidth}
        \centering
        \includegraphics[height = 8 cm]{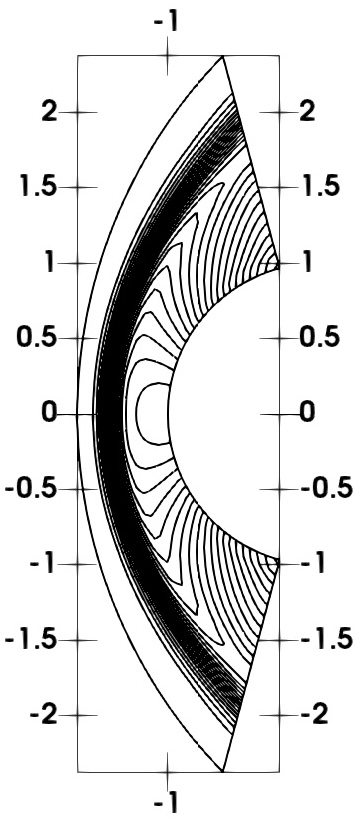}
    \end{subfigure}
    \hspace{-2 cm}
    \begin{subfigure}{0.45\textwidth}
        \centering
        \includegraphics[height = 8 cm]{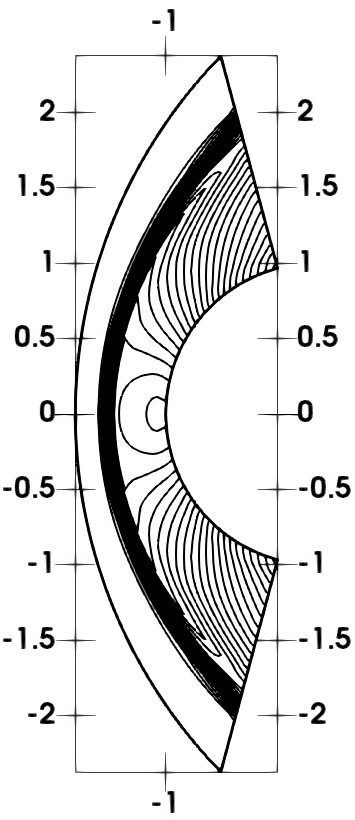}
    \end{subfigure}
    \caption{Hypersonic flow over a half cylinder; density contours (1.4:0.27:8.7); variant 2, (Left) First-order, (Right) Second-order}
    \label{2D_figure13}
\end{figure}

\begin{figure}[H]
    \centering
    \begin{subfigure}{0.45\textwidth}
        \centering
        \includegraphics[height = 8 cm]{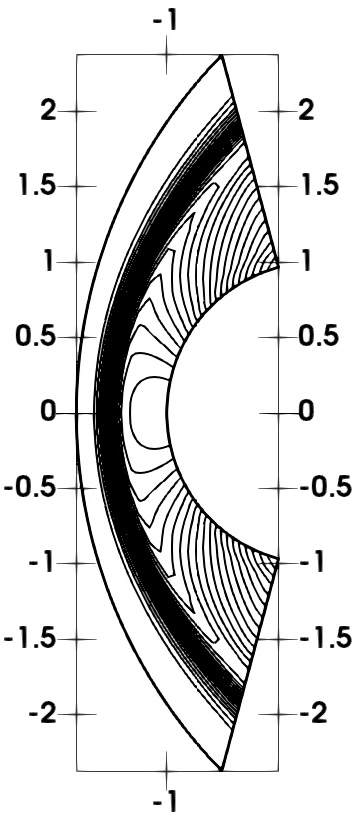}
    \end{subfigure}
    \hspace{-2 cm}
    \begin{subfigure}{0.45\textwidth}
        \centering
        \includegraphics[height = 8 cm]{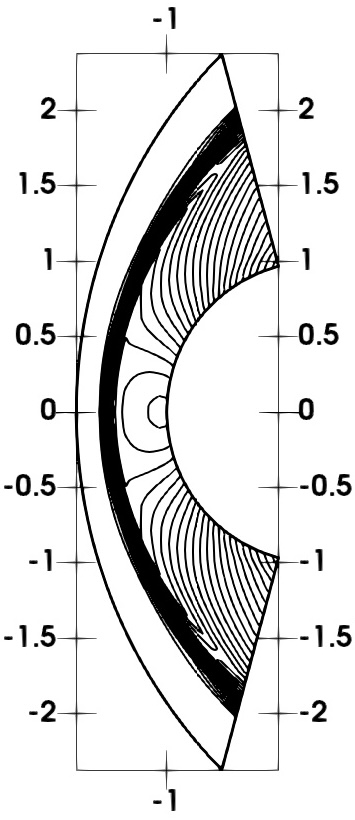}
    \end{subfigure}
    \caption{Hypersonic flow over a half cylinder; density contours (1.4:0.27:8.7); variant 3, (Left) First-order, (Right) Second-order}
    \label{2D_figure14}
\end{figure}

In this test, a half-cylinder is exposed to a hypersonic flow at Mach 20. It serves to evaluate the scheme's behavior regarding carbuncle instability \cite{Quirk}. We use the algorithm given in \cite{Huang_Wu} to generate the mesh for this test case. The computational domain is divided into $ 160 \times 20 $ cells. The initial state throughout the domain is set to the freestream conditions $ \rho = 1.4 $, $ u_1 = 20.0 $, $ u_2 = 0.0 $, $ p = 1.0 $. The supersonic inlet boundary condition is used at the left boundary of the domain, the kinetic characteristic boundary condition is used at the cylinder surface, and the zero gradient boundary condition is used at the outlet of the domain. Under steady conditions, a bow shock forms upstream of the half-cylinder. As shown in Figures \ref{2D_figure13} and \ref{2D_figure14}, the numerical results corresponding to both variants of the modified PVU scheme remain free from Carbuncle instability, in contrast to the HLLC scheme.

\subsection{Laminar flat-plate boundary layer}
\label{subsec 8.4}

\begin{figure}[!h]
    \centering
    \begin{subfigure}{0.49\textwidth}
        \centering
        \includegraphics[width=\textwidth]{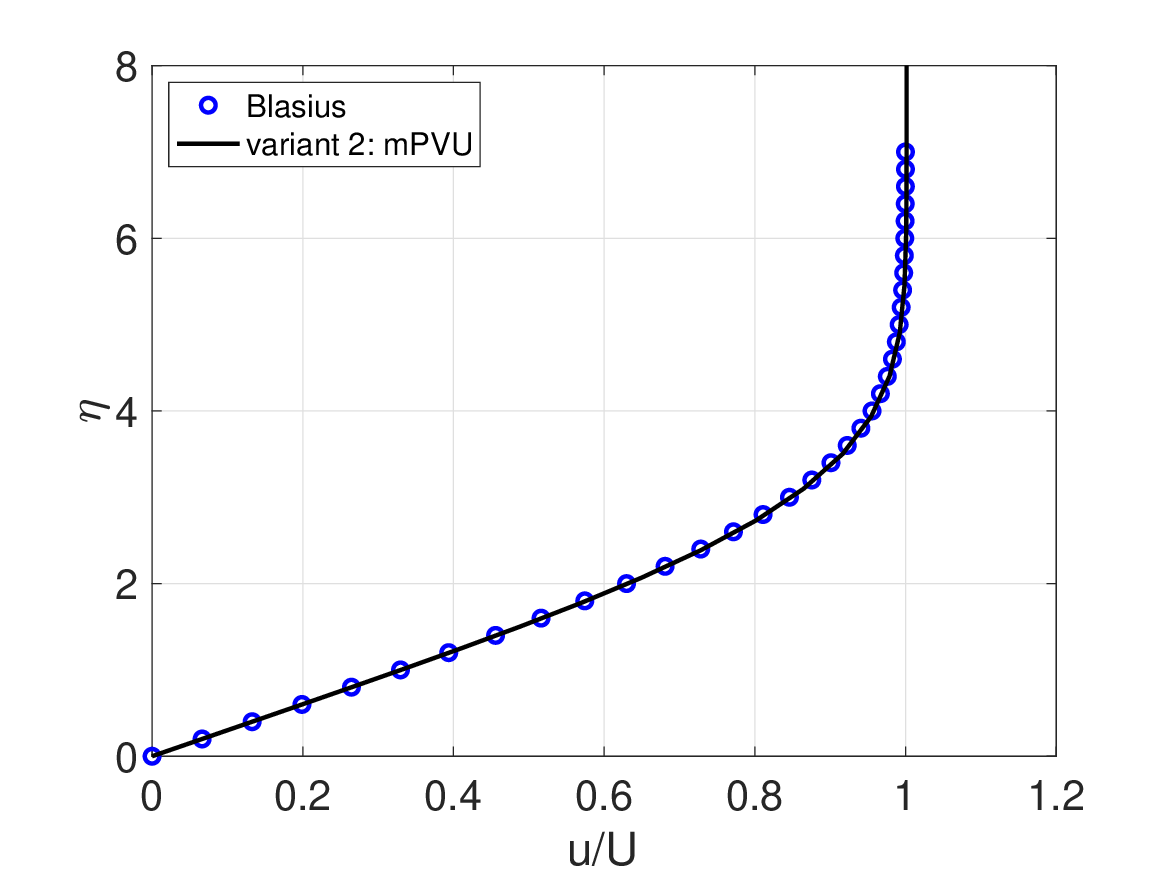}
    \end{subfigure}
    \begin{subfigure}{0.49\textwidth}
        \centering
        \includegraphics[width=\textwidth]{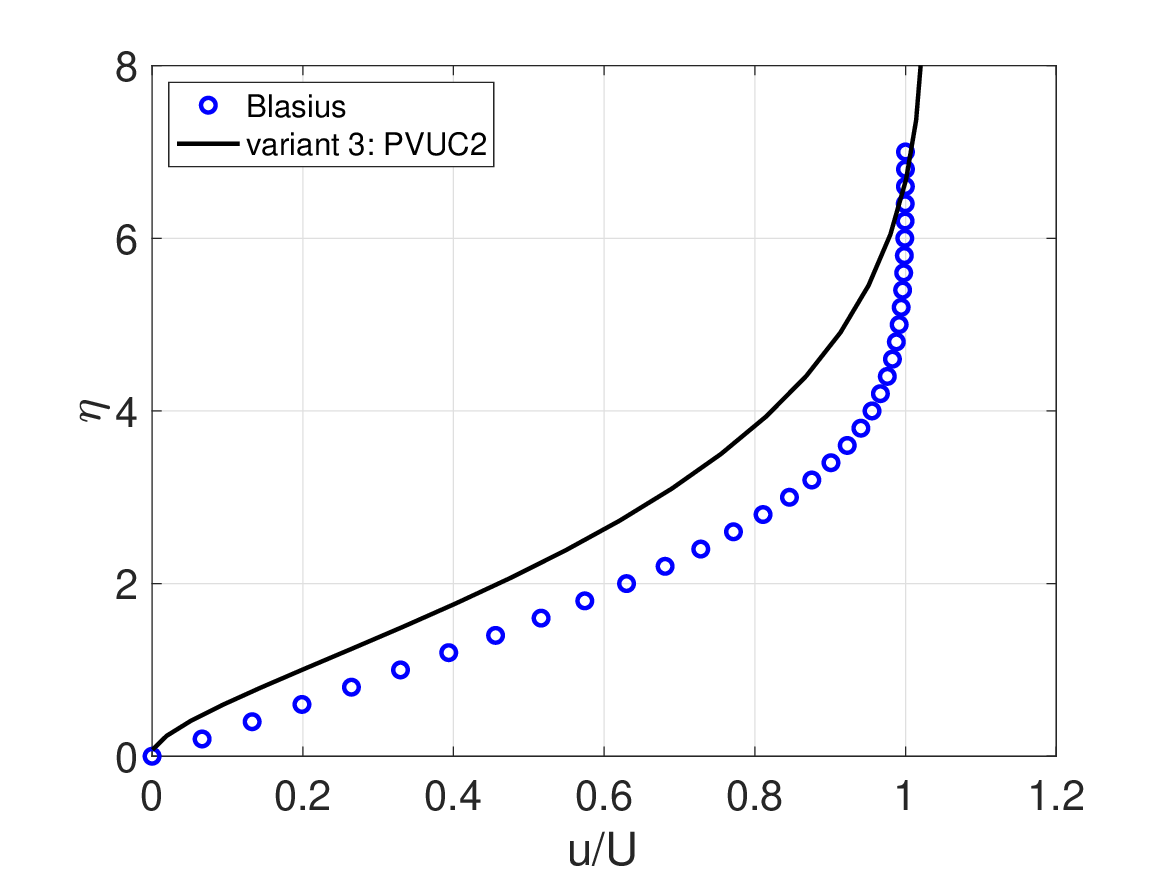}
    \end{subfigure}
    \caption{Comparison of the streamwise velocity profiles with the Blasius solution; (Left) variant 2, (Right) variant 3.}
    \label{2D_figure15}
\end{figure}
This test case considers viscous flow over a flat plate, resulting in the development of boundary layer along the plate surface. The Reynolds number, based on the plate length, is $ 10^5 $, while the free-stream Mach number is $ 0.2 $. The computational domain is a rectangle of dimensions $ [-0.5, 1.0] \times [0, 0.25] $, discretized using a Cartesian mesh of $ 300 \times 50 $ cells. The grid spacing is uniform in the x-direction, whereas the cells are geometrically stretched in the y-direction with a growth factor of $ 8 \% $. An initial section of the lower boundary is treated as a slip wall, followed by a no-slip adiabatic wall representing the flat plate. At the inlet, free-stream conditions are imposed and the numerical flux is evaluated using the prescribed free-stream state. At the upper and outlet boundaries, the free-stream pressure is specified. The computed streamwise velocity profiles are compared with the Blasius similarity solution in standard non-dimensional variables, as shown in Figure~\ref{2D_figure15}. The comparison is performed along the vertical line at $ x = 0.8 $. It is evident that variant 2 of the modified PVU scheme accurately resolves the boundary layer profile, whereas variant 3 introduces excessive numerical dissipation.

\subsection{Shock wave-Boundary layer interaction}
\label{subsec 8.5}
\begin{figure}[h]
    \centering
    \begin{subfigure}{0.48\textwidth}
        \centering
        \includegraphics[width=\textwidth]{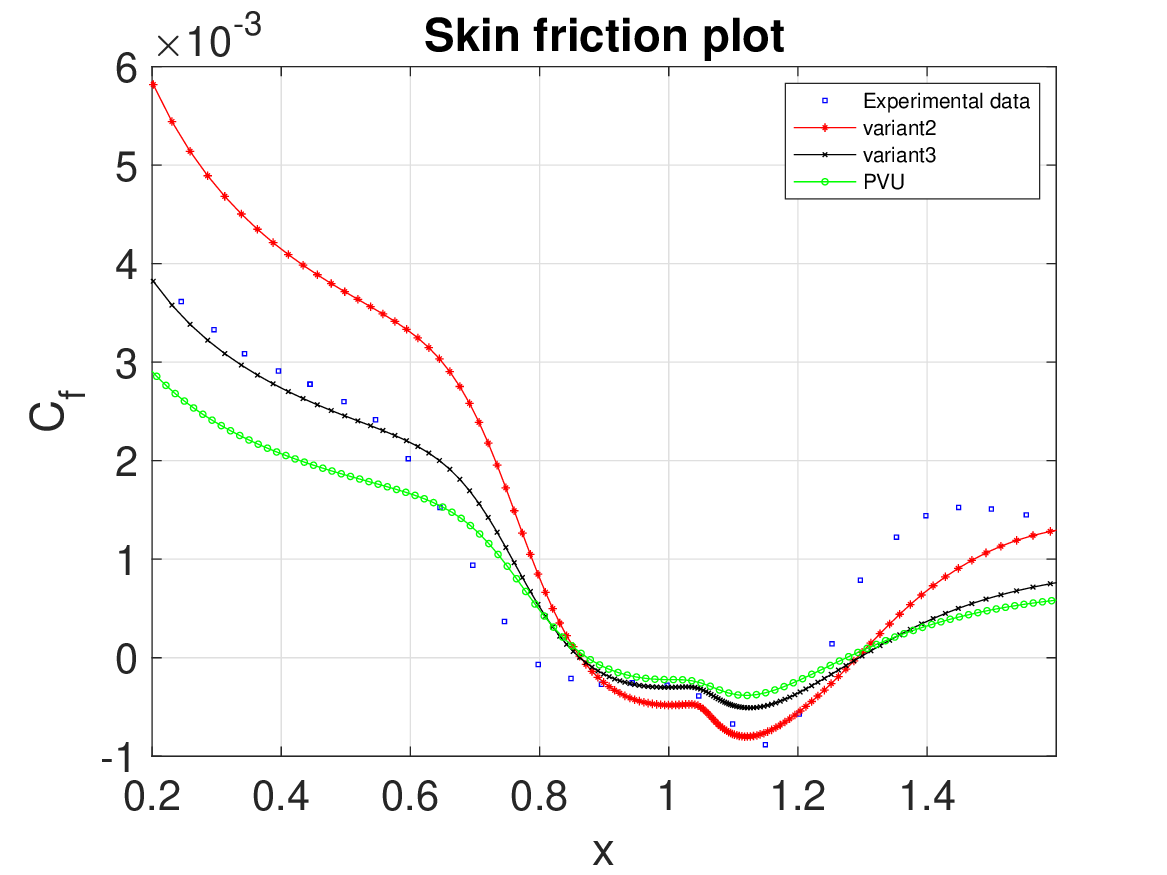}
    \end{subfigure}
    \begin{subfigure}{0.48\textwidth}
        \centering
        \includegraphics[width=\textwidth]{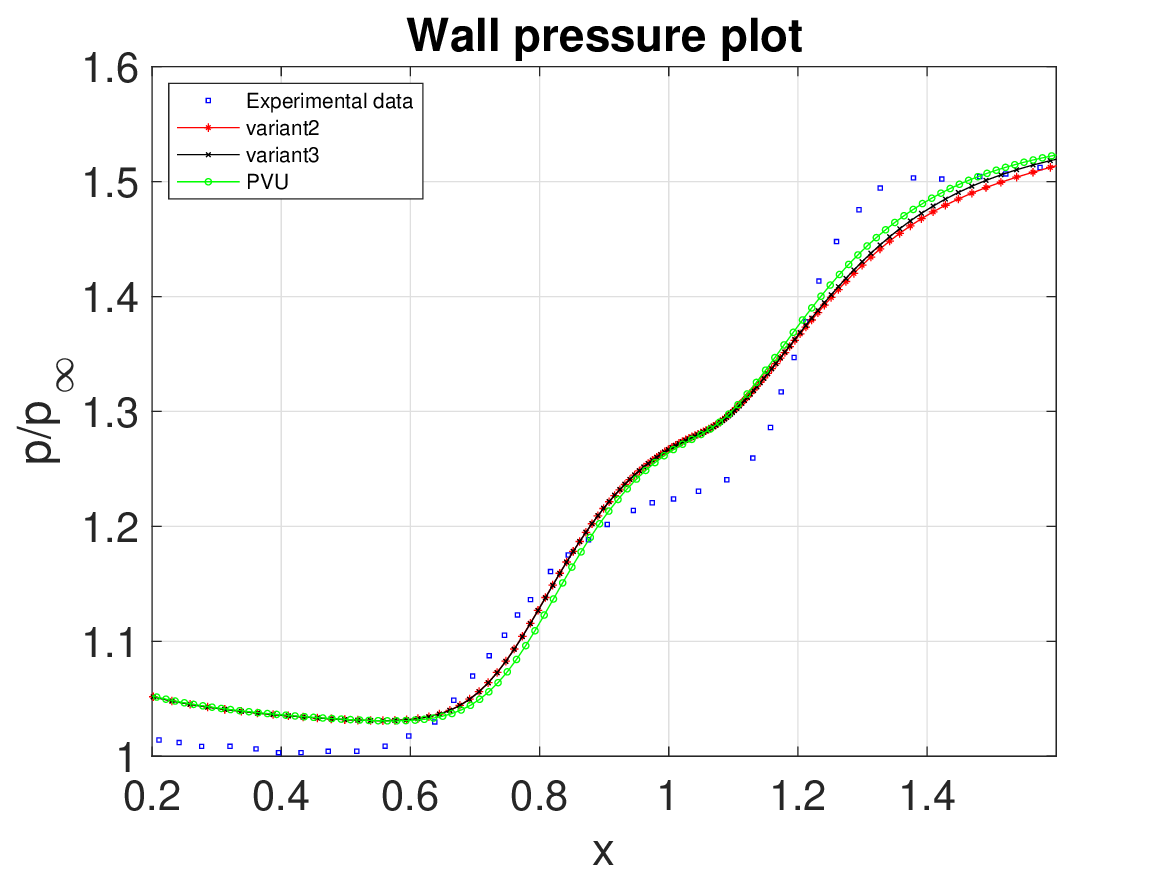}
    \end{subfigure}
    \caption{Shock wave-Boundary layer interaction test case; (Left) Skin friction plot, (Right) Wall pressure plot}
    \label{2D_figure17}
\end{figure}

The interaction between the shock wave and the boundary layer, two relatively well-understood phenomena in fluid dynamics, often leads to unexpected outcomes such as boundary layer separation. A subtle increase in pressure at the outer edge of the boundary layer leads to deceleration of the fluid particles inside the boundary layer, thus displacing the streamlines away from the wall. This leads to a further increase in pressure in the external supersonic flow, and the chain of events repeats itself, eventually separating the boundary layer \cite{Anatoly_ruban_vol_3}. A standard benchmark for assessing a numerical scheme's ability to predict negative skin friction is the self-induced boundary layer separation. In this test, an oblique shock impinges on a flat plate with a developing supersonic boundary layer. A free-stream Mach number of $ 2.15 $ is considered, with the corresponding shock angle set to $ 30.8^{\circ} $. The incoming shock interacts with the boundary layer, leading to its self-induced separation. The computational domain is a rectangle of dimensions $ [-0.2, 1.8] \times [0, 1] $. A cartesian mesh of $ 140 \times 120 $ cells covers the domain. The cell size is constant in the x-direction, whereas the y-direction features a geometric progression with a regular growth factor of $ 4.5 \% $. At the inlet, the pre-shock state is prescribed below $ y = 0.765 $, whereas the post-shock state is imposed on the remaining portion of the left boundary and throughout the right boundary. The symmetric boundary conditions are applied on the bottom wall until $ x = 0 $; the no-slip boundary condition is used on the rest of the wall. The zero-gradient boundary condition is assumed at the right boundary.
\begin{figure}[h]
    \centering
    \begin{subfigure}{0.50\textwidth}
        \includegraphics[height=3.0cm]{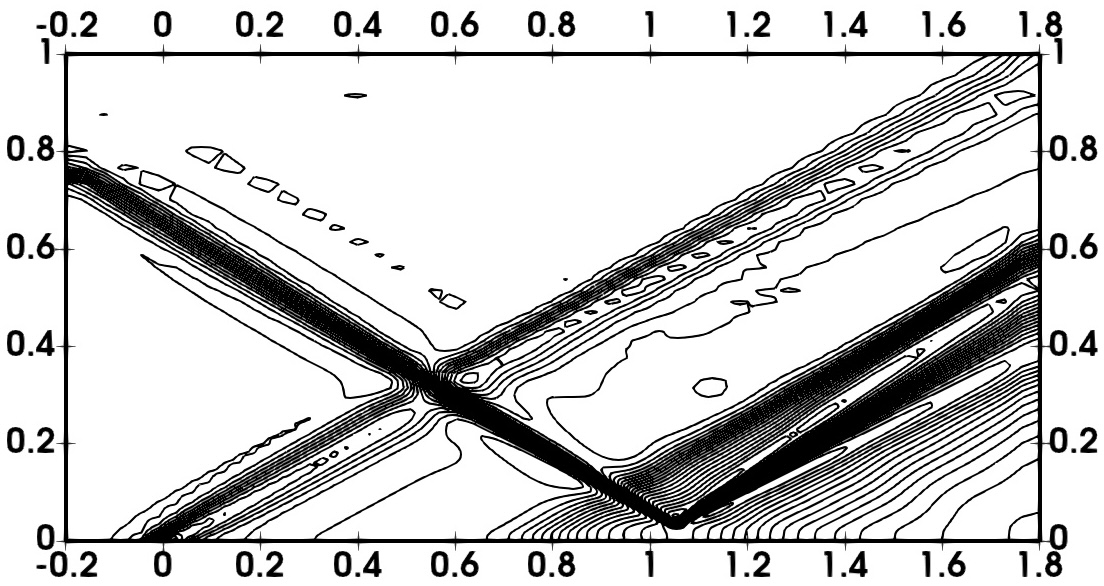}
    \end{subfigure}
    \begin{subfigure}{0.49\textwidth}
        \hspace{-1.3 cm}
        \includegraphics[height=2.4cm]{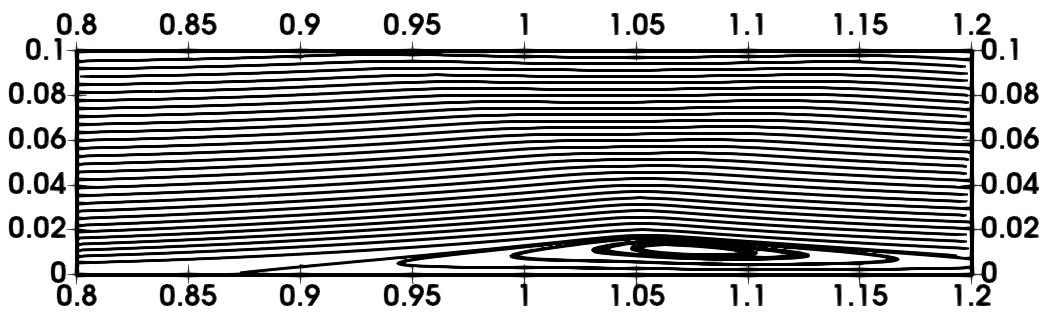}
    \end{subfigure}
    \caption{Shock wave-Boundary layer interaction test case; Pressure contours (Left), Streamlines showing the recirculation zone (Right).}
    \label{2D_figure16}
\end{figure}
Figure \ref{2D_figure16} shows the pressure contours and streamlines obtained using variant 2 of the modified PVU Boltzmann scheme. The results demonstrate accurate resolution of the reflected compression/expansion waves and the recirculation region within the separated boundary layer. Figure \ref{2D_figure17} gives the skin friction coefficient $ C_f $ and the wall pressure $ p/p_\infty $ for the shock-boundary layer interaction test.
The wall pressure distributions are nearly indistinguishable between the three schemes. In contrast, the skin friction plot clearly distinguishes the three schemes: the traditional PVU Boltzmann scheme yields the most diffusive prediction, while variant 3 -- which captures stationary contact discontinuities exactly but does not preserve slip surfaces in the inviscid case --  performs inferior to variant 2. Variant 2, which resolves both one‑dimensional stationary contact discontinuities and two‑dimensional slip surfaces exactly without numerical diffusion, provides the most accurate capture of the negative skin friction coefficient. These results demonstrate that the exact preservation of slip surfaces is crucial to accurately predict the negative $ C_f $ region in this shock–boundary layer interaction. 


\section{Conclusions}
\label{sec9}

This paper highlights the development of two contact discontinuity capturing peculiar velocity-based Boltzmann schemes. The MPDE analysis of the traditional PVU scheme shows that its overly diffused performance across stationary contact discontinuities is primarily due to the non-vanishing numerical diffusion associated with the peculiar velocity-based numerical flux. The peculiar velocity-based numerical flux is therefore designed such that the numerical diffusion pertaining to it switches off across the stationary contact discontinuities and slip surfaces, ensuring their exact capture, while maintaining non-zero diffusion elsewhere for appropriate robustness across shocks and expansion fans. Both variants of the modified PVU scheme have been tested on a variety of one- and two-dimensional compressible flow problems, providing empirical evidence of their robustness, positivity-preserving nature, and entropy-satisfying behavior. We finally conclude that the schemes which capture the steady, two-dimensional slip surfaces exactly are more effective in capturing the negative skin friction coefficient in the shock wave-boundary layer interaction test case.


\section*{CRediT author statement}

\textbf{Balwinder Singh:} Conceptualization, Methodology, Investigation, Software, Validation, Formal analysis, and preparation of the original draft.

\vspace{1 em}
\textbf{S. V. Raghurama Rao:} Conceptualization, Investigation, Formal Analysis, Supervision, and manuscript review and editing.

\section*{Declaration of competing interest}

The authors report no conflicts of interest, financial or otherwise, that could have influenced the findings of this study.

\section*{Data availability}

The study reported in this article did not involve the use of any data.

\section*{Acknowledgments}

This research did not receive specific grants from funding agencies in the public, commercial, or non-profit sectors.


\appendix
\section{Details of the 1D numerical flux}
\textbf{Variant 1:} 
\begin{equation}
    \hat{\textbf{f}}_{j + \frac{1}{2}}^{(a)} = u_j^+ \textbf{U}_j + u_{j+1}^- \textbf{U}_{j+1}
\end{equation}

\begin{equation}
    \hat{\textbf{f}}_{j + \frac{1}{2}}^{(p)} = \frac{1}{2} \left( \textbf{G}_j^{(p)} + \textbf{G}_{j+1}^{(p)} \right) - \textbf{d}_{j + \frac{1}{2}}^{(p)}
\end{equation}
where $ \textbf{d}_{j + \frac{1}{2}}^{(p)} $ denotes the pressure diffusive flux that reads

\renewcommand{\arraystretch}{2.5}
\begin{equation}
    \textbf{d}_{j + \frac{1}{2}}^{(p)} = \begin{bmatrix}
        \left( \dfrac{\gamma}{2 \pi} \right)^{\frac{1}{2}} \dfrac{1}{2 \Bar{a}} \Delta_{j + \frac{1}{2}} \left( p \right) \\
        \left( \dfrac{\gamma}{2 \pi} \right)^{\frac{1}{2}} \dfrac{1}{2 \Bar{a}} \Delta_{j + \frac{1}{2}} (pu) \\ 
        \left( \dfrac{\gamma}{2 \pi} \right)^{\frac{1}{2}} \dfrac{1}{4 \Bar{a}} \Delta_{j + \frac{1}{2}} (pu^2) + \dfrac{(\gamma + 1)}{4 (\gamma - 1)} \left( \dfrac{2}{\gamma \pi} \right)^{\frac{1}{2}} \dfrac{1}{2 \Bar{a}} \Delta_{j + \frac{1}{2}} (pa^2)
    \end{bmatrix}.
\end{equation}

\vspace{1\baselineskip}
\textbf{Variant 2:}
\begin{equation}
    \hat{\textbf{f}}_{j + \frac{1}{2}}^{(a)} = \frac{1}{2} \left( \textbf{G}_j^{(a)} + \textbf{G}_{j+1}^{(a)} \right) - \textbf{d}_{j + \frac{1}{2}}^{(a)}
\end{equation}
where $ \textbf{d}_{j + \frac{1}{2}}^{(a)} $ denotes the advective diffusive flux that reads
\begin{equation}
    \textbf{d}_{j + \frac{1}{2}}^{(a)} = \frac{1}{2} \left( \max \{ |u_j|, |u_{j + 1}| \} + \frac{\sqrt{\rho_j} u_j + \sqrt{\rho_{j + 1}} u_{j + 1}}{\sqrt{\rho_j} + \sqrt{\rho_{j + 1}}} \right) \left( \textbf{U}_{j+1} - \textbf{U}_j \right).
\end{equation}

\begin{equation}
    \hat{\textbf{f}}_{j + \frac{1}{2}}^{(p)} = \frac{1}{2} \left( \textbf{G}_j^{(p)} + \textbf{G}_{j+1}^{(p)} \right) - \textbf{d}_{j + \frac{1}{2}}^{(p)},
\end{equation}
where
\begin{equation}
    \textbf{d}_{j + \frac{1}{2}}^{(p)} = \begin{bmatrix}
        \Delta_{j+ \frac{1}{2}} \left( \dfrac{\rho}{2 \sqrt{\pi \beta}} \xi \right) \\
        \Delta_{j+ \frac{1}{2}} \left( \dfrac{\rho u}{2 \sqrt{\pi \beta}} \xi \right) \\
        \Delta_{j+ \frac{1}{2}} \left( \dfrac{\rho u^2}{4 \sqrt{\pi \beta}} \xi \right) + \dfrac{(3 - \gamma)}{4 (\gamma - 1)} \Delta_{j + \frac{1}{2}} \left( \dfrac{p}{\sqrt{\pi \beta}} \xi \right) + \Delta_{j + \frac{1}{2}} \left( \dfrac{p}{2 \sqrt{\pi \beta}} \xi^2 \right) \\
    \end{bmatrix},
\end{equation}
where $ \xi $ represents
\begin{equation}
    \xi = \dfrac{\gamma}{2} \dfrac{M^2}{\left( 1 + \dfrac{\gamma}{2} M^2 \right)}, \; M = \dfrac{u}{a}.
\end{equation}
We also present a detailed derivation of the key terms that often appear in the numerical flux of the modified peculiar velocity-based upwind Boltzmann scheme: \\
In the mass flux, we have the following term in the diffusive flux:
\begin{equation}
    \int_{-\infty}^{\infty} \, dc \, \dfrac{|c|}{2} \, \phi^{(c)} \, \bar{f}^{eq} = \dfrac{\rho}{2} \dfrac{\sqrt{\beta}}{\sqrt{\pi}} \int_{-\infty}^{\infty} \, dc \, |c| \, \exp{\left( \dfrac{-c^2}{|u|^2} \right)} \, \exp{(-\beta c^2)}
\end{equation}
Since every function appearing in the integrand is even, the integral simplifies to twice the integral over the positive half-line:
\begin{subequations}
    \begin{align}
        \int_{-\infty}^{\infty} \, dc \, \dfrac{|c|}{2} \, \phi^{(c)} \, \bar{f}^{eq} &= \rho \dfrac{\sqrt{\beta}}{\sqrt{\pi}} \int_{0}^{\infty} \, dc \, c \, \exp{\left( \dfrac{-c^2}{|u|^2} \right)} \, \exp{(-\beta c^2)} \\
        &= \dfrac{\rho}{2 \sqrt{\pi \, \beta}} \xi.
    \end{align}   
\end{subequations}
In the energy flux, one of the terms in the diffusive flux reads
\begin{subequations}
    \begin{align}
        \int_{-\infty}^{\infty} \, dc \, \dfrac{|c|}{4} \, c^2 \, \phi^{(c)} \, \bar{f}^{eq} &= \rho \dfrac{\sqrt{\beta}}{\sqrt{\pi}} \int_{0}^{\infty} \, dc \, c^3 \, \exp{\left( \dfrac{-c^2}{|u|^2} \right)} \, \exp{(-\beta c^2)} \\
        &= \dfrac{p}{2 \sqrt{\pi \, \beta}} \xi^2.
    \end{align}   
\end{subequations}

\vspace{1\baselineskip}
\textbf{Variant 3:}
The diffusive flux pertaining to the numerical pressure flux of variant 3 reads
\begin{equation}
    \textbf{d}_{j + \frac{1}{2}}^{(p)} = \begin{bmatrix}
        \left( \dfrac{\gamma}{2 \pi} \right)^{\frac{1}{2}} \dfrac{1}{2 \Bar{a}} \Delta_{j + \frac{1}{2}} \left( p \right) \\
        \left( \dfrac{\gamma}{2 \pi} \right)^{\frac{1}{2}} \dfrac{1}{2 \Bar{a}} \Delta_{j + \frac{1}{2}} (pu) \\ 
        \left( \dfrac{\gamma}{2 \pi} \right)^{\frac{1}{2}} \dfrac{1}{4 \Bar{a}} \Delta_{j + \frac{1}{2}} (pu^2) + \dfrac{(\gamma + 1)}{4 (\gamma - 1)} \left( \dfrac{\gamma}{2 \pi} \right)^{\frac{1}{2}} \Delta_{j + \frac{1}{2}} (p \Bar{a})  
    \end{bmatrix}.
\end{equation}

\section{Details of the 2D numerical flux}
\textbf{Variant 1:} 
\begin{equation}
    \hat{\textbf{f}}_{e}^{(a)} = \dfrac{1}{2} \left\{ (u_n \boldsymbol{U})_K + (u_n \boldsymbol{U})_{K_e} \right\} - \boldsymbol{d}^{(a)}_e,
\end{equation}
where $ \boldsymbol{d}_e^{(a)} $ denotes the numerical diffusion corresponding to the advective flux that reads
\begin{multline}
    \boldsymbol{d}^{(a)}_e = \dfrac{1}{2} \left( \max \{ |u_n|_K, |u_n|_{K_e} \} + \frac{\sqrt{\rho_K} (u_n)_K + \sqrt{\rho_{K_e}} (u_n)_{K_e}}{\sqrt{\rho_K} + \sqrt{\rho_{K_e}}} \right) \left( \boldsymbol{U}_{K_e} - \boldsymbol{U}_K \right).
\end{multline}

\begin{equation}
    \hat{\textbf{f}}_{e}^{(p)} = \frac{1}{2} \left( \textbf{G}_{n, K}^{(p)} + \textbf{G}_{n, K_e}^{(p)} \right) - \textbf{d}_e^{(p)},
\end{equation}
where $ \textbf{d}_{e}^{(p)} $ denotes the pressure diffusive flux that reads
\renewcommand{\arraystretch}{2.5}
\begin{equation}
    \textbf{d}_{e}^{(p)} = \begin{bmatrix}
        \left( \dfrac{\gamma}{2 \pi} \right)^{\frac{1}{2}} \dfrac{1}{2 \Bar{a}} \Delta_{e} \left( p \right) \\
        \left( \dfrac{\gamma}{2 \pi} \right)^{\frac{1}{2}} \dfrac{1}{2 \Bar{a}} \Delta_{e} \left( p u_n \right) \\
        \left( \dfrac{\gamma}{2 \pi} \right)^{\frac{1}{2}} \dfrac{1}{2 \Bar{a}} \Delta_{e} \left( p u_t \right) \\
        \left( \dfrac{\gamma}{2 \pi} \right)^{\frac{1}{2}} \dfrac{1}{4 \Bar{a}} \Delta_{e} (p | \boldsymbol{u} |^2) + \dfrac{(\gamma + 1)}{4 (\gamma - 1)} \left( \dfrac{2}{\gamma \pi} \right)^{\frac{1}{2}} \dfrac{1}{2 \Bar{a}} \Delta_{e} (p a^2)
    \end{bmatrix}.
\end{equation}

\vspace{1\baselineskip}
\textbf{Variant 2:}
The advective flux for variant 2 is the same as that of variant 1. The pressure flux reads
\begin{equation}
    \hat{\textbf{f}}_{e}^{(p)} = \frac{1}{2} \left( \textbf{G}_{n, K}^{(p)} + \textbf{G}_{n, K_e}^{(p)} \right) - \textbf{d}_{e}^{(p)},
\end{equation}
where
\begin{equation}
    \textbf{d}_{e}^{(p)} = \begin{bmatrix}
        \Delta_{e} \left( \dfrac{\rho}{2 \sqrt{\pi \beta}} \xi_n \right) \\
        \Delta_{e} \left( \dfrac{\rho u_n}{2 \sqrt{\pi \beta}} \xi_n \right) \\
        \Delta_{e} \left( \dfrac{\rho u_t}{2 \sqrt{\pi \beta}} \xi_n \right) \\
        \Delta_{e} \left( \dfrac{\rho | \boldsymbol{u} |^2}{4 \sqrt{\pi \beta}} \xi_n \right) + \dfrac{(3 - \gamma)}{4 (\gamma - 1)} \Delta_{e} \left( \dfrac{p}{\sqrt{\pi \beta}} \xi_n \right) + \Delta_{e} \left( \dfrac{p}{2 \sqrt{\pi \beta}} \xi_n^2 \right) \\
    \end{bmatrix},
\end{equation}
where $ \xi $ represents
\begin{equation}
    \xi_n = \dfrac{\gamma}{2} \dfrac{M_n^2}{\left( 1 + \dfrac{\gamma}{2} M_n^2 \right)}, \; M_n = \dfrac{u_n}{a}.
\end{equation}

\section{1D positivity analysis of traditional PVU Boltzmann scheme}
\noindent
\textbf{Theorem 1.} Assume $ f^{eq}_j = 0 $ for $ |c_j| \geq |c_j|_{max} $, then under the CFL-like condition
\begin{equation} \label{CFL-like_condition}
    (\max_{j} \{ |u_j| + |c_j|_{max} \}) \Delta t \leq \Delta x,
\end{equation}
the peculiar velocity-based upwind Boltzmann scheme (\ref{1D_finite_volume_update_formula})-(\ref{Definition_of_PV_psi}) is positive, \textit{i.e.}, $ \rho \geq 0, p \geq 0 $. 

\vspace{1\baselineskip}
\noindent
Proof: We rewrite (\ref{discretized_PV_Boltzmann_equation}) using (\ref{definition_advective_pressure_numerical_flux}) in finite difference form, which reads
\begin{equation}
    f_j^{(n+1)-} = f_j^{eq, (n)} (1 - \sigma (|u_j| + |c_j|)) + \sigma (h_{j-1}^+ + h_{j+1}^-), \, \sigma = \frac{\Delta t}{\Delta x},
\end{equation}
with
\begin{subequations} \label{Defintions_of_h+_and_h-}
    \begin{align}
        h_{j-1}^+ &= u_{j-1}^+ f_{j-1}^{eq, (n)} + c_{j-1}^+ f_{j-1}^{eq, (n)} \\
        h_{j+1}^- &= -u_{j+1}^- f_{j+1}^{eq, (n)} - c_{j+1}^- f_{j+1}^{eq, (n)}
    \end{align}
\end{subequations}
We note that
\begin{subequations}
    \begin{align}
        h_{j-1}^+ &\geq 0, \\
        h_{j+1}^- &\geq 0.
    \end{align}
\end{subequations}
We only have to consider the peculiar velocities for which $ f^{eq}_j > 0 $; these are given by
\begin{equation} \label{peculiar_velocities_for_which_f_is_positive}
    |c_j| \leq |c_j|_{max}
\end{equation}
We add $ |u_j| $ to both sides of (\ref{peculiar_velocities_for_which_f_is_positive}) and note that
\begin{equation}
    \sigma (|u_j| + |c_j|) \leq \sigma (\max_{j} \{ |u_j| + |c_j|_{max} \}) \leq 1
\end{equation}
Hence, we have $ f_j^{(n+1)-} \geq 0 $ and thus
\begin{subequations}
    \begin{align}
        \rho_j^{(n+1)} &= \int_{c} \, dc \, f_j^{(n+1)-} \geq 0, \\
        p_j^{(n+1)} &= \int_{c} \, dc \, c_j^2 \, f_j^{(n+1)-} \geq 0.
    \end{align}
\end{subequations}
and 1. is proved.

\section{2D positivity analysis of traditional PVU Boltzmann scheme}
\noindent
\textbf{Theorem 2.} Assume that $ f^{eq}_K = 0 $ for $ |(c_{n_e})_K| \geq |(c_{n_e})_K|_{max} $, where $ c_{n_e} $ denotes the peculiar velocity in a direction normal to a cell interface $ e $. Then, under the CFL-like condition,
\begin{equation} \label{sec:3.4_CFL-like_condition}
    \frac{\Delta t}{2 m(K)} \max_{K} \left\{ \sum_{e} \left( |(u_{n_e})_K| + |(c_{n_e})_K|_{max} \right)m(e) \right\} \leq 1,
\end{equation}
the two-dimensional PVU Boltzmann scheme is positive.
\vskip 1 em
\noindent
Proof: We have, from (\ref{discretized_2D_PV_Boltzmann_equation})
\begin{equation} \label{sec:3.4_update_formula_for_density}
    f^{(n + 1)-} = f_K^{eq, \, (n)} - \frac{\Delta t}{m(K)} \sum_{e} m(e) \, \hat{h}^{(n)}_e,
\end{equation}
where
\begin{equation}
    \hat{h}^{(n)}_e = \left( u_{n_e}^+ + c_{n_e}^+ \right) f^{eq, \, (n)}_K + \left( u_{n_e}^- + c_{n_e}^- \right) f^{eq, \, (n)}_{K_e}.
\end{equation}
We note that
\begin{equation}
    \boldsymbol{u} \cdot \sum_{e} \hat{\boldsymbol{n}}_e \, m(e) = 0, \, \boldsymbol{c} \cdot \sum_{e} \hat{\boldsymbol{n}}_e \, m(e) = 0.
\end{equation}
We rewrite equation \ref{sec:3.4_update_formula_for_density} in view of the above relation as follows:
\begin{multline} \label{eq:final}
    f^{(n + 1)-} = f_K^{eq, \, (n)} \left( 1 - \frac{\Delta t}{2 m(K)} \sum_{e} m(e) \, \left( | u_{n_e} | + | c_{n_e} | \right) \right) \\
    - \frac{\Delta t}{m(K)} \sum_{e} m(e) \, \left( u_{n_e}^- + c_{n_e}^- \right) \, f^{eq, \, (n)}_{K_e}.
\end{multline}
We note that
\begin{equation}
    -\frac{\Delta t}{m(K)} \sum_{e} m(e) \, \left( u_{n_e}^- + c_{n_e}^- \right) \, f^{eq, \, (n)}_{K_e} \geq 0.
\end{equation}
We only have to consider such peculiar velocities for which $ f^{eq} > 0 $, \textit{i.e.}, 
\begin{equation}
    |(c_{n_e})_K| \leq |(c_{n_e})_K|_{max}.
\end{equation}
We now add $ | (u_{n_e})_K | $ to both sides of the above equation and note that
\begin{equation}
    \sum_{e} m(e) \, \left( |(u_{n_e})_K| + |(c_{n_e})_K| \right) \leq \max_{K} \left\{ \sum_{e} \left( |(u_{n_e})_K| + |(c_{n_e})_K|_{max} \right) m(e) \right\}.
\end{equation}
Then the CFL-like condition \ref{sec:3.4_CFL-like_condition} dictates that
\begin{equation}
    f^{(n + 1)-} \geq 0,
\end{equation}
and thus
\begin{subequations}
    \begin{align}
        \rho^{(n + 1)} &= \int_{I} \, dI \int_{c_t} \, dc_t \int_{c_n} \, dc_n \, f^{(n + 1)-} \geq 0 \\
        p^{(n + 1)} &= \int_{I} \, dI \int_{c_t} \, dc_t \int_{c_n} \, dc_n \, c^2 \, f^{(n + 1)-} \geq 0.
    \end{align}
\end{subequations}

\end{document}